\documentclass[pre,twocolumn,datetime]{revtex4-1}  
\usepackage{amsfonts}
\usepackage{amsmath}
\usepackage{amssymb}
\usepackage{version}
\usepackage{cancel}
\usepackage{graphicx}%
\includeversion{New_connection}
\excludeversion{Old_connection}

\begin{document}

\noindent
Journal of Statistical Physics {\bf 191}, 137 (2024)\\
https://doi.org/10.1007/s10955-024-03345-1\\
$\;$\\
$\;$\\
$\;$





\title{On the Definition of Velocity in Discrete-Time, Stochastic Langevin Simulations}
\author{Niels Gr{\o}nbech-Jensen}
\email{ngjensen@math.ucdavis.edu}
\affiliation{Department of Mechanical \& Aerospace Engineering 
\\ Department of Mathematics\\ University of California, Davis, CA 95616, U.S.A.}

\begin{abstract}



\noindent
We systematically develop beneficial and practical velocity measures for accurate and efficient statistical simulations of the Langevin equation with direct applications to computational statistical mechanics and molecular dynamics sampling. Recognizing that the existing velocity measures for the most statistically accurate discrete-time Verlet-type algorithms are inconsistent with the simulated configurational coordinate, we seek to create and analyze new velocity companions that both improve existing methods as well as offer practical options for implementation in existing computer codes. The work is based on the set of GJ methods that, of all methods, for any time step within the stability criteria correctly reproduces the most basic statistical features of a Langevin system; namely correct Boltzmann distribution for harmonic potentials and correct transport in the form of drift and diffusion for linear potentials. Several new and improved velocities exhibiting correct drift are identified, and we expand on an earlier conclusion that, generally, only half-step velocities can exhibit correct, time-step independent Maxwell-Boltzmann distributions. Specific practical and efficient algorithms are given in familiar forms, and these are used to numerically validate the analytically derived expectations. One especially simple algorithm is highlighted, and the ability of one of the new on-site velocities to produce statistically correct averages for a particular damping value is specified.
\end{abstract}

\maketitle
\section{Introduction}
\label{sec:intro}
Numerical methods for simulating the Langevin equation have evolved considerably over the past decade. One particular topical driver of this interest has been computational statistical mechanics, exemplified by molecular dynamics in areas of, e.g., materials science, soft matter, and biomolecular modeling \cite{AllenTildesley,Frenkel,Rapaport,Hoover_book,Leach}, where the balance between computational efficiency and accuracy is a core concern due to the complexity of simulation requirements.
The recent advances in algorithmic techniques for simulating the Langevin equation are found not in fundamentally new methods, but instead in better understanding the opportunities and features presented by tuning the parameters of the stochastically perturbed Newton-St{\o}rmer-Verlet \cite{Toxsvaerd,Stormer_1921,Verlet} discrete-time approximation to continuous time Langevin behavior. This understanding has led to algorithm revisions that allow for simulations that maintain statistical accuracy using considerably larger time steps than previous methods offered. In the following we will for brevity refer to the Newton-St{\o}rmer-Verlet method as simply the Verlet method. Various strategies to optimize the numerical methods for thermal systems include {\it i}) Strang splitting \cite{Strang,Ricci,LM} of the inertial, interactive, and thermodynamic operators of the evolution, {\it ii}) instantaneous application \cite{BBK,Pastor_88} of noise and friction to each time-step of the Verlet method, {\it iii}) one-time-step integration of the Langevin equation \cite{SS,vgb_1982,GJF1}, and {\it iv}) direct parameter fitting \cite{2GJ,GJ} of a generalized Verlet-type expression to optimize pre-determined objectives. More thorough lists of various methods and derivations can be found in, e.g., Refs.~\cite{Finkelstein_1,GJ,Sivak}. These different kinds of derivation techniques lead to very similar stochastic Verlet-type methods, only distinguished by the coefficients to the same kinds of terms, and all converging to the standard Verlet method when the damping coefficient vanishes. Yet, the features of the resulting algorithms, and therefore the simulation outcomes, can exhibit significantly different errors for increasing time step, as can be seen in, e.g., Refs.~\cite{2GJ,Finkelstein_1,Josh_2020}. When approximating continuous-time systems, the inevitable discrete-time errors of the chosen algorithm should therefore be understood when selecting an appropriate time-step, which on one hand must be chosen large enough for time-efficient simulations, and on the other hand chosen small enough to make the simulation results meaningful in the context of the application. For studies in computational statistical mechanics, i.e., the Langevin equation, it is commonly the case that individual simulated trajectories are not of direct interest, but instead only their statistical ensembles have significance. For such cases, a significant development came with the GJF method \cite{GJF1}, which in discrete time reproduces precise configurational Boltzmann statistics for the noisy, damped harmonic oscillator as well as correct drift and diffusion transport on linear potentials, regardless of the applied time step for as long as the simulation is within the stability conditions; even if each individually sampled trajectory may be highly inaccurate for large time steps. Later, a very similar method, VRORV  \cite{Sivak}, with the same statistical configurational properties as the GJF method, was developed based on a time-scale revision to a previously developed split-operator method \cite{LM}, which has time-step dependent transport results. The two methods, GJF and VRORV, were recognized in Ref.~\cite{GJ} as belonging to a set (GJ) of very similar methods, all with the mentioned desirable statistical properties of the configurational coordinate, and it was argued that this set includes all stochastic Verlet-type methods with these configurational properties. We therefore limit the focus of this paper to the GJ methods since these methods possess the most of the basic statistically important features of a simulated configurational coordinate.

Common for discrete-time simulations is that not only do errors appear in the simulated configurational coordinate, but they, of course, also appear in the associated kinetic coordinate, the velocity. Additionally, these errors tend to be inconsistent, such that the erroneous velocity measure is not only incorrect, but also incorrectly measuring the velocity of the incorrect configurational coordinate; see, e.g., appendix in Ref.~\cite{2GJ} for an example of the harmonic oscillator. One should therefore not expect that correct statistics of the configurational coordinate implies that correct statistics of the associated kinetic coordinate follows. Indeed, the statistical measures of the configurational GJ methods, of which GJF is one special case and VRORV is another, generally yield incorrect drift velocities, both when measured by on-site and half-step measures relative to the configurational coordinate. Only the GJF method measures the drift correctly by the on-site velocity, which is shown to not yield correct Maxwell-Boltzmann statistics. For the half-step velocities, which yield correct Maxwell-Boltzmann statistics, only one of the methods in question has yielded correct drift, and this method, GJ-III, is different from the GJF method. Thus, the accompanying velocities are not statistically consistent with the trajectories. As argued in Ref.~\cite{2GJ}, discrete-time velocity measures are inherently ambiguous, allowing many different measures with different properties to be considered as companions to a given configurational evolution. One should therefore consider the definition of discrete-time velocity as an attachment to an existing algorithm for the configurational coordinate, such that it can be designed to yield certain results.

It is the aim of this work to systematically explore the possibilities for defining both on-site and half-step GJ associated velocities that simultaneously measure correct drift and correct Maxwell-Boltzmann statistics.
This is done as follows. Section~\ref{sec:Background} reviews and contextualizes key background material to provide the tools for the upcoming analysis and development as well as exemplifies the issue at hand. After listing key continuous-time expectations, the basic Verlet formalism with emphasis on velocity definitions is reviewed in Sec.~\ref{sec:Verlet}, and then, in Sec.~\ref{sec:GJ}, we give the essence of the stochastic GJ Verlet-type methods, which will be used as the foundation for the subsequent development of velocity measures. Sections~\ref{sec:Half-step_1} and \ref{sec:Half-step_2} describe the requirements and development of possible half-step velocities, whereas Sec.~\ref{sec:On-site} describes the requirements and development of possible on-site velocities. The sections include functional algorithms in different forms that can be readily implemented in existing computer codes. Finally, in Sec.~\ref{sec:Discussion}, the results are discussed and a simple combined algorithm is given for convenient time-step evolution that produces any or all of the derived velocity options. One simple expression of the algorithms is highlighted in Appendix~\ref{sec:Appendix_A}, where also a particular choice of the damping parameter produces an algorithm, in which an on-site velocity, measured at the same time as the corresponding statistically correct coordinate, produces statistically correct kinetic energies.

\section{Background}
\label{sec:Background}
The physical, continuous-time system of interest is modeled by the Langevin equation \cite{Langevin,Langevin_Eq} ,
\begin{eqnarray}
m\dot{v}+\alpha\dot{r} & = & f+\beta \; , \label{eq:Langevin}
\end{eqnarray}
for an object with mass $m>0$, location (configurational coordinate) $r$, and velocity $v=\dot{r}$. The mass is subject to a force $f=-\nabla E_p$, where $E_p(r)$ is a potential energy surface, and a linear friction force $-\alpha v$ given by the damping coefficient $\alpha$. The associated thermal noise force, $\beta$, is guided by the fluctuation-dissipation relationship that constrains the first two moments \cite{Parisi},
\begin{subequations}
\begin{eqnarray}
\langle\beta(t)\rangle & = & 0 \\
\langle\beta(t)\beta(t^\prime)\rangle & = & 2\,\alpha\, k_BT\, \delta(t-t^\prime) \; , 
\end{eqnarray}
\label{eq:FD}\noindent
\end{subequations}
where $T$ is the temperature of the heat bath, $k_B$ is Boltzmann's constant, $\delta(t)$ is Dirac's delta function, and $\langle\cdot\rangle$ represents a statistical average.

This work makes use of key results from statistical mechanics for linear systems \cite{Langevin_Eq}. These are, for a tilted potential, $f={\rm const}$ ($\alpha>0$), the drift velocity:
\begin{subequations}
\begin{eqnarray}
\frac{\langle r(t+s)-r(t)\rangle}{s} & = & \frac{f}{\alpha}\; \; , \; s\neq0 \label{eq:Drift_r}\\
\langle v(t) \rangle & = & \frac{f}{\alpha}\, ;  \label{eq:Drift_v}
\end{eqnarray}
for a Hooke spring, $f=-\kappa r$ ($\kappa>0$), the thermal correlations:
\begin{eqnarray}
\langle r(t)r(t)\rangle & = & \frac{k_BT}{\kappa}   \label{eq:Corr_rr} \\
\langle v(t)v(t)\rangle & = & \frac{k_BT}{m}  \label{eq:Corr_vv}\\
\langle v(t)r(t+s)\rangle & = & \frac{k_BT}{m\Omega_\alpha}e^{-|s|\alpha/2m}\sin{\Omega_\alpha s} \, , \label{eq:Corr_vr}
\end{eqnarray}
where $\Omega_\alpha^2=\Omega_0^2-(\frac{\alpha}{2m})^2$ with $\Omega_0=\sqrt{\kappa/m}$ being the natural frequency of the oscillator; and, for a flat potential, $f=0$ ($\alpha>0$), the Einstein diffusion:
\begin{eqnarray}
D_E & = & \lim_{s\rightarrow\infty}\frac{\left\langle(r(t+s)-r(t))^2\right\rangle}{2s} \; = \; \frac{k_BT}{\alpha}\, . \label{eq:Diff_rr}
\end{eqnarray}\label{eq:Cont_time_quantities}\noindent
\end{subequations}
For use later in this presentation, we notice that, because both $r(t)$ and $v(t)$ in Eqs.~(\ref{eq:Corr_rr})-(\ref{eq:Corr_vr}) are Gaussian stochastic variables for the harmonic potential with $\kappa>0$, we can relatively easily derive
\begin{eqnarray}
\langle v^2(t)r^2(t+s)\rangle & = & \frac{(k_BT)^2}{m\kappa}+2\,\left\langle v(t)r(t+s)\right\rangle^2\, . \label{eq:Corr_v2r2}
\end{eqnarray}
Thus, with the cross-correlation $\rho_{vr}(s)$ given by
\begin{eqnarray}
\rho_{vr}(s) & = & \frac{\langle v(t)r(t+s)\rangle-\langle v(t)\rangle\langle r(t)\rangle}{\sqrt{[\langle r(t)r(t)\rangle-\langle r(t)\rangle^2][\langle v(t)v(t)\rangle-\langle v(t)\rangle^2]}}\nonumber \\
& = &  \frac{\Omega_0}{\Omega_\alpha}e^{-|s|\alpha/2m}\sin{\Omega_\alpha s} \, , \label{eq:Cont_corr_vr}
\end{eqnarray}
we have the correlation
\begin{eqnarray}
\rho_{v^2r^2}(s) & = & \rho_{E_kE_p}(s) \; = \; \rho^2_{vr}(s)\, ,  \label{eq:Cont_corr_v2r2}
\end{eqnarray}
where $\rho_{E_kE_p}(s)$ is the correlation between kinetic and potential energies for the stochastic harmonic oscillator. Equation~(\ref{eq:Cont_corr_v2r2}) points to the importance of making sure that a simulated value of $\rho_{vr}(0)=0$, since a simulated value, $\rho_{E_kE_p}(0)=0$, is necessary for, e.g., obtaining reliable approximations to certain thermodynamic quantities in molecular modeling \cite{AllenTildesley}.
This work explores the possibilities for defining discrete-time velocities that can mimic the core relationships of Eqs.~(\ref{eq:Drift_v}), (\ref{eq:Corr_vv}), and (\ref{eq:Corr_vr}) given that the discrete-time GJ methods satisfy Eqs.~(\ref{eq:Drift_r}), (\ref{eq:Corr_rr}), and (\ref{eq:Diff_rr}); i.e., explore GJ velocity definitions that seek to simultaneously provide correct measures of drift, temperature and fluctuations, and cross-correlation with the configurational coordinate.

\subsection{Verlet Velocity Properties}
\label{sec:Verlet}
In discrete time, the Verlet method approximates the configurational solution to the frictionless ($\alpha=0$) dynamics through the second order difference equation with second order accuracy in $\Delta{t}$ \cite{Toxsvaerd,Stormer_1921,Verlet}
\begin{subequations}
\begin{eqnarray}
r^{n+1} & = & 2r^n-r^{n-1}+\frac{\Delta{t}^2}{m}f^n\, ,  \label{eq:Stormer}
\end{eqnarray}
where the superscript $n$ of the discrete-time variables $r^n$ and $f^n=f(t_n,r^n)$ denotes the discrete time $t_n=t_0+n\,\Delta{t}$, $\Delta{t}$ being the time step of the method. Defining discrete-time velocity can be done in at least two reasonable ways. One is the second order, central-difference expression for the {\it on-site} velocity \cite{AllenTildesley,Swope,Beeman}
\begin{eqnarray}
v_o^n & = & \frac{r^{n+1}-r^{n-1}}{2\,\Delta{t}}\, . \label{eq:on-site_org}
\end{eqnarray}
Another is also a second order, central-difference expression, but applied over a single time step, yielding the {\it half-step} velocity \cite{AllenTildesley,Buneman,Hockney}
\begin{eqnarray}
v_o^{n+\frac{1}{2}} & = & \frac{r^{n+1}-r^n}{\Delta{t}}\, . \label{eq:half-step_org}
\end{eqnarray}
\label{eq:Verlet}\noindent
\end{subequations}
Reference \cite{2GJ} emphasizes two important observations from Eq.~(\ref{eq:Verlet}). First, unlike in continuous time, discrete-time evolution progresses without the existence of a velocity, as implied by Eq.~(\ref{eq:Stormer}). Second, the two velocity definitions, Eqs.~(\ref{eq:on-site_org}) and (\ref{eq:half-step_org}), do not only yield different results, but they are also both inconsistent with the configurational coordinate in that neither of them are the conjugated variable to $r^n$ even for a simple harmonic oscillator, $f^n=-\kappa r^n$, $\kappa>0$ being a spring constant. Only in the limit $\Delta{t}\rightarrow0$ do the velocity definitions converge and become physically meaningful. However, computational efficiency does not favor this limit, and practical simulations with appreciable time steps within the stability limit are therefore left with fundamental inconsistencies between the simulated configurational and kinetic measures as the time step is increased. These inconsistencies lead to, e.g., time-dependent total energy of a simulated harmonic oscillator, as measured from the discrete-time variables in Eq.~(\ref{eq:Verlet}), and overall depressed kinetic measures for convex potentials in both on-site and half-step velocities, even if the half-step velocity is more accurate than the on-site variable. Thus, discrete-time velocity is an inherently ambiguous quantity that should be addressed separately from the associated configurational coordinate.

For future reference, we parenthetically note that the Verlet method in Eq.~(\ref{eq:Verlet}) can be conveniently written in the practical and compact form \cite{Tuckerman}
\begin{subequations}
\allowdisplaybreaks\begin{eqnarray}
v_o^{n+\frac{1}{2}} & = & v_o^n + \frac{\Delta{t}}{2m}f^n\, , \label{eq:half-step_tuck}\\
r^{n+1} & = & r^n + \Delta{t}\,v_o^{n+\frac{1}{2}}\, , \label{eq:Stormer_tuck}\\
v_o^{n+1} & = & v_o^{n+\frac{1}{2}}+\frac{\Delta{t}}{2m}f^{n+1}\, , \label{eq:on-site_tuck}
\end{eqnarray}\label{eq:tuck}\noindent
\end{subequations}
that uses the natural initial conditions, $r^n$ and $v^n$, at the top of the time step, and produces the natural conclusion, $r^{n+1}$ and $v^{n+1}$, at the bottom of the step. Equation~(\ref{eq:tuck}) combines the so-called {\it velocity-Verlet} \cite{Swope,Beeman} and {\it leap-frog} \cite{Buneman,Hockney} forms of the Verlet method into one compact form with direct access to the results of both velocity definitions.

\subsection{Discrete-Time Langevin Simulations}
\label{sec:GJ}
We now reintroduce fluctuations and dissipation ($\alpha>0$) to the problem in Eq.~(\ref{eq:Langevin}).
As outlined above in Eq.~(\ref{eq:Verlet}), the properties of a definition of velocity depends on the configurational behavior, which in turn can be considered independent of the associated velocity. Thus, before analyzing the possibilities for useful velocity definitions, we must first settle on an optimal configurational integrator. We adopt the established set of GJ methods \cite{GJ}, which has the most beneficial statistical properties of any Verlet-type algorithm for Langevin simulations.
In analogy with Eq.~(\ref{eq:Stormer}), the stochastic configurational coordinate of the GJ methods is described by
\begin{eqnarray}
r^{n+1} & = & 2c_1r^n-c_2r^{n-1}+\frac{c_3\Delta{t}}{m}(\Delta{t}f^n+\frac{\beta^n+\beta^{n+1}}{2})\nonumber \\ \label{eq:Stormer_GJ}
\end{eqnarray}
with
\begin{subequations}
\begin{eqnarray}
2c_1 & = & 1+c_2\label{eq:c1}\\
\frac{\alpha\Delta{t}}{m}c_3 & = & 1-c_2\, .\label{eq:c3}
\end{eqnarray}\label{eq:c1c3}\noindent
\end{subequations}
The functional parameter $c_2$ is a decaying function of the known quantity $\frac{\alpha\Delta{t}}{m}$ such that $c_2\rightarrow1-\frac{\alpha\Delta{t}}{m}$ for $\frac{\alpha\Delta{t}}{m}\rightarrow0$. Notice that coefficients, such as $c_i$ (as well as $\gamma_i$ and $\mu_i$ introduced later in Eqs.~(\ref{eq:uel_ansatz}) and (\ref{eq:vel_ansatz})), of a general purpose algorithm should depend only on known quantities; i.e., not on, e.g., the curvature of the potential surface as expressed by the force $f^n$, which changes and is not generally predictable in nonlinear and complex systems. The accumulated noise over the time interval $(t_n,t_{n+1}]$ is
\begin{eqnarray}
\beta^{n+1} & = & \int_{t_n}^{t_{n+1}}\beta\,dt \, , \label{eq:int_noise}
\end{eqnarray}
where the two lowest moments of the integrated fluctuations follow directly from Eq.~(\ref{eq:FD})
\begin{subequations}
\begin{eqnarray}
\left\langle \beta^n\right\rangle & = & 0 \label{eq:FD_discrete_1} \\
\left\langle \beta^n\beta^\ell \right\rangle & = & 2\,\alpha\Delta{t}\,k_BT\,\delta_{n,\ell} \, , \label{eq:FD_discrete_2}
\end{eqnarray}\label{eq:FD_discrete}\noindent
\end{subequations}
$\delta_{n,\ell}$ being the Kronecker delta function. As noted in Ref.~\cite{Gauss_noise}, and consistent with the central limit theorem \cite{central_limit} applied to the integral Eq.~(\ref{eq:int_noise}), $\beta^n$ must be chosen from a Gaussian distribution in order to ensure proper statistics.
The earliest special case of these methods, the GJF method \cite{GJF1} (GJ-I), was the first method to produce correct, time-step-independent drift and diffusive transport for motion on a flat surface ($f^n=0$) as well as correct Boltzmann statistics for the noisy, damped harmonic oscillator ($f^n=-\kappa r^n$, $\kappa>0$). Inserting such linear force into Eq.~(\ref{eq:Stormer_GJ}) reads
\begin{subequations}
\begin{eqnarray}
r^{n+1} & = & 2c_1Xr^n-c_2r^{n-1}+\frac{c_3\Delta{t}}{2m}(\beta^n+\beta^{n+1})\label{eq:Stormer_GJ_lin}
\end{eqnarray}
with
\begin{eqnarray}
X & = & 1-\frac{c_3}{c_1}\frac{\Omega_0^2\Delta{t}^2}{2} \, , \label{eq:X}
\end{eqnarray}\label{eq:Stormer_lin_complete}\noindent
\end{subequations}
where, again, the natural frequency is given by $\Omega_0^2=\kappa/m$. Stability analysis on the linearized system in Eq.~(\ref{eq:Stormer_GJ_lin}) yields the stability range \cite{GJ}
\begin{eqnarray}
\sqrt{\frac{c_3}{c_1}}\Omega_0\Delta{t} & < & 2\, . \label{eq:stability}
\end{eqnarray}

The configurational variance for the linear oscillator in Eq.~(\ref{eq:Stormer_GJ_lin}) is
\begin{subequations}
\begin{eqnarray}
\left\langle r^nr^n\right\rangle & = & \frac{k_BT}{\kappa} \; , \; \; f^n=-\kappa r^n \, . \label{eq:r2_corr}
\end{eqnarray}
The two other crucial features of the method are correct time-step-independent configurational drift velocity, $v_d$, and diffusion, $D_E$, on tilted and flat surfaces, respectively
\begin{eqnarray}
v_d & = & \frac{\left\langle r^{n+1}-r^n\right\rangle}{\Delta{t}} \; = \; \frac{f}{\alpha} \; , \; \; f^n=f={\rm const} \label{eq:drift}\\
D_E & = & \lim_{n\rightarrow\infty}\frac{\left\langle(r^{q+n}-r^q)^2\right\rangle_q}{2n\Delta{t}} \; = \; \frac{k_BT}{\alpha} \; , \;  \; f^n=f=0\, .  \nonumber \\ \label{eq:D_E}
\end{eqnarray}\label{eq:GJ_config}\noindent
\end{subequations}
The GJF method was later generalized to the complete and exclusive GJ set of methods with the configurational properties in Eq.~(\ref{eq:GJ_config}) \cite{GJ}. Notice that even if the statistical properties given in Eq.~(\ref{eq:GJ_config}) from the trajectories of Eq.~(\ref{eq:Stormer_GJ}) are correct, this does not imply that the individual trajectories are correct. We further list correlations that will become necessary for the development of velocity variables. For the linear oscillator Eq.~(\ref{eq:Stormer_GJ_lin}), these are  \cite{GJ}
\begin{subequations}
\begin{eqnarray}
\left\langle r^nr^{n+1}\right\rangle & = & \frac{k_BT}{\kappa}(1-c_1(1-X)) \label{eq:r2p1}\\
\left\langle r^{n-1}r^{n+1}\right\rangle & = & \frac{k_BT}{\kappa}(1-2c_1(1+c_1)(1-X) \nonumber \\
&&+2c_1^2(1-X)^2) \label{eq:r2m1p1}\\
\left\langle \beta^nr^n\right\rangle & = & \frac{k_BT}{\kappa}2c_1\alpha(1-X) \label{eq:br}\\
\left\langle \beta^{n-1}r^n\right\rangle & = &\left\langle \beta^nr^n\right\rangle(1+2c_1-2c_1(1-X)) \, , \label{eq:brm1p1}
\end{eqnarray}\label{eq:rbr_corr}\noindent
where we add the following two relationships for future reference:
\begin{eqnarray}
\langle r^{n-2}r^{n+1}\rangle & = & 2c_1X\langle r^{n-1}r^{n+1}\rangle - c_2\langle r^nr^n\rangle \label{eq:rm2rp1_corr}\\
\langle \beta^{n-1}r^{n+1}\rangle & = & 2c_1X\langle \beta^{n}r^{n+1}\rangle - c_2\langle \beta^nr^n\rangle\, .  \label{eq:bm1rp1_corr}
\end{eqnarray} \noindent
\end{subequations}

\begin{figure}[t]
\centering
\scalebox{0.5}{\centering \includegraphics[trim={2.0cm 5.0cm 2cm 6.5cm},clip]{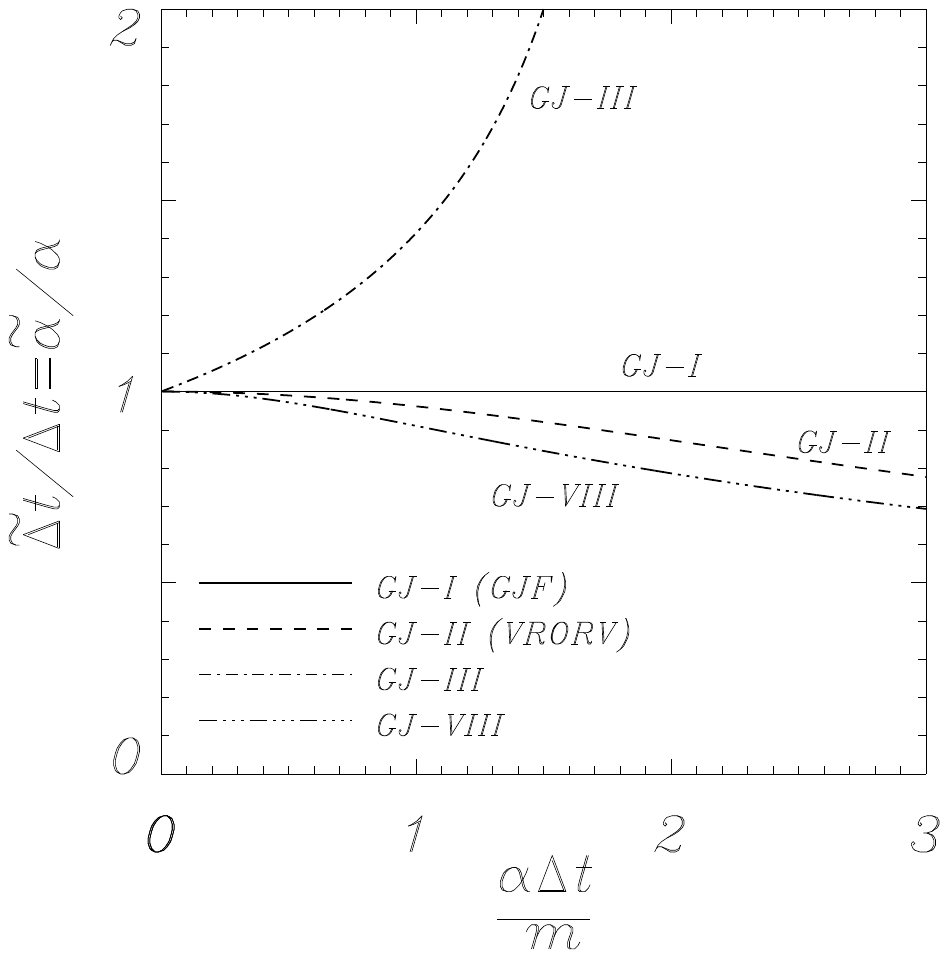}}
\caption{Scaling of time and damping from Eq.~(\ref{eq:scaling}) for methods, GJ-I, GJ-II, GJ-III, and GJ-VIII, given in Eqs.~(\ref{eq:c2_GJ_1_2_3}) and (\ref{eq:c2_GJ8}).
}
\label{fig:fig_1}
\end{figure}

A few key examples of GJ methods are:
\begin{subequations}
\begin{eqnarray}
{\rm GJ\!\!-\!\!I\; (GJF)}: \;\; c_2 & = & \frac{1-\frac{\alpha\Delta{t}}{2m}}{1+\frac{\alpha\Delta{t}}{2m}} \; \Leftrightarrow \; c_3=c_1 \label{eq:c2_GJ1}\\
{\rm GJ\!\!-\!\!II\; (VRORV)}: \;\; c_2 & = & \exp(-\frac{\alpha\Delta{t}}{m}) \label{eq:c2_GJ2}\\
{\rm GJ\!\!-\!\!III}: \;\; c_2 & = & 1-\frac{\alpha\Delta{t}}{m} \; \Leftrightarrow \; c_3=1 \, ,  \label{eq:c2_GJ3}
\end{eqnarray}\label{eq:c2_GJ_1_2_3}\noindent
\end{subequations}
originally mentioned in Refs.~\cite{GJF1}, \cite{Sivak}, and \cite{GJ}, respectively.

We notice that all GJ methods can be described in the revised GJ-I form ($c_3=c_1$)
\begin{eqnarray}
r^{n+1} & = & 2c_1r^n-c_2r^{n-1}+\frac{c_1\widetilde{\Delta{t}}^2}{m}f^n\nonumber \\
&& +\frac{c_1\widetilde{\Delta{t}}}{2m}\sqrt{2\,\widetilde{\alpha}\widetilde{\Delta{t}}\,k_BT}\,({\cal N}^n+{\cal N}^{n+1})\, ,
\end{eqnarray}
where ${\cal N}^n\in N(0,1)$ are independent and Gaussian distributed, and where rescaled damping and time are
\begin{subequations}
\begin{eqnarray}
\widetilde{\alpha} & = & \sqrt{\frac{c_3}{c_1}}\,\alpha \label{eq:alpha_tilde}\\
\widetilde{\Delta{t}} & = & \sqrt{\frac{c_3}{c_1}}\,\Delta{t} \, , \label{eq:dt_tilde}
\end{eqnarray}\label{eq:scaling}\noindent
\end{subequations}
respectively. Figure~\ref{fig:fig_1} shows the scaling of time and damping as a function of reduced time step for the GJ methods of Eq.~(\ref{eq:c2_GJ_1_2_3}) as well as the GJ-VIII method introduced later in Eq.~(\ref{eq:c2_GJ8}).

On-site and half-step velocities that accompany the GJ trajectory in Eq.~(\ref{eq:Stormer_GJ}) were developed in Ref.~\cite{GJ} such that the algorithm for obtaining both position and associated velocity requires only a single new stochastic number per time-step per degree of freedom. 

\begin{description}
\item[On-site velocity] The velocity measure, 
\begin{eqnarray}
v_1^n & = & \frac{r^{n+1}-(1-c_2)r^n-c_2r^{n-1}}{2\sqrt{c_1c_3}\,\Delta{t}}+\sqrt{\frac{c_3}{c_1}}\frac{\beta^n-\beta^{n+1}}{4m}\, , \nonumber \\
\label{eq:OS_def_GJ}
\end{eqnarray}
at times $t_n$ was determined to be the best possible three-point finite-difference approximation. While it was found that no three-point velocity can measure kinetic properties correctly, this velocity is asymptotically correct such that the measured kinetic temperature is a second order approximation in $\Delta{t}$. The drift velocity \cite{GJ},
\begin{eqnarray}
\langle v_1^n\rangle & = & \sqrt{\frac{c_1}{c_3}}\,\frac{f}{\alpha} \, , \label{eq:Drift_OS_1}
\end{eqnarray}
is also generally measured incorrectly as the time step is increased, but it measures drift correctly for the GJ-I method, Eq.~(\ref{eq:c2_GJ1}); see Fig.~\ref{fig:fig_2}. In comparison, the velocity measure of Eq.~(\ref{eq:on-site_org}) has correct drift velocity, but generally produces a first order error in $\Delta{t}$ for kinetic statistics, e.g., kinetic energy and temperature. 

\begin{figure}[t]
\centering
\scalebox{0.5}{\centering \includegraphics[trim={2.5cm 5.0cm 2.0cm 6.5cm},clip]{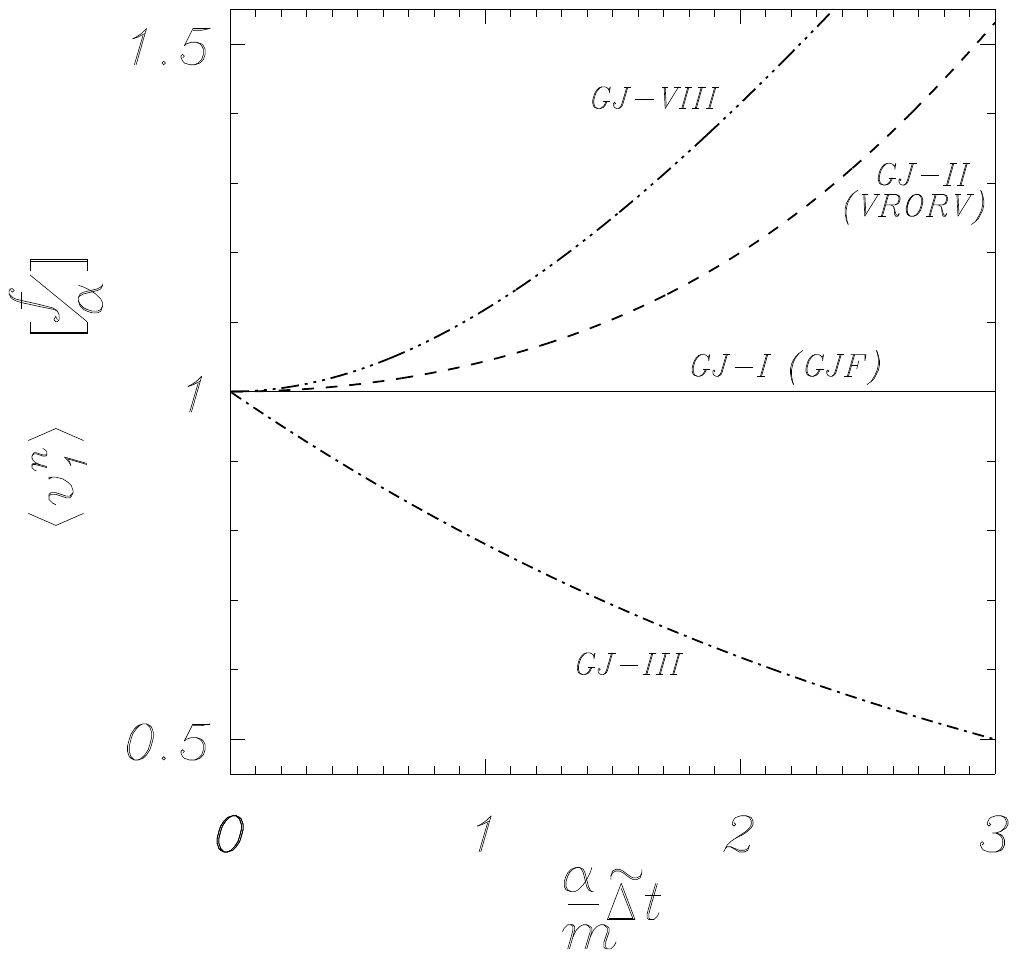}}
\caption{Drift velocity, Eq.~(\ref{eq:Drift_OS_1}), for constant force, $f\neq0$, as a function of reduced, scaled time step for $v_1^{n}$ in Eq.~(\ref{eq:OS_def_GJ}) for methods given in Eqs.~(\ref{eq:c2_GJ_1_2_3}) and (\ref{eq:c2_GJ8}).
}
\label{fig:fig_2}
\end{figure}

\item[Half-step velocity] The velocity measure,
\begin{eqnarray}
v_1^{n+\frac{1}{2}} & = & \frac{r^{n+1}-r^n}{\sqrt{c_3}\,\Delta{t}} \, , \label{eq:HS_def_GJ}
\end{eqnarray}
at times $t_{n+\frac{1}{2}}$ was developed to always produce correct kinetic statistics (for linear systems), but it measures the drift velocity \cite{GJ},
\begin{eqnarray}
\langle v_1^{n+\frac{1}{2}} \rangle & = & \frac{1}{\sqrt{c_3}}\,\frac{f}{\alpha} \, , \label{eq:Drift_HS_1}
\end{eqnarray}
incorrectly, except for the GJ-III method (Eq.~(\ref{eq:c2_GJ3})), where it is measured correctly; see Fig.~\ref{fig:fig_3}. In contrast, the half-step velocity of Eq.~(\ref{eq:half-step_org}) generally produces a first order error for kinetic temperature (except for the GJ-III method), while it measures the drift velocity correctly.
 
\begin{figure}[t]
\centering
\scalebox{0.5}{\centering \includegraphics[trim={2.5cm 5.0cm 2.0cm 6.5cm},clip]{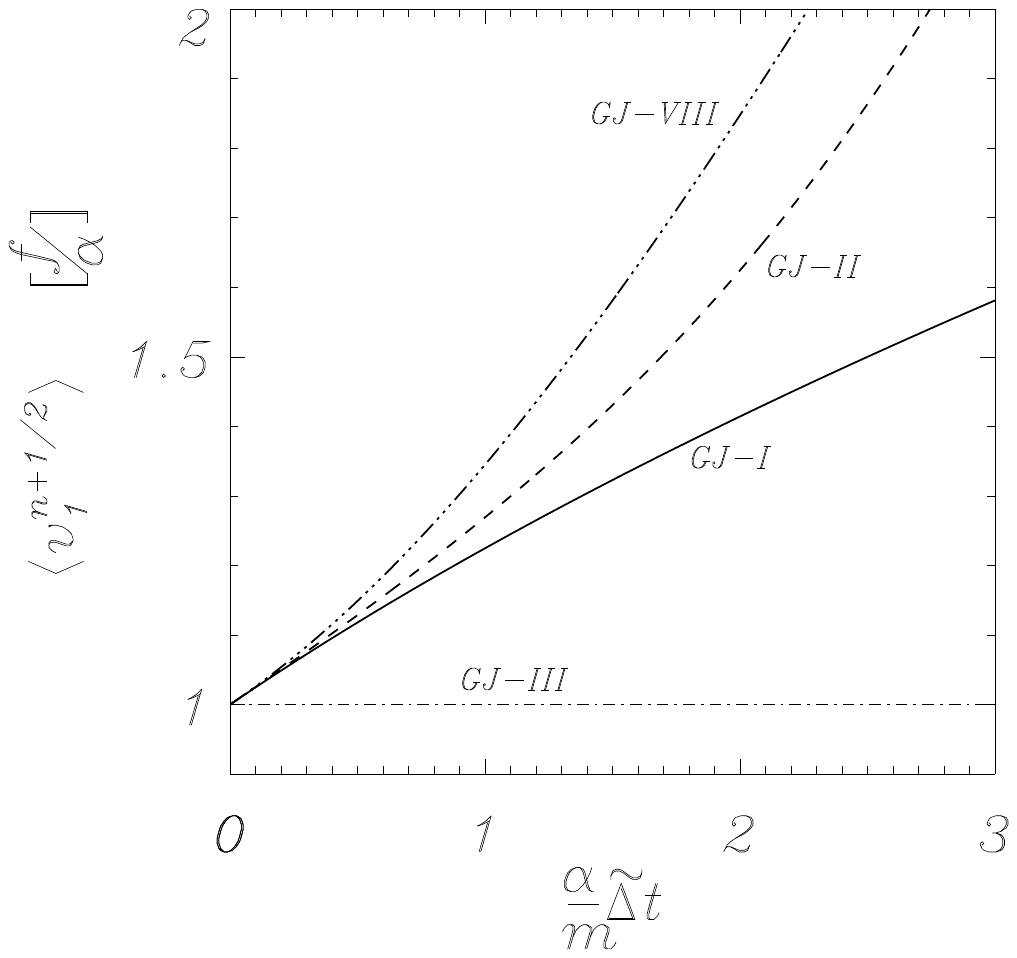}}
\caption{Drift velocity, Eq.~(\ref{eq:Drift_HS_1}), for constant force, $f\neq0$, as a function of reduced, scaled time step for $v_1^{n+\frac{1}{2}}$ in Eq.~(\ref{eq:HS_def_GJ}) for methods given in Eqs.~(\ref{eq:c2_GJ_1_2_3}) and (\ref{eq:c2_GJ8}).
}
\label{fig:fig_3}
\end{figure}

\end{description}

Thus, it is of interest to identify velocity measures that can simultaneously provide correct drift velocity and statistical kinetic measures.

The GJ methods have been formulated in Ref.~\cite{GJ} for the above-mentioned velocities in the velocity-Verlet and leap-frog forms. For use in this paper, we outline the algorithm in the revised splitting and compact forms that include both half-step and on-site velocities.

The revised splitting form of the GJ metods is \cite{Gauss_noise}
\begin{subequations}
\begin{eqnarray}
v^{n+\frac{1}{4}} & = & v_1^n+\sqrt{\frac{c_3}{c_1}}\frac{\Delta{t}}{2m}f^n \label{eq:A_v4}\\
r^{n+\frac{1}{2}} & = & r^n+\sqrt{\frac{c_3}{c_1}}\frac{\Delta{t}}{2}v^{n+\frac{1}{4}} \label{eq:A_r2}\\
v_1^{n+\frac{1}{2}} 
& = & \sqrt{c_1}\,v^{n+\frac{1}{4}} + \frac{\sqrt{c_3}}{2m}\,\beta^{n+1}\label{eq:A_v2}\\
v^{n+\frac{3}{4}}  & = & \frac{c_2}{\sqrt{c_1}}\,v_1^{n+\frac{1}{2}} + \sqrt{\frac{c_3}{c_1}}\frac{1}{2m}\,\beta^{n+1} \label{eq:A_v34}\\
r^{n+1} & = & r^{n+\frac{1}{2}}+\sqrt{\frac{c_3}{c_1}}\frac{\Delta{t}}{2}v^{n+\frac{3}{4}} \label{eq:A_r1}\\
v_1^{n+1} & = & v^{n+\frac{3}{4}}+\sqrt{\frac{c_3}{c_1}}\frac{\Delta{t}}{2m}f^{n+1}\, .  \label{eq:A_v1}
\end{eqnarray}\label{eq:A_Split_GJ}\noindent
\end{subequations}
There are several deviations in Eq.~(\ref{eq:A_Split_GJ}) from the restrictions of a standard ABO splitting formulation (see, e.g., Ref.~\cite{LM}). One difference (for all methods except GJ-I) is that the interactive, Eqs.~(\ref{eq:A_v4}) and (\ref{eq:A_v1}), and inertial, Eqs.~(\ref{eq:A_r2}) and (\ref{eq:A_r1}), operations generally include the rescaled time step of Eq.~(\ref{eq:scaling}), which depends on the friction coefficient, $\alpha$. Only for the GJ-I method is the time step unscaled. The ABO splitting formulation would include the friction dependence only in the thermodynamic operations of Eqs.~(\ref{eq:A_v2}) and (\ref{eq:A_v34}). Another departure (for all methods except for GJ-II) is that the functional parameter $c_2$ is here not restricted to the exponential function given in Eq.~(\ref{eq:c2_GJ2}), but allowed to be a broad class of monotonic functions with certain limiting features. A third deviation is that the two thermodynamic operators of Eqs.~(\ref{eq:A_v2}) and (\ref{eq:A_v34}) are different (e.g., geometrically asymmetric in the damping), whereas ABO splitting would build the algorithm from identical operators in those two steps. Finally, for ABO-type splitting expressions, the thermodynamic operations would each express a new realization of the stochastic variable, whereas the GJ algorithm in Eq.~(\ref{eq:A_Split_GJ}) maintains the same stochastic value throughout the time step. The last two of the mentioned deviations from ABO methods are addressed further in Sec.~\ref{sec:Half-step_2} below, where multiple stochastic realizations as well as thermodynamic operator symmetry are considered within the GJ structure.

Compacting the operations of Eqs.~(\ref{eq:A_v4}), (\ref{eq:A_r2}), and (\ref{eq:A_v2}), as well as those of Eqs.~(\ref{eq:A_v34}) and (\ref{eq:A_r1}), the algorithm of Eq.~(\ref{eq:A_Split_GJ}) can be written \cite{GJ}
\begin{subequations}
\begin{eqnarray}
v_1^{n+\frac{1}{2}} & = & \sqrt{c_1}v_1^n+\frac{\sqrt{c_3}\Delta{t}}{2m}f^n+\frac{\sqrt{c_3}}{2m}\beta^{n+1} \label{eq:A_Cu}\\
r^{n+1} & = & r^n+\sqrt{c_3}\Delta{t}\,v_1^{n+\frac{1}{2}} \label{eq:A_Cr}\\
v_1^{n+1} & = & \frac{c_2}{\sqrt{c_1}} v_1^{n+\frac{1}{2}} + \sqrt{\frac{c_3}{c_1}}\frac{\Delta{t}}{2m}f^{n+1}+\sqrt{\frac{c_3}{c_1}}\frac{1}{2m}\beta^{n+1}\, ,  \nonumber \\ \label{eq:A_Cv}
\end{eqnarray}\label{eq:A_Compact_GJ}\noindent
\end{subequations}
which, consistent with Eq.~(\ref{eq:tuck}), expresses only the relevant variables.

We finally note that bypassing the explicit half-step velocity in Eq.~(\ref{eq:A_v2}) yields the combined thermodynamic operation,
\begin{eqnarray}
v^{n+\frac{3}{4}} & = & c_2\,v^{n+\frac{1}{4}}+\frac{\sqrt{c_1c_3}}{m}\,\beta^{n+1}\, , \label{eq:A_Comb_Thermo}
\end{eqnarray}
in lieu of Eqs.~(\ref{eq:A_v2}) and (\ref{eq:A_v34}). Equation~(\ref{eq:A_Comb_Thermo}) is valid for all GJ methods that are based on the on-site velocity, $v_1^{n}$, regardless of the half-step velocity of interest.

Given these options for velocity measures, we will in the following use the GJ configurational coordinate $r^n$ of Eq.~(\ref{eq:Stormer_GJ}) or (\ref{eq:Stormer_GJ_lin}), to develop a variety of new velocity definitions that can be integrated with the GJ methods as kinetic companions in the formats provided by Eqs.~(\ref{eq:A_Split_GJ}) and (\ref{eq:A_Compact_GJ}).

\section{Half-Step Velocities -- sIngle noise value per time-step}
\label{sec:Half-step_1}
We first investigate velocity expressions involving a single random variable per time step. With $r^n$ given by Eq.~(\ref{eq:Stormer_GJ}), the stochastic, finite-difference ansatz is
\begin{eqnarray}
v^{n+\frac{1}{2}} & = & \frac{\gamma_1r^{n+1}+\gamma_2r^n+\gamma_3r^{n-1}+\gamma_6r^{n-2}}{\Delta{t}} \nonumber \\
&& + \frac{\gamma_5\beta^{n+1}+\gamma_4\beta^{n}+\gamma_7\beta^{n-1}}{m} \, , \label{eq:uel_ansatz}
\end{eqnarray}
where the variable $v^{n+\frac{1}{2}}$ represents the velocity at time $t_{n+\frac{1}{2}}$. Equation~(\ref{eq:uel_ansatz}) is an expanded ansatz compared to the one used in Ref.~\cite{GJ}, which included the $\gamma_i$ terms only for $i=1,2,3,4,5$. The goal is to determine the parameters $\gamma_i$, as a function of only the known parameter $\frac{\alpha\Delta{t}}{m}$, in order to best meet key objectives for $v^{n+\frac{1}{2}}$. We require that each term is well behaved  for all relevant parameters, and that the necessary condition,
\begin{eqnarray}
\gamma_1+\gamma_2+\gamma_3+\gamma_6 & = & 0 \, , \label{eq:uel_cond_0}
\end{eqnarray}
is satisfied, so that the velocity measure can be consistent with, e.g., a particle at rest, $r^n={\rm const}$, and that the finite-difference component of $v^{n+\frac{1}{2}}$ has a well-defined limit for $\Delta{t}\rightarrow0$.
The analysis will be conducted mostly for linear systems: Harmonic ($f=-\kappa r$), tilted ($f={\rm const}$), and flat ($f=0$) potentials.

\subsection{Conditions for being half-step}
A half-step velocity must be antisymmetric in its correlations with the configurational coordinates, $r^n$ and $r^{n+1}$, as evident from covariance and correlation, Eqs.~(\ref{eq:Corr_vr}) and (\ref{eq:Cont_corr_vr}), respectively. Thus, we enforce the statistical antisymmetry, which can be written in two identical forms \cite{GJ}
\begin{eqnarray}
\langle r^nv^{n+\frac{1}{2}}\rangle+\langle r^nv^{n-\frac{1}{2}}\rangle & = & \langle r^{n+1}v^{n+\frac{1}{2}}\rangle+\langle r^nv^{n+\frac{1}{2}}\rangle \; = \; 0 \, . \nonumber \\\label{eq:uel_cross_corr}
\end{eqnarray}
Inserting Eq.~(\ref{eq:uel_ansatz}) into Eq.~(\ref{eq:uel_cross_corr}), then using Eqs.~(\ref{eq:uel_cond_0}), (\ref{eq:rm2rp1_corr}), and (\ref{eq:bm1rp1_corr}), gives the expression
\begin{eqnarray}
&&\langle r^nv^{n+\frac{1}{2}}\rangle+\langle r^nv^{n-\frac{1}{2}}\rangle \; = \; \frac{\gamma_1+\gamma_2}{\Delta{t}}\langle r^nr^n\rangle-\frac{2c_1\gamma_6}{\Delta{t}}\langle r^nr^{n+1}\rangle \nonumber \\
&&+\frac{\gamma_3+\gamma_6+2c_1X\gamma_6}{\Delta{t}}\langle r^{n-1}r^{n+1}\rangle + \frac{\gamma_4+\gamma_5-c_2\gamma_7}{m}\langle\beta^nr^n\rangle \nonumber \\
&&+\frac{\gamma_4+\gamma_7+2c_1X\gamma_7}{m}\langle\beta^{n-1}r^n\rangle\, .  \label{eq:half-step_cross_corr}
\end{eqnarray}
Through the relationships of Eq.~(\ref{eq:rbr_corr}) and $1-X=\frac{c_3}{2c_1}(\Omega_0\Delta{t})^2$ we can write this expression as
\begin{eqnarray}
\langle r^nv^{n+\frac{1}{2}}\rangle+\langle r^nv^{n-\frac{1}{2}}\rangle & = & \frac{k_BT}{\sqrt{m\kappa}}\sum_{k=0}^3(\Omega_0\Delta{t})^{2k-1}\Lambda_{2k-1}\, ,  \nonumber \\ \label{eq:half-step_Lambda}
\label{eq:Half-step_Lambda}
\end{eqnarray}
where $\Lambda_{2k-1}=\Lambda_{2k-1}(\frac{\alpha\Delta{t}}{m})$ does not depend on $\Omega_0\Delta{t}$. It follows from Eq.~(\ref{eq:uel_cond_0}) that $\Lambda_{-1}=0$. The remaining coefficients are given by
\begin{subequations}
\begin{eqnarray}
\frac{1}{c_3}\Lambda_1 & = & -(1+c_1)(\gamma_3+\gamma_6-2\frac{\alpha\Delta{t}}{m}\gamma_4)+2\frac{\alpha\Delta{t}}{m}\gamma_5\nonumber \\
& &- (1+c_1+2c_1^2)(\gamma_6-2\frac{\alpha\Delta{t}}{m}\gamma_7)\label{eq:Lambda_1}\\
\frac{1}{c_3^2}\Lambda_3 & = & \frac{1}{2}(\gamma_3+\gamma_6-2\frac{\alpha\Delta{t}}{m}\gamma_4)\nonumber \\
& & +(2c_1+1)(\gamma_6-2\frac{\alpha\Delta{t}}{m}\gamma_7)\label{eq:Lambda_3}\\
\frac{1}{c_3^3}\Lambda_5 & = & -\frac{1}{2}(\gamma_6-2\frac{\alpha\Delta{t}}{m}\gamma_7)\, . \label{Lambda_5}
\end{eqnarray}\label{eq:Lambda_135}\noindent
\end{subequations}
The half-step conditions, $\Lambda_1=\Lambda_3=\Lambda_5=0$, thus become
\begin{subequations}
\begin{eqnarray}
\gamma_5 & = & 0\label{eq:uel_cond_1}\\
\gamma_3+\gamma_6 & = & 2\frac{\alpha\Delta{t}}{m}\gamma_4\label{eq:uel_cond_2}\\
\gamma_6 & = & 2\frac{\alpha\Delta{t}}{m}\gamma_7\, , \label{eq:uel_cond_3}
\end{eqnarray}\label{eq:uel_cond_123}
\end{subequations}
and enforcing these conditions yields the covariance
\begin{eqnarray}
\left\langle v^{n\mp\frac{1}{2}}r^n\right\rangle & = & \pm\frac{k_BT}{\sqrt{m\kappa}}\frac{\Omega_0\Delta{t}}{2}c_3\gamma_1\, ,  \label{eq:HS_corr_vr}
\end{eqnarray}
which meets the objective of antisymmetry.
This expression also correctly reproduces the corresponding continuous-time covariance in Eq.~(\ref{eq:Corr_vr}) for $s=\frac{1}{2}\Delta{t}\rightarrow0$ if $\gamma_1\rightarrow1$ in that limit. We will see below in Eq.~(\ref{eq:uel_limit_gamma_1}) that, indeed, $\gamma_1\rightarrow1$ for $\Delta{t}\rightarrow0$ for the resulting velocity measures.

\subsection{Condition for Maxwell-Boltzmann distribution}
The half-step velocity variance $\langle v^{n+\frac{1}{2}}v^{n+\frac{1}{2}}\rangle$ for Eq.~(\ref{eq:uel_ansatz}), combined with Eqs.~(\ref{eq:rm2rp1_corr}) and (\ref{eq:bm1rp1_corr}), is
\begin{subequations}
\begin{eqnarray}
\left\langle v^{n+\frac{1}{2}}v^{n+\frac{1}{2}}\right\rangle & = & \frac{\gamma_1^2+\gamma_2^2+\gamma_3^2+\gamma_6^2}{\Delta{t}^2} \langle r^nr^n\rangle \nonumber \\
&+ & 2\frac{\gamma_1\gamma_2+\gamma_2\gamma_3+\gamma_3\gamma_6-c_2\gamma_1\gamma_6}{\Delta{t}^2} \langle r^nr^{n+1}\rangle\nonumber \\
&+ & 2 \frac{\gamma_1\gamma_3+\gamma_2\gamma_6+2c_1X\gamma_1\gamma_6}{\Delta{t}^2} \langle r^{n-1}r^{n+1}\rangle\nonumber \\
&+ & 2\frac{\gamma_1\gamma_5+\gamma_2\gamma_4+\gamma_3\gamma_7-c_2\gamma_1\gamma_7}{m\Delta{t}}\langle\beta^nr^n\rangle\nonumber \\
&+ & 2\frac{\gamma_1\gamma_4+\gamma_2\gamma_7+2c_1X\gamma_1\gamma_7}{m\Delta{t}}\langle\beta^{n-1}r^n\rangle\nonumber \\
&+ & \frac{\gamma_4^2+\gamma_5^2+\gamma_7^2}{m^2}\langle\beta^n\beta^n\rangle \, , \label{eq:u2_corr_0}
\end{eqnarray}
which, given Eqs.~(\ref{eq:FD_discrete_2}), (\ref{eq:X}), and (\ref{eq:r2p1})-(\ref{eq:brm1p1}), can be written
\begin{eqnarray}
\left\langle v^{n+\frac{1}{2}} v^{n+\frac{1}{2}}\right\rangle & = & \frac{k_BT}{m}\sum_{k=-1}^2(\Omega_0\Delta{t})^{2k}\Lambda_{2k} \, , \label{eq:u2_corr_1}
\end{eqnarray}\noindent
\end{subequations}
where $\Lambda_{2k}=\Lambda_{2k}(\frac{\alpha\Delta{t}}{m})$ does not depend on $\Omega_0\Delta{t}$, consistent with $\Lambda_{2k-1}$ given in Eq.~(\ref{eq:half-step_Lambda}). Equation~(\ref{eq:uel_cond_0}) ensures that $\Lambda_{-2}=0$. The functional coefficients are
\begin{subequations}
\begin{eqnarray}
\frac{\Lambda_0}{c_3}
& = & \gamma_1^2+\frac{1}{2}\frac{(\gamma_1+\gamma_2)^2+\gamma_6^2}{1-c_2} \label{eq:Lambda_0}\\
\frac{\Lambda_2}{c_3^2} & = & \gamma_1(\gamma_3+\gamma_6-2\frac{\alpha\Delta{t}}{m}\gamma_4) \nonumber \\
&+ & (\gamma_1+\gamma_2+4c_1\gamma_1)(\gamma_6-2\frac{\alpha\Delta{t}}{m}\gamma_7) \label{eq:Lambda_2}\\
\frac{\Lambda_4}{c_3^3} & = & -\gamma_1(\gamma_6-2\frac{\alpha\Delta{t}}{m}\gamma_7)\, ,  \label{eq:Lambda_4}
\end{eqnarray}\label{eq:Lambda_024}\noindent
\end{subequations}
where we have used Eq.~(\ref{eq:uel_cond_0}) as well as all three conditions in Eq.~(\ref{eq:uel_cond_123}) to simplify the expression for $\Lambda_0$ in Eq.~(\ref{eq:Lambda_0}). The correct, time-step-independent variance of the half-step velocity is
\begin{eqnarray}
\left\langle v^{n+\frac{1}{2}}v^{n+\frac{1}{2}}\right\rangle & = & \frac{k_BT}{m} \, , \label{eq:cond_HS_MB}
\end{eqnarray}
implying the two conditions,  $\Lambda_2=\Lambda_4=0$, which are redundant with $\Lambda_3=\Lambda_5=0$, given in Eqs.~(\ref{eq:uel_cond_2}) and (\ref{eq:uel_cond_3}). The only additional contribution to determining the $\gamma_i$ parameters from enforcing the Maxwell-Botzmann distribution is therefore the condition
\begin{eqnarray}
\Lambda_0 & = & 1\, , \label{eq:uel_cond_4a}
\end{eqnarray}
which, with Eq.~(\ref{eq:Lambda_0}), implies that $\gamma_6\rightarrow0$ and $\gamma_2\rightarrow-\gamma_1$ for $\frac{\alpha\Delta{t}}{m}\rightarrow0$ ($c_2\rightarrow1$). Thus, by Eq.~(\ref{eq:uel_cond_0}), this, in turn, implies that $\gamma_3\rightarrow0$ for $\frac{\alpha\Delta{t}}{m}\rightarrow0$, consistent with the half-step central-difference velocity approximation Eq.~(\ref{eq:half-step_org}) for a lossless system. Clearly, one must expect that $\gamma_1\rightarrow-\gamma_2\rightarrow1$ in that limit. Indeed, Eq.~(\ref{eq:uel_limit_gamma_1}) below will confirm that assertion.

A special case that satisfies the condition of Eq.~(\ref{eq:uel_cond_4a}) is for $\gamma_3=\gamma_6=0$, which results in the previously developed half-step velocity given in Eq.~(\ref{eq:HS_def_GJ}) for $\gamma_1=1/\sqrt{c_3}$. Notice that $\gamma_1=-1/\sqrt{c_3}$ is also representing a half-step velocity with correct Maxwell-Boltzmann distribution. However, the drift velocity will in this case have a sign opposite to $v_d$ in Eq.~(\ref{eq:drift}); i.e., indicating a velocity in the opposite direction of the applied constant force $f$.

\subsection{Correct half-step drift velocity}
As can be seen directly from the ansatz in Eq.~(\ref{eq:uel_ansatz}), the condition for obtaining correct drift velocity, $\langle v^{n+\frac{1}{2}}\rangle=v_d$, can be expressed by 
\begin{eqnarray}
\gamma_2 &=& 1-2\gamma_1+\gamma_6 \, . \label{eq:Cond_uel_drift}
\end{eqnarray}
Inserting this condition into Eq.~(\ref{eq:Lambda_0}), with the constraint of Eq.~(\ref{eq:uel_cond_4a}), yields
\begin{eqnarray}
\frac{\Lambda_0}{c_3} & = & \gamma_1^2+\frac{1}{2}\frac{(\gamma_1-1-\gamma_6)^2+\gamma_6^2}{1-c_2} \; = \; \frac{1}{c_3}\, .  \label{eq:Lambda_0b} \end{eqnarray}
This polynomial in $\gamma_1$ is well behaved with two distinct real solutions for moderate values of $\gamma_6$. Thus, it is possible to construct a general half-step velocity for the GJ methods with correct, time-step-independent thermal measure and correct drift velocity.

\subsection{Constructing half-step velocities with a single noise value per time-step}
\begin{figure}[t]
\centering
\scalebox{0.5}{\centering \includegraphics[trim={2.5cm 5.0cm 2.0cm 6.5cm},clip]{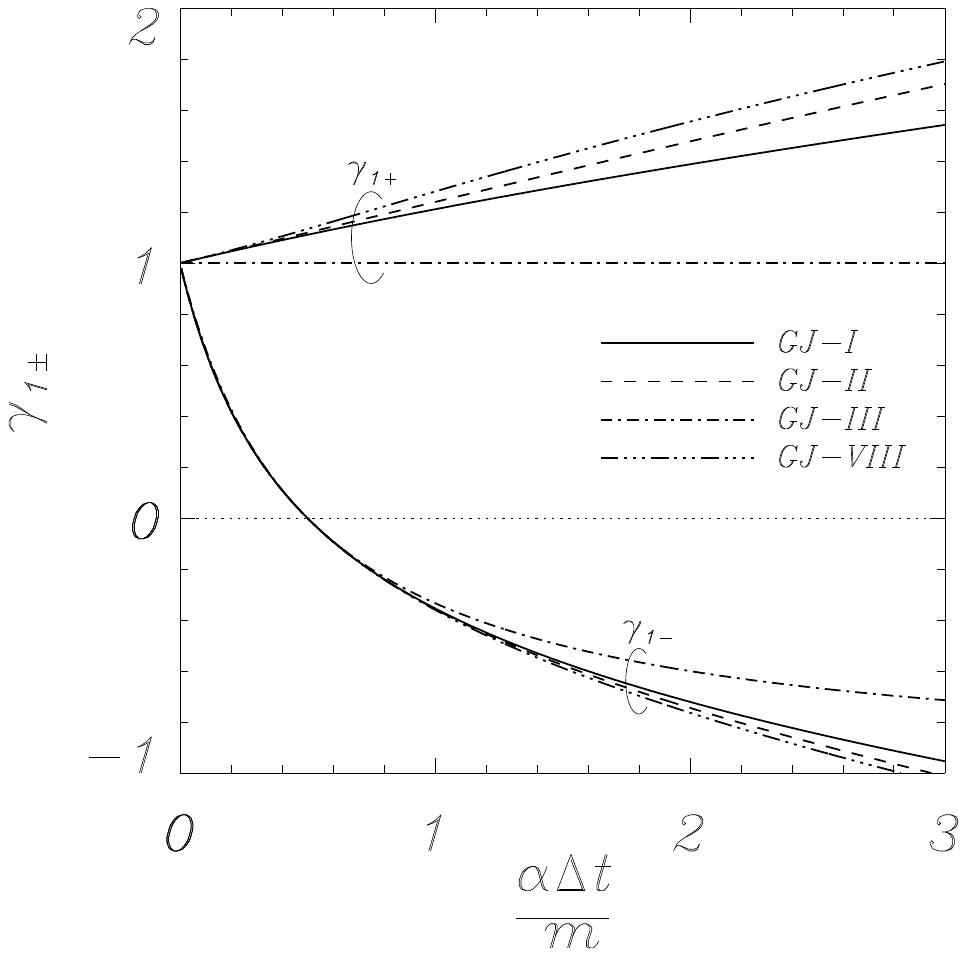}}
\caption{Lead coefficients, $\gamma_{1_\pm}$, of Eq.~(\ref{eq:half-step_gamma_1}), for the half-step velocities, Eq.~(\ref{eq:HS_one_noise}), as a function of reduced time step for methods given in Eqs.~(\ref{eq:c2_GJ_1_2_3}) and (\ref{eq:c2_GJ8}).
}
\label{fig:fig_4}
\end{figure}

Given that the six independent conditions, Eqs.~(\ref{eq:uel_cond_0}), (\ref{eq:uel_cond_123}), (\ref{eq:uel_cond_4a}), and (\ref{eq:Cond_uel_drift}), on the seven parameters in the ansatz of Eq.~(\ref{eq:uel_ansatz}) are enough to accomplish the stated goal of obtaining a general half-step velocity with correct drift and Maxwell-Boltzmann distribution, it is reasonable to simplify the half-step velocity ansatz by removing $\gamma_6$ and $\gamma_7$ (thereby making the condition in Eq.~(\ref{eq:uel_cond_3}) obsolete); i.e., we proceed with
\begin{eqnarray}
\gamma_6 \; = \; \gamma_7 & = & 0\, ,  \label{eq:half-step_g6g7}
\end{eqnarray}
which, when inserted in Eq.~(\ref{eq:Lambda_0b}), yields the two real solutions
\begin{eqnarray}
\gamma_{1_\pm} & = & \frac{1\pm\sqrt{2\frac{\alpha\Delta{t}}{m}}\sqrt{2(1-c_2)+1-c_3}}{3-2c_2} \, . \label{eq:half-step_gamma_1}
\end{eqnarray}
These two lead coefficients are visually exemplified in Fig.~\ref{fig:fig_4} for several GJ methods. The remaining half-step velocity parameters are then given by the two relationships, Eqs.~(\ref{eq:uel_cond_0}) and (\ref{eq:Cond_uel_drift}), which give
\begin{subequations}
\begin{eqnarray}
\gamma_2 & = & 1-2\gamma_1 \label{eq:uel_gamma_2_drift}\\
\gamma_3 & = & \gamma_1-1 \, . \label{eq:uel_gamma_3_drift}
\end{eqnarray}\label{eq:uel_gamma_23_drift}\noindent
\end{subequations}
These, along with Eqs.~(\ref{eq:uel_cond_1}) and (\ref{eq:uel_cond_2}), yield the expression,
\begin{eqnarray}
v_\pm^{n+\frac{1}{2}} & = & \frac{\gamma_{1_\pm}(r^{n+1}-r^n)-\gamma_{3_\pm}(r^n-r^{n-1})}{\Delta{t}} \nonumber \\
&& + \frac{c_3}{2m}\frac{\gamma_{3_\pm}}{1-c_2}\,\beta^n \, ,  \label{eq:HS_one_noise}
\end{eqnarray}
for a pair of velocities given by the two identified values, $\gamma_{1_\pm}$.  The two half-step velocities, $v_\pm^{n+\frac{1}{2}}$, are different, but yield the same correct drift velocity for a linear potential, and the same correct statistics for a harmonic potential. The differences are found in, e.g., the antisymmetric correlations to the configurational coordinate, $r^n$; see Eq.~(\ref{eq:HS_corr_vr}).

Notice that the small-time-step limit for $\gamma_{1_\pm}$ shows the expected result,
\begin{eqnarray}
\gamma_{1_\pm} & \rightarrow & 1-2\frac{\alpha\Delta{t}}{m}\left(1\mp\sqrt{1+\frac{1}{4}c_2^{\prime\prime}(0)}\right) \; \rightarrow \; 1_\pm  \, , \label{eq:uel_limit_gamma_1}
\end{eqnarray}
for $\frac{\alpha\Delta{t}}{m}\rightarrow0$.  The second derivative of $c_2$ with respect to $\frac{\alpha\Delta{t}}{m}$ for $\frac{\alpha\Delta{t}}{m}\rightarrow0$ is denoted $c_2^{\prime\prime}(0)$.

We here outline a formulation of the GJ algorithms that reproduces the GJ configurational coordinate given in Eq.~(\ref{eq:Stormer_GJ}), but with the particular improved half-step velocity definitions of Eqs.~(\ref{eq:half-step_gamma_1}), (\ref{eq:uel_gamma_3_drift}), and (\ref{eq:HS_one_noise}), and with the original GJ on-site velocity of Ref.~\cite{GJ} (see Eq.~(\ref{eq:OS_def_GJ})):
\begin{subequations}
\begin{eqnarray}
&&  v^{n+\frac{1}{4}} \; = \; v_1^n+\sqrt{\frac{c_3}{c_1}}\frac{\Delta{t}}{2m}f^n \nonumber \\
&&r^{n+\frac{1}{2}} \; = \; r^n+\sqrt{\frac{c_3}{c_1}}\frac{\Delta{t}}{2}v^{n+\frac{1}{4}} \nonumber \\
v_\pm^{n+\frac{1}{2}} 
& = &  \frac{c_2\gamma_{1_\pm}-\gamma_{3_\pm}}{c_2}\sqrt{c_1c_3}\,v^{n+\frac{1}{4}} + \frac{c_3}{2m}\gamma_{1_\pm}\beta^{n+1} \nonumber \\
&+&\frac{\gamma_{3_\pm}}{c_2m}\left(\sqrt{\frac{c_3}{c_1}}\Delta{t}f^n+ \frac{c_3}{2}\frac{1}{1-c_2}\beta^n\right)\label{eq:Z_v2}\\
v^{n+\frac{3}{4}} 
& = & \frac{c_2^2v_\pm^{n+\frac{1}{2}}+\frac{c_3}{2m}(c_2-\gamma_{3_\pm})\beta^{n+1}}{\sqrt{c_1c_3}(c_2\gamma_{1_\pm}-\gamma_{3_\pm})} \label{eq:Z_v34} \\
&-& \frac{c_2\gamma_{3_\pm}}{c_2\gamma_{1_\pm}-\gamma_{3_\pm}}\sqrt{\frac{c_3}{c_1}}\frac{1}{m}\left(\Delta{t}f^n+\frac{1}{1-c_2}\frac{1}{2}\beta^n\right)\nonumber  \\
&& r^{n+1} \; = \; r^{n+\frac{1}{2}}+\sqrt{\frac{c_3}{c_1}}\frac{\Delta{t}}{2}v^{n+\frac{3}{4}} \nonumber \\
&& v_1^{n+1} \; = \; v^{n+\frac{3}{4}}+\sqrt{\frac{c_3}{c_1}}\frac{\Delta{t}}{2m}f^{n+1} \,  , \nonumber
\end{eqnarray}\label{eq:Z_Split_GJ}\noindent
\end{subequations}
where the unlabeled equations are already given in Eq.~(\ref{eq:A_Split_GJ}). A significant simplification of this algorithm is to replace Eq.~(\ref{eq:Z_v34}) with Eq.~(\ref{eq:A_Comb_Thermo}), which applies for as long as the on-site velocity is $v_1^n$.
We note paranthetically that one of the velocities, $v_+^{n+\frac{1}{2}}$, identified by Eqs.~(\ref{eq:half-step_gamma_1}) and (\ref{eq:HS_one_noise}) coincides with the half-step velocity, $v_1^{n+\frac{1}{2}}$ from Eq.~(\ref{eq:HS_def_GJ}), developed in Ref.~\cite{GJ} for the special case of the GJ-III method, where $c_3=1\Rightarrow\gamma_{1_+}=1$ (see Eq.~(\ref{eq:c2_GJ3})). The parameters for this special case are $\gamma_{1_+}=-\gamma_{2_+}=1$ and $\gamma_{3_+}=\gamma_{4_+}=0$.

\begin{figure}[t]
\centering
\scalebox{0.5}{\centering \includegraphics[trim={2.5cm 6.0cm 1.5cm 5.5cm},clip]{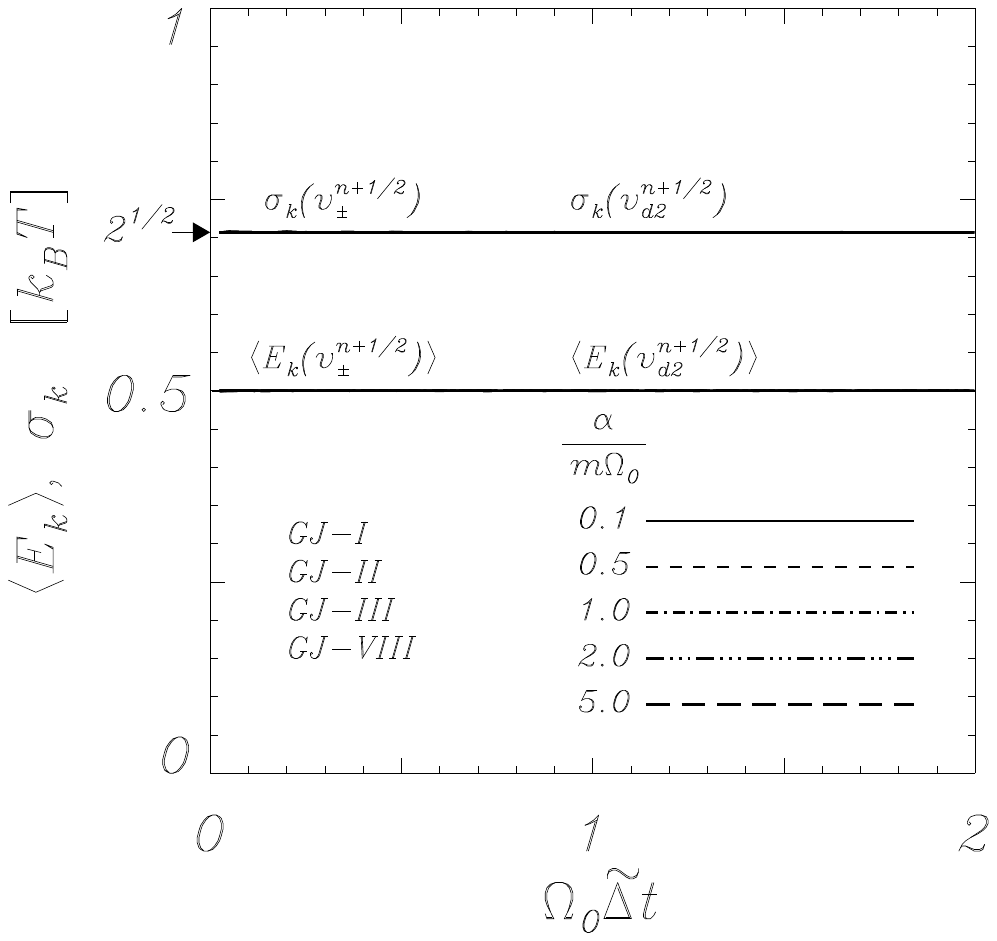}}
\caption{Kinetic statistics for a noisy harmonic oscillator with damping parameter as shown in the figure for the half-step velocities, $v_\pm^{n+\frac{1}{2}}$ and $v_{d2}^{n+\frac{1}{2}}$ of Eqs.~(\ref{eq:HS_one_noise}) and (\ref{eq:HS_vb2_from_v1}), as a function of reduced time step for methods given in Eqs.~(\ref{eq:c2_GJ_1_2_3}) and (\ref{eq:c2_GJ8}). All results for kinetic energies coincide, as do the results for their fluctuations. Simulation details given in Sec.~\ref{sec:HS_one_numerical}.
}
\label{fig:fig_5}
\end{figure}

\begin{figure}[t]
\centering
\scalebox{0.475}{\centering \includegraphics[trim={2.5cm 1.5cm 0.0cm 3.0cm},clip]{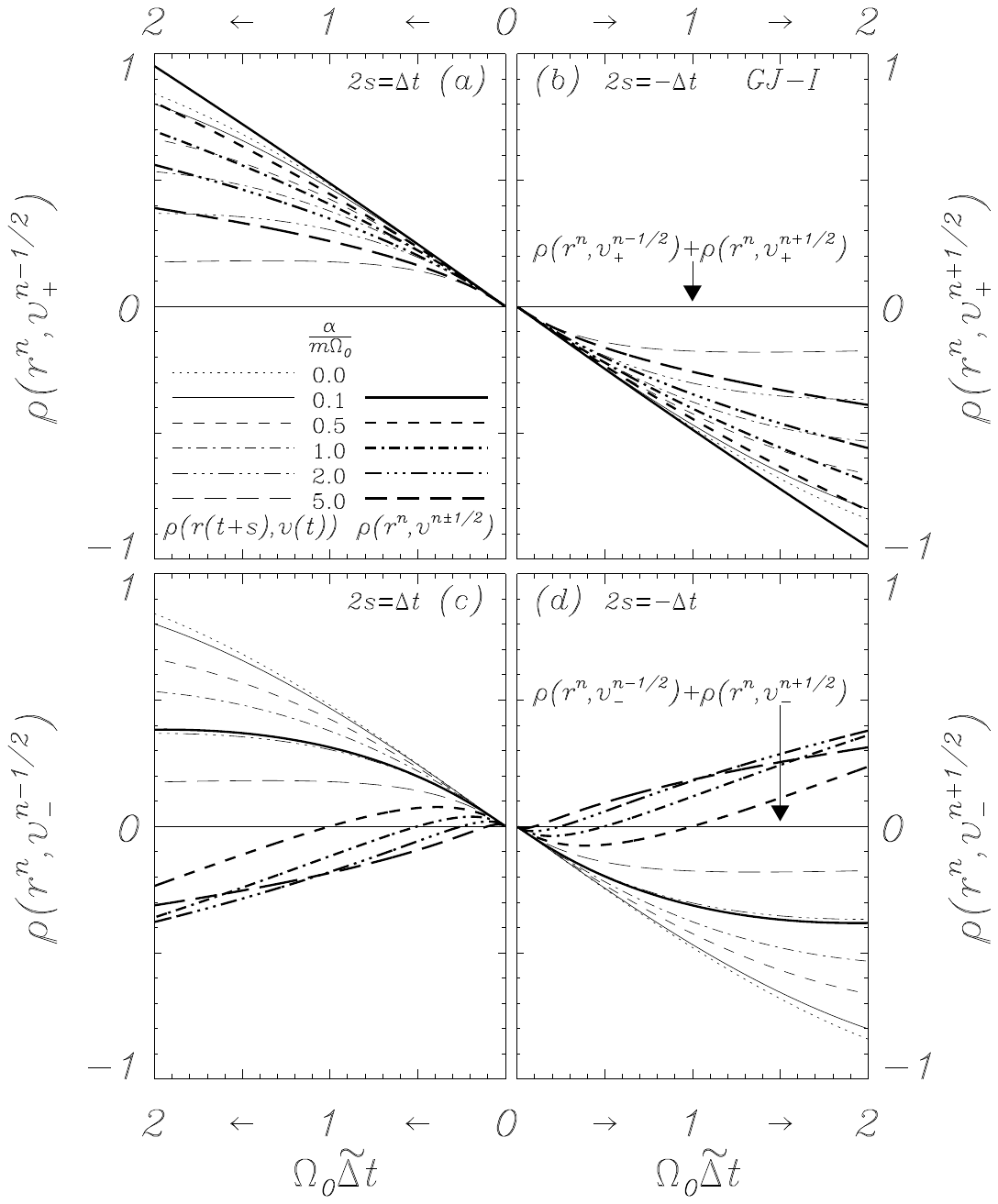}}
\caption{Cross-correlations between the configurational coordinate and the half-step velocities, $v_\pm^{n+\frac{1}{2}}$ for the GJ-I method with several different damping parameters. Thick curves represent the analytical result, Eq.~(\ref{eq:cross_corr}), as well as simulations of Eq.~(\ref{eq:Z2_Split_GJ}), which coincide with the analysis. Thin curves represent the continuous-time expectation, Eq.~(\ref{eq:Cont_corr_vr}) with $2s=\Delta{t}$. Simulation details given in Sec.~\ref{sec:HS_one_numerical}.
}
\label{fig:fig_6}
\end{figure}

\begin{figure}[t]
\centering
\scalebox{0.475}{\centering \includegraphics[trim={2.5cm 1.5cm 0.0cm 3.0cm},clip]{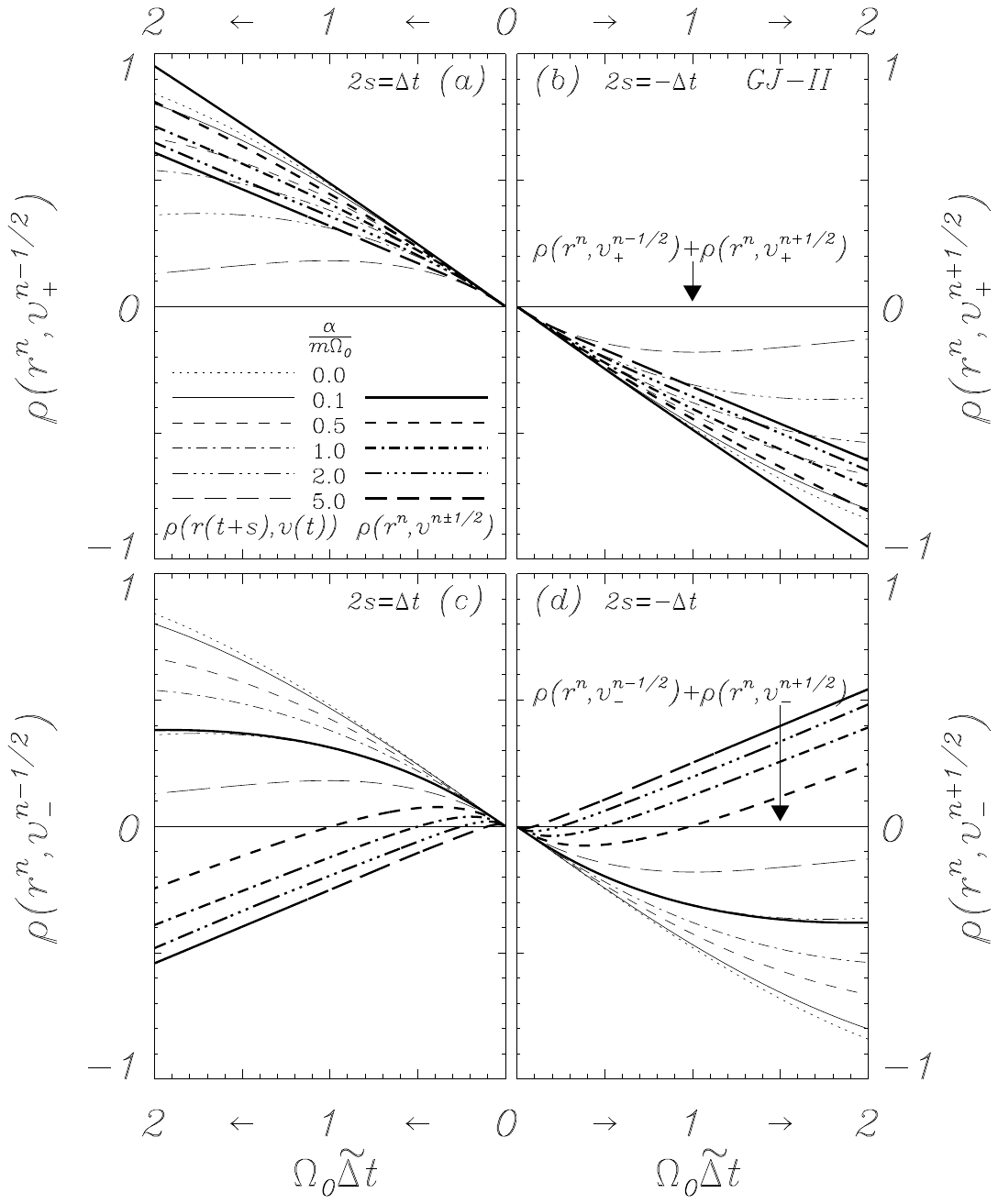}}
\caption{Cross-correlations between the configurational coordinate and the half-step velocities, $v_\pm^{n+\frac{1}{2}}$ for the GJ-II method with several different damping parameters. Thick curves represent the analytical result, Eq.~(\ref{eq:cross_corr}), as well as simulations of Eq.~(\ref{eq:Z2_Split_GJ}), which coincide with the analysis. Thin curves represent the continuous-time expectation, Eq.~(\ref{eq:Cont_corr_vr}) with $2s=\Delta{t}$. Simulation details given in Sec.~\ref{sec:HS_one_numerical}.
}
\label{fig:fig_7}
\end{figure}

\begin{figure}[t]
\centering
\scalebox{0.475}{\centering \includegraphics[trim={2.5cm 1.5cm 0.0cm 3.0cm},clip]{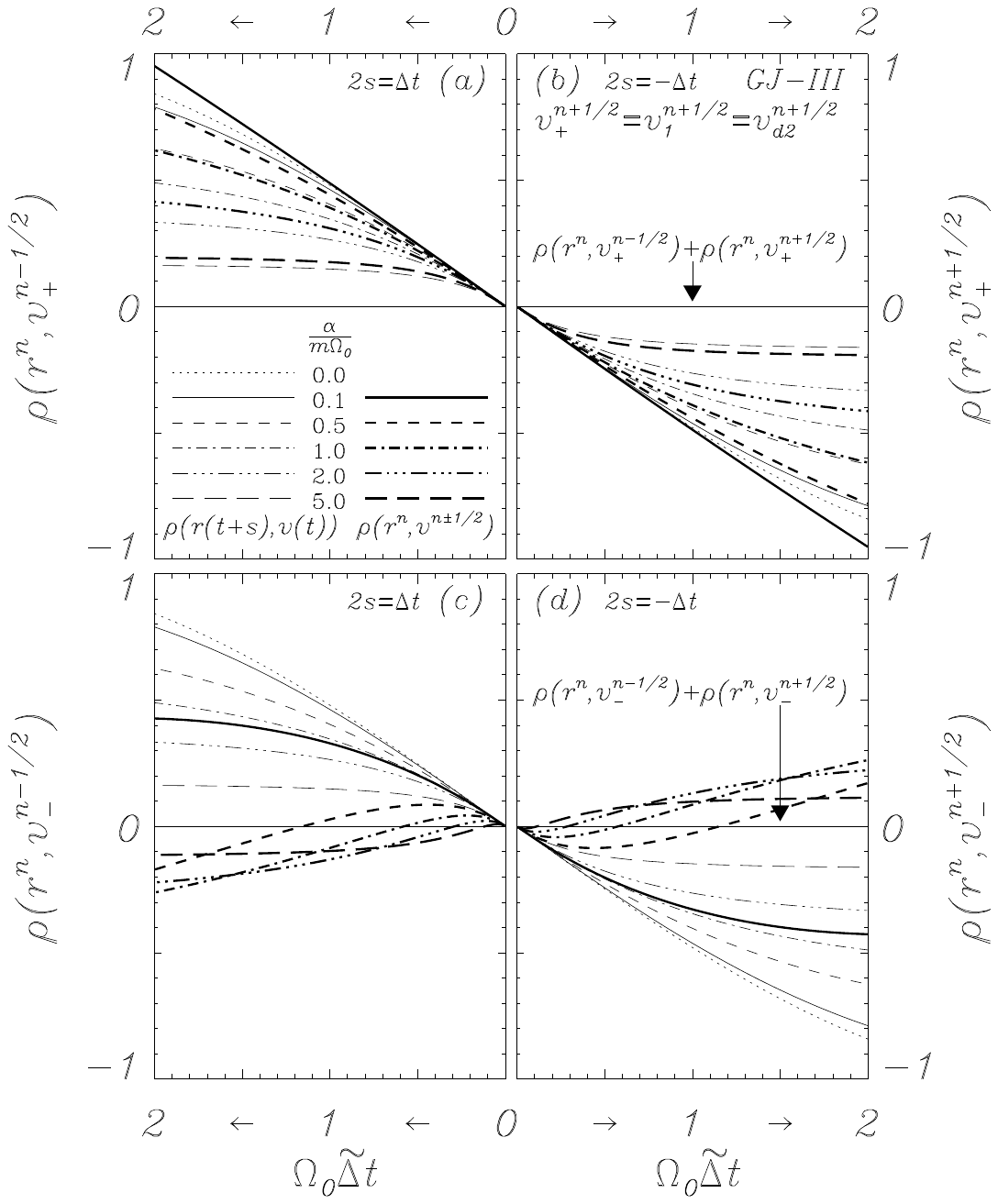}}
\caption{Cross-correlations between the configurational coordinate and the half-step velocities, $v_\pm^{n+\frac{1}{2}}$ for the GJ-III method with several different damping parameters. Thick curves represent the analytical result, Eq.~(\ref{eq:cross_corr}), as well as simulations of Eq.~(\ref{eq:Z2_Split_GJ}), which coincide with the analysis. Thin curves represent the continuous-time expectation, Eq.~(\ref{eq:Cont_corr_vr}) with $2s=\Delta{t}$. Simulation details given in Sec.~\ref{sec:HS_one_numerical}.
}
\label{fig:fig_8}
\end{figure}

The two thermodynamic operations, Eqs.~(\ref{eq:Z_v2}) and (\ref{eq:Z_v34}), appear to have rather complicated coefficients. However, since all coefficients are constants for any one given time step, this formulation is no more complicated to execute than the standard GJ method in Eq.~(\ref{eq:A_Split_GJ}), except for the presence of the noise variable, $\beta^n$, from the previous time step. Alternatively, one may include the half-step velocities $v_\pm^{n+\frac{1}{2}}$ in a slightly simpler manner by writing Eq.~(\ref{eq:HS_one_noise}) as
\begin{eqnarray}
v_\pm^{n+\frac{1}{2}} & = & \sqrt{c_3}\left(\gamma_{1_\pm}v_1^{n+\frac{1}{2}}-\gamma_{3_\pm}v_1^{n-\frac{1}{2}}\right)
+ \frac{c_3}{2m}\frac{\gamma_{3_\pm}}{1-c_2}\,\beta^n \, , \nonumber \\ \label{eq:HS2_one_noise}
\end{eqnarray}
which, when incorporated into Eq.~(\ref{eq:A_Split_GJ}), reads
\allowdisplaybreaks\begin{subequations}
\begin{eqnarray}
&& v^{n+\frac{1}{4}} \; = \; v_1^n+\sqrt{\frac{c_3}{c_1}}\frac{\Delta{t}}{2m}f^n \nonumber \\
&& r^{n+\frac{1}{2}} \; = \; r^n+\sqrt{\frac{c_3}{c_1}}\frac{\Delta{t}}{2}v^{n+\frac{1}{4}} \nonumber\\
&& v_1^{n+\frac{1}{2}}  \; = \; \sqrt{c_1}\,v^{n+\frac{1}{4}} + \frac{\sqrt{c_3}}{2m}\,\beta^{n+1}\nonumber\\
v_\pm^{n+\frac{1}{2}} & = & \sqrt{c_3}\gamma_{1_\pm}v_1^{n+\frac{1}{2}}-\gamma_{3_\pm}v_*^{n-\frac{1}{2}} \label{eq:Z2_vpm2}\\
v_*^{n+\frac{1}{2}} & = &  \sqrt{c_3}v_1^{n+\frac{1}{2}}-\frac{c_3}{2m}\frac{1}{1-c_2}\,\beta^{n+1} \label{eq:Z2_vs2}\\
&& v^{n+\frac{3}{4}}  \; = \; \frac{c_2}{\sqrt{c_1}}\,v_1^{n+\frac{1}{2}} + \sqrt{\frac{c_3}{c_1}}\frac{1}{2m}\,\beta^{n+1}\nonumber\\
&& r^{n+1} \; = \; r^{n+\frac{1}{2}}+\sqrt{\frac{c_3}{c_1}}\frac{\Delta{t}}{2}v^{n+\frac{3}{4}}\nonumber\\
&& v_1^{n+1} \; = \; v^{n+\frac{3}{4}}+\sqrt{\frac{c_3}{c_1}}\frac{\Delta{t}}{2m}f^{n+1}\, ,  \nonumber
\end{eqnarray}\label{eq:Z2_Split_GJ}\noindent
\end{subequations}
where only the labeled equations differ from Eq.~(\ref{eq:A_Split_GJ}), and where the single value, $v_*^{n+\frac{1}{2}}$, in Eq.~(\ref{eq:Z2_vs2}) is used in Eq.~(\ref{eq:Z2_vpm2}) to generate both half-step velocities, $v_\pm^{n+\frac{1}{2}}$, in the following time step. Obviously, this formulation produces all three half-step velocities, $v_1^{n+\frac{1}{2}}$ and $v_\pm^{n+\frac{1}{2}}$, in each time step.

\subsection{Numerical simulations and validation}
\label{sec:HS_one_numerical}

While the correct drift velocity of the derived measures, $v_\pm^{n+\frac{1}{2}}$ in Eq.~(\ref{eq:HS_one_noise}), is trivially given by inserting Eq.~(\ref{eq:uel_gamma_3_drift}) into Eq.~(\ref{eq:HS_one_noise}), we wish to numerically validate the other, less trivially observed, key properties that we have sought to acquire; namely the kinetic statistics and the cross-correlations to the configurational coordinate.

Validating the harmonic oscillator expressions, we simulate Eq.~(\ref{eq:Z2_Split_GJ}) with $f^n=-\kappa r^n$, using the RANMAR noise generator \cite{LAMMPS-Manual} for uniformly distributed stochastic numbers, and a Box-Muller transformation \cite{random} into a Normal distribution. Simulations are conducted for the four methods given in Eqs.~(\ref{eq:c2_GJ_1_2_3}) and (\ref{eq:c2_GJ8}), for $\frac{\alpha}{m\Omega_0}=0.1, 0.5, 1.0, 2.0, 5.0$, and statistics for each data point is acquired as an average of 10$^3$ simulations, each consisting of 10$^6$ time steps, starting with random initial conditions drawn from Boltzmann distributions. Time steps are chosen to span the entire stability range. The results for the kinetic energy, $\langle E_k\rangle$, and its standard deviation, $\sigma_k$, for the two velocities $v_\pm^{n+\frac{1}{2}}$, are shown in Fig.~\ref{fig:fig_5}. The kinetic energy fluctuations are given by
\begin{eqnarray}
\sigma_k(v_\pm^{n+\frac{1}{2}}) & = & \frac{1}{2}m\sqrt{\langle(v_\pm^{n+\frac{1}{2}})^4\rangle-\langle(v_\pm^{n+\frac{1}{2}})^2\rangle^2}\, ,  \label{eq:stand_dev_sim_HS}
\end{eqnarray}
where the averages are taken over all time steps of each set of simulations.
Both the average kinetic energy and its fluctuations, which are expected to be $\sigma_k=\sqrt{2}k_BT$ for a Gaussian distribution, are in perfect statistical agreement, time-step independent, and in accordance with analytical expectations for all relevant time steps. All simulation results are indistinguishable from the analytically derived expectations on the shown scale.

Cross-correlations,
\begin{eqnarray}
\rho(r^n,v^{n\pm\frac{1}{2}}) & = & \frac{\langle r^nv^{n\pm\frac{1}{2}}\rangle-\langle r^n\rangle\langle v^{n\pm\frac{1}{2}}\rangle}{\sigma_p(r^n)\sigma_k(v^{n\pm\frac{1}{2}})} \, , \label{eq:cross_corr}
\end{eqnarray}
where, for the Hooke potential,
\begin{eqnarray}
\sigma_p(r^n) & = & \frac{1}{2}\kappa\sqrt{\langle(r^n)^4\rangle-\langle(r^n)^2\rangle^2}\, ,  \label{eq:stand_dev_sim_r}
\end{eqnarray}
are shown in Figs.~\ref{fig:fig_6} (GJ-I), \ref{fig:fig_7} (GJ-II), and \ref{fig:fig_8} (GJ-III) for the half-step velocities $v_\pm^{n+\frac{1}{2}}$. Results are shown for different damping strengths. Simulation results are compared to,  and coincide with, the correlations derived from Eq.~(\ref{eq:HS_corr_vr}). These figures also show the continuous-time expectations given in Eq.~(\ref{eq:Cont_corr_vr}). The figures visualize the desirable half-step design feature of the velocity measures through the perfect antisymmetry, but, not surprisingly, it is also apparent that the continuous-time cross-correlation is not easily matched, especially for the velocity $v_-^{n+\frac{1}{2}}$, which, following from the sign change in $\gamma_{1_-}$ of Eq.~(\ref{eq:half-step_gamma_1}) and Fig.~\ref{fig:fig_4}, changes the sign of the correlation for moderate to strong damping at moderate time steps. Yet, the critical accomplishment of these velocities is that they possess the antisymmetric correlation with the configurational coordinate, respond with the correct drift velocity, and that they provide correct and time-step-independent kinetic statistics, such as the kinetic temperature and its fluctuations. The overall impression is that the GJ-III method for $v_+^{n+\frac{1}{2}}$ gives the best agreement with the continuous-time correlation.

\section{Half-Step Velocities -- two noise values per time-step}
\label{sec:Half-step_2}
As shown in Refs.~\cite{GJ,Josh_2020}, the configurational coordinate can produce all three statistical objectives of correct Boltzmann distribution, drift, and diffusion given in Eq.~(\ref{eq:GJ_config}), {\it only} if a single new noise variable per time step contributes to the fluctuations that affect the coordinate. This feature must be maintained when including an accompanying velocity to an algorithm. Therefore, one can consider additional noise contributions to a defined velocity measure only for as long as these additional fluctuations do not interfere with the configurational (on-site) evolution of $r^n$. An example of this is the ABO splitting methodology \cite{LM} (O, here referring to an operator of the same temporal duration as the A and B operators), where each of the two thermodynamic operations of a complete time step carries a separate noise contribution, such that the arrangement, e.g., BAO$^2$AB, yields one linear combination of the two noise terms for the configurational coordinate, and another for the associated half-step velocity, given by the application of the first sequence of ABO. This was realized for the time-scale revised BAO$^2$AB method, named VRO$^2$RV \cite{Sivak}, where the VRO generated half-step velocity is subject to a different (and independent) noise contribution from the single noise contribution that influences the configurational coordinate by the completion of the time step. We therefore introduce, similarly to $\beta^{n}$ in Eq.~(\ref{eq:FD_discrete}), a Gaussian variable, $\beta^{n+\frac{1}{2}}\in N(0,2\alpha\Delta{t}k_BT)$, for which
\begin{eqnarray}
\langle\beta^{n+\frac{1}{2}}\beta^{\frac{\ell+1}{2}}\rangle & = & 2\,\alpha\Delta{t}\,k_BT\,\delta_{2n,\ell} \, .  \label{eq:beta_half}
\end{eqnarray}
Using this independent stochastic variable, we generalize the feature of orthogonality between a separate, independent noise value for the half-step velocity from that of the configurational coordinate, by adding the term $\frac{\gamma_8}{m}\beta^{n+\frac{1}{2}}$ to the half-step velocity ansatz in Eq.~(\ref{eq:uel_ansatz}). It is clear that the condition given in Eq.~(\ref{eq:uel_cond_0}) still applies, and Eq.~(\ref{eq:Stormer_GJ}) implies that $\langle\beta^{n+\frac{1}{2}}r^\ell\rangle=0$, since $r^\ell$ is independent of $\beta^{n+\frac{1}{2}}$ for all $\ell$. It thereby follows that the half-step conditions, Eq.~(\ref{eq:uel_cond_123}), are also valid for the augmented ansatz with $\frac{1}{m}\gamma_8\beta^{n+\frac{1}{2}}$ included. The only revision to the functional $\gamma_i$-parameters arises for the condition in Eq.~(\ref{eq:uel_cond_4a}), with $\Lambda_0$ given in Eq.~(\ref{eq:Lambda_0}), such that
\begin{eqnarray}
\Lambda_0 & = & c_3\gamma_1^2+\frac{c_3}{2}\frac{(\gamma_1+\gamma_2)^2+\gamma_6^2}{1-c_2}+2\frac{1-c_2}{c_3}\gamma_8^2 \; = \; 1 \, , \nonumber \\ \label{eq:Lambda_08}
\end{eqnarray}
where we note that the correct drift velocity has not yet been enforced. With ample opportunity to satisfy the half-step and Maxwell-Boltzmann conditions, we simplify this equation by setting $\gamma_3=\gamma_6=0$, which, by Eq.~(\ref{eq:uel_cond_0}), implies that $\gamma_1=-\gamma_2$. It follows from Eq.~(\ref{eq:uel_cond_123}) that $\gamma_4=\gamma_5=\gamma_7=0$. Inserting this into Eq.~(\ref{eq:Lambda_08}) yields
\begin{eqnarray}
\gamma_8^2 & = & \frac{c_3}{2}\frac{1-c_3\gamma_1^2}{1-c_2} \, . \label{eq:uel_gamma_8}
\end{eqnarray}
We can then write the two-point half-step velocities with correct Maxwell-Boltzmann distribution as
\begin{eqnarray}
v_d^{n+\frac{1}{2}} & = & \gamma_1\frac{r^{n+1}-r^n}{\Delta{t}}+\frac{\sqrt{c_3}}{\sqrt{2}m}\sqrt{\frac{1-c_3\gamma_1^2}{1-c_2}}\,\beta^{n+\frac{1}{2}}\, ,\label{eq:uel_v*}
\end{eqnarray}
where $\gamma_1$ has not yet been determined. Implementing this half-step expression into the GJ algorithm of Eq.~(\ref{eq:A_Split_GJ}) gives
\begin{subequations}
\begin{eqnarray}
&&  v^{n+\frac{1}{4}} \; = \; v_1^n+\sqrt{\frac{c_3}{c_1}}\frac{\Delta{t}}{2m}f^n \nonumber \\
&&r^{n+\frac{1}{2}} \; = \; r^n+\sqrt{\frac{c_3}{c_1}}\frac{\Delta{t}}{2}v^{n+\frac{1}{4}} \nonumber \\
v_d^{n+\frac{1}{2}} 
& = &  \gamma_1\sqrt{c_1c_3}\,v^{n+\frac{1}{4}} + \frac{\sqrt{c_3}}{\sqrt{2}m}\sqrt{\frac{1-c_1c_3\gamma_1^2}{1-c_2}}\,\beta_+^{n} \label{eq:D1_v2}\\
v^{n+\frac{3}{4}} & = &\frac{c_2}{\gamma_1\sqrt{c_1c_3}}v_d^{n+\frac{1}{2}} +\frac{1}{\gamma_1m}\frac{c_1c_3\gamma_1^2-c_2}{\sqrt{1-c_2}\sqrt{1-c_1c_3\gamma_1^2}}\beta_+^n \nonumber \\ & + & \frac{1}{m}\sqrt{c_1c_3}\sqrt{\frac{1-c_3\gamma_1^2}{1-c_1c_3\gamma_1^2}}\beta_-^{n+1}  \label{eq:D1_v34} \\
&& r^{n+1} \; = \; r^{n+\frac{1}{2}}+\sqrt{\frac{c_3}{c_1}}\frac{\Delta{t}}{2}v^{n+\frac{3}{4}} \nonumber \\
&& v_1^{n+1} \; = \; v^{n+\frac{3}{4}}+\sqrt{\frac{c_3}{c_1}}\frac{\Delta{t}}{2m}f^{n+1} \, , \nonumber
\end{eqnarray}\label{eq:D1_Split_GJ}\noindent
\end{subequations}
where we have introduced
\begin{subequations}
\begin{eqnarray}
\beta_+^n & = & \frac{\gamma_1\sqrt{\frac{c_3}{2}}\beta^{n+1}+\sqrt{\frac{1-c_3\gamma_1^2}{1-c_2}}\beta^{n+\frac{1}{2}}}{\sqrt{\frac{1-c_1c_3\gamma_1^2}{1-c_2}}} \label{eq:D1_noise_p}\\
\beta_-^{n+1} & = & \frac{\sqrt{\frac{1-c_3\gamma_1^2}{1-c_2}}\beta^{n+1}-\gamma_1\sqrt{\frac{c_3}{2}}\beta^{n+\frac{1}{2}}}{\sqrt{\frac{1-c_1c_3\gamma_1^2}{1-c_2}}} \, , \label{eq:D1_noise_m}
\end{eqnarray}\label{eq:Noise_HS_v3}\noindent
\end{subequations}
such that $\beta_-^{n+1}$ does not appear in Eq.~(\ref{eq:D1_v2}) for $v_d^{n+\frac{1}{2}}$, and such that $\beta_+^n$ and $\beta_-^{n+1}$ are mutually independent, both being $N(0,2\alpha\Delta{t}k_BT)$. The significant simplification, replacing Eq.~(\ref{eq:D1_v34}) with Eq.~(\ref{eq:A_Comb_Thermo}), may also apply here.

There are at least two notable special cases:\\

\begin{figure}[t]
\centering
\scalebox{0.45}{\centering \includegraphics[trim={2.0cm 5.0cm 1cm 6.5cm},clip]{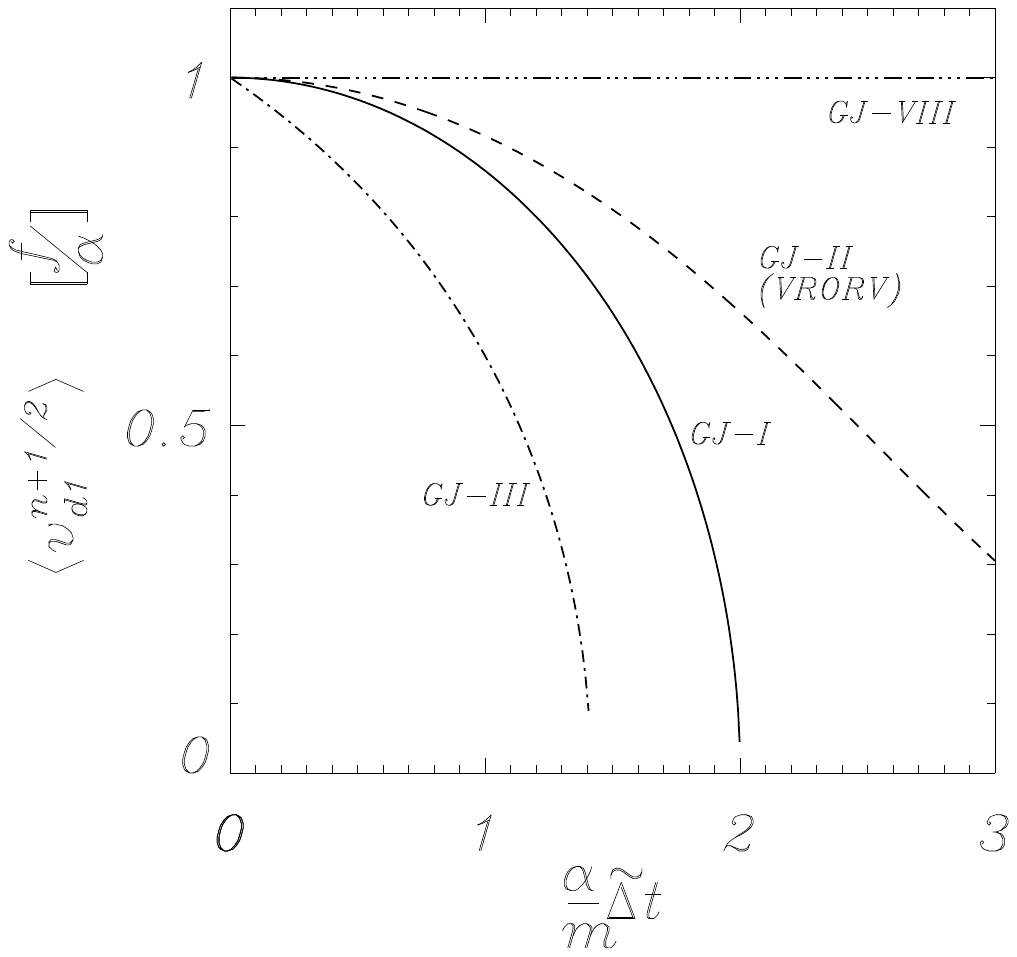}}
\caption{Drift velocity, Eq.~(\ref{eq:drift_d1}), for constant force, $f\neq0$, as a function of reduced, scaled time step for the fully symmetric split-operator methods of Eq.~(\ref{eq:D2_Split_GJ}). Velocity is $v_{d1}^{n+\frac{1}{2}}$ of Eq.~(\ref{eq:uel_v*}) with $\gamma_{1_\pm}$ from Eq.~(\ref{eq:gamma_1_symm}) for methods given in Eqs.~(\ref{eq:c2_GJ_1_2_3}) and (\ref{eq:c2_GJ8}). Special case, GJ-II, is the half-step velocity of the VRORV method \cite{Sivak}.
}
\label{fig:fig_9}
\end{figure}

\noindent
{\underline{\bf 1. Split-Operator Symmetry:}} Equating the velocity attenuation factors in the two thermodynamic operations, Eq.~(\ref{eq:D1_Split_GJ}), yields
\begin{eqnarray}
\gamma_1 & = & \sqrt{\frac{c_2}{c_1c_3}} \label{eq:gamma_1_symm}
\end{eqnarray}
which, when inserted into Eq.~(\ref{eq:D1_Split_GJ}), reads
\begin{subequations}
\begin{eqnarray}
&&  v^{n+\frac{1}{4}} \; = \; v_1^n+\sqrt{\frac{c_3}{c_1}}\frac{\Delta{t}}{2m}f^n \nonumber \\
&&r^{n+\frac{1}{2}} \; = \; r^n+\sqrt{\frac{c_3}{c_1}}\frac{\Delta{t}}{2}v^{n+\frac{1}{4}} \nonumber \\
v_{d1}^{n+\frac{1}{2}} 
& = &  \sqrt{c_2}\,v^{n+\frac{1}{4}} + \frac{\sqrt{c_3}}{\sqrt{2}m}\,\beta_+^{n} \label{eq:D2_v2}\\
v^{n+\frac{3}{4}} & = &\sqrt{c_2}\,v_{d1}^{n+\frac{1}{2}} + \frac{\sqrt{c_3}}{\sqrt{2}m}\,\beta_-^{n+1}  \label{eq:D2_v34} \\
&& r^{n+1} \; = \; r^{n+\frac{1}{2}}+\sqrt{\frac{c_3}{c_1}}\frac{\Delta{t}}{2}v^{n+\frac{3}{4}} \nonumber \\
&& v_1^{n+1} \; = \; v^{n+\frac{3}{4}}+\sqrt{\frac{c_3}{c_1}}\frac{\Delta{t}}{2m}f^{n+1} \, , \nonumber
\end{eqnarray}\label{eq:D2_Split_GJ}\noindent
\end{subequations}
where, for this choice of $\gamma_1$, the noise relationships of Eq.~(\ref{eq:Noise_HS_v3}) become
\begin{subequations}
\begin{eqnarray}
\beta_+^n & = & \frac{\sqrt{c_2}\beta^{n+1}+\beta^{n+\frac{1}{2}}}{\sqrt{1+c_2}} \label{eq:D2_noise_p}\\
\beta_-^{n+1} & = & \frac{\beta^{n+1}-\sqrt{c_2}\beta^{n+\frac{1}{2}}}{\sqrt{1+c_2}} \, . \label{eq:D2_noise_m}
\end{eqnarray}\label{eq:Noise_HS_vb1}\noindent
\end{subequations}

This method is perfectly symmetric with one noise component, $\beta_+^n$, being applied only to the first half of the time step, and the other, $\beta_-^{n+1}$, being applied only to the second. This set of methods is a GJ-generalization of the VRORV method \cite{Sivak} (see Eq.~(\ref{eq:c2_GJ2})). As is the case for VRORV, these methods generally do not provide correct measure of drift in either half-step or on-site velocity. The latter is given by the fact that this set of methods share its on-site velocity with the $v_1^n$ velocity of the GJ set, and the former is seen by, e.g., inserting $\gamma_1$ from Eq.~(\ref{eq:gamma_1_symm}) into the velocity expression in Eq.~(\ref{eq:uel_v*}), from where we see that
\begin{eqnarray}
\langle v_{d1}^{n+\frac{1}{2}}\rangle & = & \gamma_1\frac{f}{\alpha} \; = \; \sqrt{\frac{c_2}{c_1c_3}}\frac{f}{\alpha} \label{eq:drift_d1}
\end{eqnarray}
for $f={\rm const}$. As exemplified in Fig.~\ref{fig:fig_9}, this is generally incorrect, except for one special case, $c_2=c_1c_3$, for which this symmetrical method always measures correct drift velocity in $v_{d1}^{n+\frac{1}{2}}$; i.e., 
\begin{eqnarray}
c_2 & = & \sqrt{(\frac{\alpha\Delta{t}}{m})^2+1}-\frac{\alpha\Delta{t}}{m} \,.  \label{eq:c2_GJ8}
\end{eqnarray}
We may denote this method GJ-VIII, following the numbering of other GJ methods in Refs.~\cite{GJ,Josh_3}.

The generalized double-noise split-operator expressions of Eq.~(\ref{eq:D2_Split_GJ}) with symmetry in all operations, interactive/inertial/thermodynamic,  exemplify the difficulty of the split-operator formalism to simultaneously accomplish several key algorithmic objectives, such as correct drift in the configurational coordinate, requiring the attenuation parameter to appear outside the thermodynamic operations, and correct drift in the half-step velocity not being possible with repeated applications of the same thermodynamic operator.
\\

\noindent
{\underline{\bf 2. Correct Drift Velocity:}}
Directly enforcing the half-step drift velocity condition, Eq.~(\ref{eq:Cond_uel_drift}), and combining this with Eq.~(\ref{eq:uel_cond_0}) for $\gamma_3=\gamma_6=0$, yields
\begin{eqnarray}
\gamma_1 & = & 1\, ,  \label{eq:gamma_1_HS1b}
\end{eqnarray}
which from Eq.~(\ref{eq:uel_v*}) is easily validated to give correct drift with Eq.~(\ref{eq:uel_gamma_8}) ensuring correct Maxwell-Boltzmann distribution.
Inserting this into Eq.~(\ref{eq:D1_Split_GJ}) gives
\begin{subequations}
\begin{eqnarray}
&&  v^{n+\frac{1}{4}} \; = \; v_1^n+\sqrt{\frac{c_3}{c_1}}\frac{\Delta{t}}{2m}f^n \nonumber \\
&&r^{n+\frac{1}{2}} \; = \; r^n+\sqrt{\frac{c_3}{c_1}}\frac{\Delta{t}}{2}v^{n+\frac{1}{4}} \nonumber \\
v_{d2}^{n+\frac{1}{2}} 
& = &  \sqrt{c_1c_3}\,v^{n+\frac{1}{4}} + \frac{\sqrt{c_3}}{\sqrt{2}m}\sqrt{\frac{1-c_1c_3}{1-c_2}}\,\beta_+^{n} \label{eq:D3_v2}\\
v^{n+\frac{3}{4}}  & = &\frac{c_2}{\sqrt{c_1c_3}}v_{d2}^{n+\frac{1}{2}} +\frac{1}{m}\frac{c_1c_3-c_2}{\sqrt{1-c_2}\sqrt{1-c_1c_3}}\beta_+^n \nonumber \\ & + & \frac{1}{m}\sqrt{c_1c_3}\sqrt{\frac{1-c_3}{1-c_1c_3}}\beta_-^{n+1}  \label{eq:D3_v34} \\
&& r^{n+1} \; = \; r^{n+\frac{1}{2}}+\sqrt{\frac{c_3}{c_1}}\frac{\Delta{t}}{2}v^{n+\frac{3}{4}} \nonumber \\
&& v_1^{n+1} \; = \; v^{n+\frac{3}{4}}+\sqrt{\frac{c_3}{c_1}}\frac{\Delta{t}}{2m}f^{n+1}\, ,  \nonumber
\end{eqnarray}\label{eq:D3_Split_GJ}\noindent
\end{subequations}
where the split-operator stochastic values become
\begin{subequations}
\begin{eqnarray}
\beta_+^n & = & \frac{\sqrt{\frac{c_3}{2}}\beta^{n+1}+\sqrt{\frac{1-c_3}{1-c_2}}\beta^{n+\frac{1}{2}}}{\sqrt{\frac{1-c_1c_3}{1-c_2}}} \label{eq:D3_noise_p}\\
\beta_-^{n+1} & = & \frac{\sqrt{\frac{1-c_3}{1-c_2}}\beta^{n+1}-\sqrt{\frac{c_3}{2}}\beta^{n+\frac{1}{2}}}{\sqrt{\frac{1-c_1c_3}{1-c_2}}}\, .  \label{eq:D3_noise_m}
\end{eqnarray}\label{eq:Noise_HS_v4}\noindent
\end{subequations}

\begin{figure}[t]
\centering
\scalebox{0.5}{\centering \includegraphics[trim={2.5cm 11.5cm 0cm 3.5cm},clip]{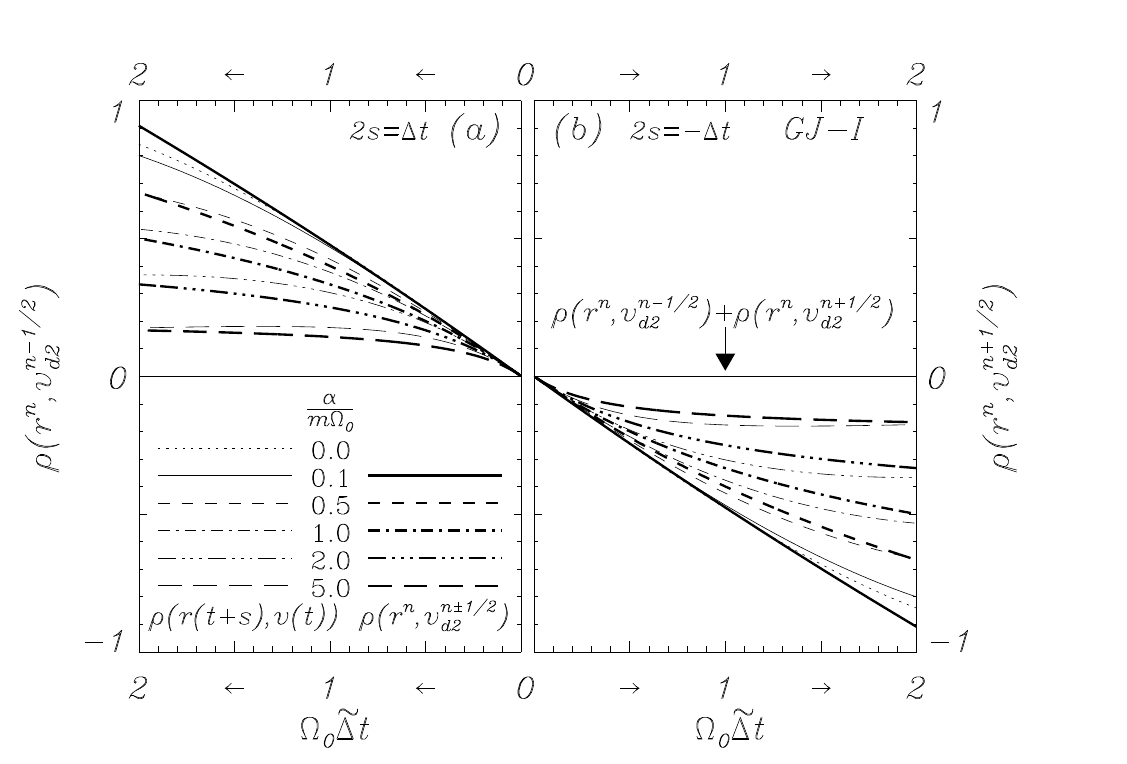}}
\caption{Cross-correlations between the configurational coordinate and the half-step velocity, $v_{d2}^{n+\frac{1}{2}}$ of Eq.~(\ref{eq:HS_vb2_from_v1}), for the GJ-I method with several different damping parameters. Thick curves represent the expression of the form in Eq.~(\ref{eq:cross_corr}), using the covariance of Eq.~(\ref{eq:double_noise_correlation}), as well as simulations of Eq.~(\ref{eq:D32_Split_GJ}), which coincide with the analysis. Thin curves represent the continuous-time expectation, Eq.~(\ref{eq:Cont_corr_vr}) with $2s=\Delta{t}$. Simulation details given in Sec.~\ref{sec:HS_one_numerical}.
}
\label{fig:fig_10}
\end{figure}

As noted for other expressed algorithms above, one may choose to replace Eq.~(\ref{eq:D3_v34}) with Eq.~(\ref{eq:A_Comb_Thermo}) for simplicity. We may also entirely bypass the noise values $\beta_\pm^n$ by writing
\begin{eqnarray}
v_{d2}^{n+\frac{1}{2}} & = & \sqrt{c_3}v_1^{n+\frac{1}{2}}+\frac{\sqrt{c_3}}{\sqrt{2}m}\sqrt{\frac{1-c_3}{1-c_2}}\,\beta^{n+\frac{1}{2}} \, , \label{eq:HS_vb2_from_v1}
\end{eqnarray}
then include this into the standard GJ set from Eq.~(\ref{eq:A_Split_GJ}) as follows
\begin{subequations}
\begin{eqnarray}
&&  v^{n+\frac{1}{4}} \; = \; v_1^n+\sqrt{\frac{c_3}{c_1}}\frac{\Delta{t}}{2m}f^n \nonumber \\
&&r^{n+\frac{1}{2}} \; = \; r^n+\sqrt{\frac{c_3}{c_1}}\frac{\Delta{t}}{2}v^{n+\frac{1}{4}} \nonumber \\
&& v_1^{n+\frac{1}{2}} \; = \; \sqrt{c_1}\,v^{n+\frac{1}{4}} + \frac{\sqrt{c_3}}{2m}\,\beta^{n+1} \nonumber \\
v_{d2}^{n+\frac{1}{2}} & = & \sqrt{c_3}v_1^{n+\frac{1}{2}}+\frac{\sqrt{c_3}}{\sqrt{2}m}\sqrt{\frac{1-c_3}{1-c_2}}\,\beta^{n+\frac{1}{2}}\\
&& v^{n+\frac{3}{4}}  \; = \; \frac{c_2}{\sqrt{c_1}}\,v_1^{n+\frac{1}{2}} + \sqrt{\frac{c_3}{c_1}}\frac{1}{2m}\,\beta^{n+1} \nonumber  \\
&& r^{n+1} \; = \; r^{n+\frac{1}{2}}+\sqrt{\frac{c_3}{c_1}}\frac{\Delta{t}}{2}v^{n+\frac{3}{4}} \nonumber \\
&& v_1^{n+1} \; = \; v^{n+\frac{3}{4}}+\sqrt{\frac{c_3}{c_1}}\frac{\Delta{t}}{2m}f^{n+1}\, ,  \nonumber
\end{eqnarray}\label{eq:D32_Split_GJ}\noindent
\end{subequations}
where both half-step velocities, $v_1^{n+\frac{1}{2}}$ and $v_{d2}^{n+\frac{1}{2}}$, are obtained in each time step. Of course, one can further incorporate Eqs.~(\ref{eq:Z2_vpm2}) and (\ref{eq:Z2_vs2}) as well, thereby integrating the Langevin equation with several different measures of the half-step velocity in each time step. From the defining expression, Eq.~(\ref{eq:D3_v2}), of $v_{d2}^{n+\frac{1}{2}}$ with Eqs.~(\ref{eq:r2_corr}) and (\ref{eq:r2p1}), we easily find the covariances
\begin{figure}[t]
\centering
\scalebox{0.5}{\centering \includegraphics[trim={2.5cm 11.5cm 0cm 3.5cm},clip]{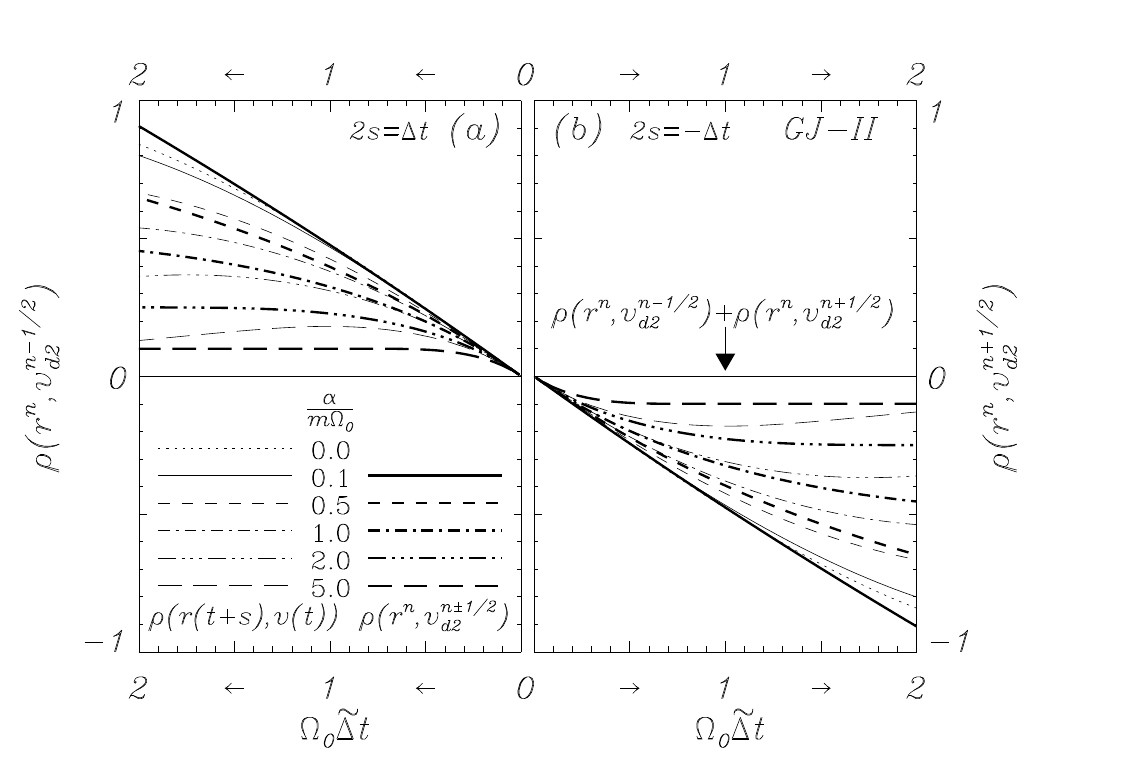}}
\caption{Cross-correlations between the configurational coordinate and the half-step velocity, $v_{d2}^{n+\frac{1}{2}}$ of Eq.~(\ref{eq:HS_vb2_from_v1}), for the GJ-II method with several different damping parameters. Thick curves represent the expression of the form in Eq.~(\ref{eq:cross_corr}), using the covariance of Eq.~(\ref{eq:double_noise_correlation}), as well as simulations of Eq.~(\ref{eq:D32_Split_GJ}), which coincide with the analysis. Thin curves represent the continuous-time expectation, Eq.~(\ref{eq:Cont_corr_vr}) with $2s=\Delta{t}$. Simulation details given in Sec.~\ref{sec:HS_one_numerical}.
}
\label{fig:fig_11}
\end{figure}
\begin{eqnarray}
\left\langle r^nv_{d2}^{n\mp\frac{1}{2}}\right\rangle & = & \pm\frac{k_BT}{\sqrt{\kappa m}}\frac{c_3}{2}\Omega_0\Delta{t} \, . \label{eq:double_noise_correlation}
\end{eqnarray}
Using this expression to generate the cross-correlation of the form given in Eq.~(\ref{eq:Cont_corr_vr}) yields Figs.~\ref{fig:fig_10} and \ref{fig:fig_11} (as well as Figs.~\ref{fig:fig_8}ab) for the three methods given in Eq.~(\ref{eq:c2_GJ_1_2_3}). Simulation results from Eq.~(\ref{eq:D32_Split_GJ}) are visually indistinguishable from the analysis, and the perfect antisymmetry is confirmed, as is the correct and time-step-independent kinetic statistics shown in Fig.~\ref{fig:fig_5}.

\section{On-Site Velocities}
\label{sec:On-site}\noindent
Similarly to the ansatz for the half-step velocity, Eq.~(\ref{eq:uel_ansatz}), we write the on-site velocity $v^n$ at time $t_n$.\\
\begin{eqnarray}
v^{n} & = & \frac{\mu_1r^{n+1}+\mu_2r^n+\mu_3r^{n-1}+\mu_6r^{n-2}}{\Delta{t}} \nonumber \\
&& + \frac{\mu_5\beta^{n+1}+\mu_4\beta^{n}+\mu_7\beta^{n-1}}{m}\, ,  \label{eq:vel_ansatz}
\end{eqnarray}
where we, in analogy with the half-step velocity ansatz in Eq.~(\ref{eq:uel_ansatz}), seek to determine the seven functional parameters, $\mu_i=\mu_i(\frac{\alpha\Delta{t}}{m})$, such that $v_n$ gives the best possible statistical measure of the continuous-time velocity, given the GJ trajectory, $r^n$, from Eq.~(\ref{eq:Stormer_GJ}). 
As mentioned for the half-step velocity in Eq.~(\ref{eq:uel_ansatz}), Eq.~(\ref{eq:vel_ansatz}) is an expanded ansatz compared to the one used in Ref.~\cite{GJ}, which included the $\mu_i$ terms only for $i=1,2,3,4,5$.
Analogous to Eq.~(\ref{eq:uel_cond_0}), we have 
\begin{eqnarray}
\mu_1+\mu_2+\mu_3+\mu_6 & = & 0\,  ,   \label{eq:vel_cond_0}
\end{eqnarray}
which is necessary to correctly represent the velocity of a particle at rest, as well as for obtaining meaningful velocities for $\Delta{t}\rightarrow0$.

\subsection{Conditions for being on-site}
\label{sec:HS_On-site}
A key property of an on-site velocity is given from Eq.~(\ref{eq:Corr_vr}); i.e., $v^n$ must satisfy the condition \cite{GJ}
\begin{eqnarray}
&& \left\langle r^nv^n\right\rangle \; = \; 0 \; \Rightarrow \label{eq:on-site_cond}\\
&& \frac{\mu_1+\mu_3}{\Delta{t}}\left\langle r^nr^{n+1}\right\rangle+\frac{\mu_2}{\Delta{t}}\left\langle r^nr^n\right\rangle+\frac{\mu_6}{\Delta{t}}\left\langle r^{n-1}r^{n+1}\right\rangle \label{eq:on-site_corr-2}\\
&& + \frac{1}{m}\Big({\mu_5\cancel{\left\langle\beta^{n+1}r^n\right\rangle}}+\mu_4\left\langle\beta^nr^n\right\rangle+\mu_7\left\langle\beta^{n-1}r^n\right\rangle\Big)\; = \; 0 \nonumber
\end{eqnarray}
where $\langle\beta^{n+1}r^n\rangle=0$, since Eq.~(\ref{eq:int_noise}) shows that $r^n$ must be independent of $\beta^{n+1}$. 

Inserting Eqs.~(\ref{eq:r2_corr}) and (\ref{eq:rbr_corr}) into Eq.~(\ref{eq:on-site_corr-2}), then using Eq.~(\ref{eq:vel_cond_0}), we rewrite Eq.~(\ref{eq:on-site_cond}) as
\begin{eqnarray}
\langle v^nr^n\rangle & = & \frac{k_BT}{\sqrt{\kappa m}}\sum_{k=0}^{2}(\Omega_0\Delta{t})^{2k-1}\Gamma_{2k-1}\; = \; 0 \, ,  \label{eq:on-site_Gamma}
\end{eqnarray}
where $\Gamma_{2k-1}=\Gamma_{2k-1}(\frac{\alpha\Delta{t}}{m})$ does not depend on $\Omega_0\Delta{t}$, and
Eq.~(\ref{eq:vel_cond_0}) ensures that $\Gamma_{-1}=0$. Given $(1-X)$ from Eq.~(\ref{eq:X}) and the relationships of Eq.~(\ref{eq:rbr_corr}), we can then write
\begin{eqnarray}
\Gamma_1 & = & \frac{c_3}{2}\left(\mu_2+2\frac{\alpha\Delta{t}}{m}\mu_4\right)-c_3\frac{1+2c_1}{2}\left(\mu_6-2\frac{\alpha\Delta{t}}{m}\mu_7\right) \nonumber \\ \label{eq:Gamma_1} \\
\Gamma_3 & = & \frac{c_3^2}{2}\left(\mu_6-2\frac{\alpha\Delta{t}}{m}\mu_7\right) \, .\label{eq:Gamma_3}
\end{eqnarray}
The conditions,  $\Gamma_1=\Gamma_3=0$, for the velocity to be on-site, are therefore
\begin{subequations}
\begin{eqnarray}
-\mu_2 & = & 2\frac{\alpha\Delta{t}}{m}\mu_4 \label{eq:vel_cond_1}\\
\mu_6 & = & 2\frac{\alpha\Delta{t}}{m}\mu_7\, ,  \label{eq:vel_cond_2}
\end{eqnarray}\label{eq:vel_cond_12}\noindent
\end{subequations}
linking noise coefficients to the finite-difference part of the velocity expression.

\subsection{Causality condition for on-site velocity}
\label{sec:OS_causality}

Successively inserting Eq.~(\ref{eq:Stormer_GJ}) into Eq.~(\ref{eq:vel_ansatz}) for $v^{n+1}$ gives
\begin{eqnarray}
&&v^{n+1} \; = \; c_2\,v^n+\frac{c_3\Delta{t}}{m}[\mu_1f^{n+1}+(\mu_1+\mu_2)f^n-\mu_6f^{n-1}]\nonumber \\
&&- \frac{\mu_6\frac{c_3}{2}+c_2\mu_7}{m}\beta^{n-1}+\frac{\mu_7-c_2\mu_4+(\mu_1+\mu_2-\mu_6)\frac{c_3}{2}}{m}\beta^n \nonumber \\
&&+ \frac{\mu_4-c_2\mu_5+\mu_1c_3+\mu_2\frac{c_3}{2}}{m}\beta^{n+1}+\underbrace{\frac{\mu_5+\mu_1\frac{c_3}{2}}{m}}_{=0}\beta^{n+2} \, , \label{eq:vel_velm1}
\end{eqnarray}
where Eq.~(\ref{eq:vel_cond_0}) has been used to remove explicit representation of $\mu_3$ from the expression, and where
\begin{eqnarray}
\mu_5 & = & -\mu_1\frac{c_3}{2} \label{eq:vel_cond_3}
\end{eqnarray}
since, according to Eq.~(\ref{eq:int_noise}), the on-site velocity, $v^{n+1}$, at time $t_{n+1}$ cannot depend on the integrated noise $\beta^{n+2}$ over $(t_{n+1},t_{n+2}]$. We note parenthetically that the first term on the right hand side of Eq.~(\ref{eq:vel_velm1}) shows that the pivotal parameter $c_2$ is the one-time-step velocity attenuation factor \cite{GJ}. We also note that Ref.~\cite{GJ} did not include the parameters, $\mu_6$ and $\mu_7$, and it imposed a constraint $c_3\mu_3=-2c_2\mu_4$ in order to avoid integrated noise prior to the beginning of a time step, which in Eq.~(\ref{eq:vel_velm1}) starts at $t_n$ (see the term $\propto\beta^n$ for $\mu_6=\mu_7=0$). This was done for aesthetic reasons, and in order to keep the resulting algorithm in the usual Verlet format in which no information prior to the beginning of a time step is explicitely represented in the equations. Such optional constraint is not imposed here.

The three conditions, Eqs.~(\ref{eq:vel_cond_12}) and (\ref{eq:vel_cond_3}), link the three coefficients of the noise component in the velocity ansatz to the finite-difference coefficients. Thus, we can write Eq.~(\ref{eq:vel_velm1}) as
\begin{eqnarray}
&&v^{n+1} \; = \; c_2\,v^n+\frac{c_3\Delta{t}}{m}[\mu_1f^{n+1}+(\mu_1+\mu_2)f^n-\mu_6f^{n-1}]\nonumber \\
&&-\frac{c_3}{2m}\frac{\mu_6}{1-c_2}\,\beta^{n-1}+\frac{c_3}{2m}\left(\frac{\mu_2+c_2\mu_6}{1-c_2}+\mu_1\right)\beta^n \nonumber \\
&&+ \frac{c_3}{2m}\left((1+2c_1)\mu_1-\frac{c_2\mu_2}{1-c_2}\right)\beta^{n+1} \; . \label{eq:vel_velm2}
\end{eqnarray}

\subsection{Conditions for Maxwell-Boltzmann distribution}
\label{sec:OS_MB}
Given the necessary parameter constraints, Eqs.~(\ref{eq:vel_cond_0}), (\ref{eq:vel_cond_12}), and (\ref{eq:vel_cond_3}), for $v^n$ to be an on-site velocity that is not influenced by future fluctuations, we can now investigate the properties of the second moment $\left\langle v^nv^n\right\rangle$ for Eq.~(\ref{eq:vel_ansatz}) combined with Eq.~(\ref{eq:Stormer_lin_complete}):

\begin{subequations}
\begin{eqnarray}
\left\langle v^nv^n\right\rangle & = & \frac{\mu_1^2+\mu_2^2+\mu_3^2+\mu_6^2}{\Delta{t}^2} \langle r^nr^n\rangle \nonumber \\
&+ & 2\frac{\mu_1\mu_2+\mu_2\mu_3+\mu_3\mu_6-c_2\mu_1\mu_6}{\Delta{t}^2} \langle r^nr^{n+1}\rangle\nonumber \\
&+ & 2 \frac{\mu_1\mu_3+\mu_2\mu_6+2c_1X\mu_1\mu_6}{\Delta{t}^2} \langle r^{n-1}r^{n+1}\rangle\nonumber \\
&+ & 2\frac{\mu_1\mu_5+\mu_2\mu_4+\mu_3\mu_7-c_2\mu_1\mu_7}{m\Delta{t}}\langle\beta^nr^n\rangle\nonumber \\
&+ & 2\frac{\mu_1\mu_4+\mu_2\mu_7+2c_1X\mu_1\mu_7}{m\Delta{t}}\langle\beta^{n-1}r^n\rangle\nonumber \\
&+ & \frac{\mu_4^2+\mu_5^2+\mu_7^2}{m^2}\langle\beta^n\beta^n\rangle\, .  \label{eq:v2_corr_0}
\end{eqnarray}
Given that $1-X=\frac{c_3}{2c_1}(\Omega_0\Delta{t})^2$ (from Eq.~(\ref{eq:X})), and using Eqs.~(\ref{eq:FD_discrete_2}) and (\ref{eq:rbr_corr}), we can see that $\langle v^nv^n\rangle$ can be written
\begin{eqnarray}
\left\langle v^nv^n\right\rangle & = & \frac{k_BT}{m}\sum_{k=-1}^{2}(\Omega_0\Delta{t})^{2k}\,\Gamma_{2k}\, ,  \label{eq:v2_corr_1}
\end{eqnarray} \label{eq:v2_corr_12}\noindent
\end{subequations}\noindent
where $\Gamma_{2k}=\Gamma_{2k}(\frac{\alpha\Delta{t}}{m})$ is a single-variable function of $\frac{\alpha\Delta{t}}{m}$ that does not depend on $\Omega_0\Delta{t}$, consistent with $\Gamma_{2k-1}$ given in Eq.~(\ref{eq:on-site_Gamma}). Ideally, the coefficients should be $\Gamma_{-2}=\Gamma_2=\Gamma_4=0$ and $\Gamma_0=1$ in order for the on-site velocity variable to be in agreement with the desired Maxwell-Boltzmann distribution, $\langle v^nv^n\rangle=\frac{k_BT}{m}$. Identifying the components of Eq.~(\ref{eq:v2_corr_0}) that match each of the four terms in Eq.~(\ref{eq:v2_corr_1}), we first find that 
\begin{subequations}
\begin{eqnarray}
\Gamma_{-2} & = & 0 \label{eq:Gamma_-2}
\end{eqnarray}
from Eq.~(\ref{eq:vel_cond_0}). Further, after some work, we find that combining Eq~(\ref{eq:v2_corr_12}) with Eqs.~(\ref{eq:vel_cond_0}), (\ref{eq:vel_cond_12}), and (\ref{eq:vel_cond_3}) yields
\begin{eqnarray}
\Gamma_{0} & = & c_3\left(2(1+c_1)\mu_1^2+\mu_1\mu_2\right)+2\frac{1-c_2}{c_3}(\mu_4^2-\mu_5^2+\mu_7^2) \nonumber \\
& = & \frac{c_3}{2}\left((5+3c_2)\mu_1^2+2\mu_1\mu_2+\frac{\mu_2^2+\mu_6^2}{1-c_2}\right) \, ; \label{eq:Gamma_0}
\end{eqnarray}
using Eqs.~(\ref{eq:vel_cond_0}) and (\ref{eq:vel_cond_12}) in Eq.~(\ref{eq:v2_corr_12}) gives
\begin{eqnarray}
\Gamma_2 & = & -c_3^2\mu_1^2 \label{eq:Gamma_2}\, ;
\end{eqnarray}
and combining Eq~(\ref{eq:v2_corr_12}) with Eq.~(\ref{eq:vel_cond_2}) gives
\begin{eqnarray}
\Gamma_{4} & = & c_3^2\mu_1(2\frac{\alpha\Delta{t}}{m}\mu_7-\mu_6) \; = \; 0\, .  \label{eq:Gamma_4}
\end{eqnarray}\label{eq:Gamma}\noindent
\end{subequations}
Thus, the two new conditions on $\mu_i$ from the Maxwell-Boltzmann requirement are
\begin{subequations}
\begin{eqnarray}
\Gamma_0 & = & 1 \label{eq:vel_cond_4}\\
\Gamma_2 & = & 0 \label{eq:vel_cond_5}
\end{eqnarray}\label{eq:vel_cond_45}\noindent
\end{subequations}
with Eqs.~(\ref{eq:Gamma_0}) and (\ref{eq:Gamma_2}), since Eq.~(\ref{eq:Gamma_4}) is redundant with Eq.~(\ref{eq:vel_cond_2}).

\subsection{Possibility for a correct on-site velocity}
\label{sec:OS-Possibility}

Summarizing the search for an on-site velocity that satisfies the fluctuation-dissipation balance, we have formulated the six independent conditions, Eqs.~(\ref{eq:vel_cond_0}), (\ref{eq:vel_cond_12}), (\ref{eq:vel_cond_3}), and (\ref{eq:vel_cond_45}) for determining the seven $\mu_i$ parameters in the ansatz Eq.~(\ref{eq:vel_ansatz}). Equations~(\ref{eq:Gamma_-2}) and (\ref{eq:Gamma_4}) are trivially satisfied for on-site velocities, and do therefore not contribute further to the determination of $\mu_i$.

Equations (\ref{eq:Gamma_0}) and (\ref{eq:vel_cond_4}) show that $\mu_2$, $\mu_6$ $\rightarrow0$ for $c_2\rightarrow1$ (i.e., in the continuous-time limit, $\frac{\alpha\Delta{t}}{m}\rightarrow0$). By Eq.~(\ref{eq:vel_cond_0}), this implies that $\mu_3\rightarrow-\mu_1$ for $\frac{\alpha\Delta{t}}{m}\rightarrow0$, which is consistent with the on-site central-difference velocity approximation Eq.~(\ref{eq:on-site_org}) for a lossless system. Since a velocity expression would be disconnected from the trajectory $r^n$ if all coefficients $\mu_i$ ($i=1,2,3,6$) were zero, we conclude that neither $\mu_1$ nor $\mu_3$ are generally zero, and certainly not in the limit $\frac{\alpha\Delta{t}}{m}\rightarrow0$, where $\mu_2$ and $\mu_6$ vanish, and where one therefore must expect that $\mu_1\rightarrow\frac{1}{2}$. This expectation will be substantiated in Eq.~(\ref{eq:gamma_1_limit0}) below. With that in mind, we will proceed under the assumption that, generally, $\mu_1\neq0$, except for the possibility of isolated point-values of $\frac{\alpha\Delta{t}}{m}>0$ (see the last part of Appendix~\ref{sec:Appendix_A} for a practical case where $\mu_1=0$ for $\frac{\alpha\Delta{t}}{m}>0$). However, attempting to enforce the condition Eq.~(\ref{eq:vel_cond_5}) with $\Gamma_2$ given by Eq.~(\ref{eq:Gamma_2}) implies that 
either $\mu_1$ or $c_3$ must be zero. Since $c_3>0$ for all relevant values of $c_2$ ($|c_2|\le1$ for stability), and since we have just argued that $\mu_1$ is generally non-zero, except perhaps in isolated points, we must conclude that the condition Eq.~(\ref{eq:vel_cond_5}) {\it cannot} generally be satisfied. 
In particular, for a Hooke potential, any of the possible on-site velocity measures will yield the statistical average of the kinetic energy
\begin{eqnarray}
\langle E_k \rangle & = & \frac{m}{2}\langle v^nv^n\rangle \; = \; \frac{k_BT}{2}\left(1-c_3^2\mu_1^2\Omega_0^2\Delta{t}^2\right)\, .  \label{eq:on-site_energy}
\end{eqnarray}
This expression illuminates the inevitable depression of the thermodynamic on-site kinetic measure for convex potentials, as given by the fact that the condition, Eq.~(\ref{eq:vel_cond_5}) with Eq.~(\ref{eq:Gamma_2}), cannot generally be enforced for on-site velocities.
Thus, we expand a conclusion of Ref.~\cite{GJ} that {\it it is not possible to construct a finite-difference, on-site velocity that for general time steps satisfies the Maxwell-Boltzmann distribution in discrete time}, to the addition of the parameters $\mu_6$ and $\mu_7$, and without the aesthetic constraint, $c_3\mu_3=-2c_2\mu_4$, of Ref.~\cite{GJ}.

Given that the condition in Eq.~(\ref{eq:vel_cond_5}) cannot be generally satisfied in a meaningful way, we focus on condition Eq.~(\ref{eq:vel_cond_4}) and accept the second order error term represented by the coefficient $\Gamma_2$ in Eq.~(\ref{eq:Gamma_2}).

\subsection{Correct on-site drift velocity}
\label{sec:OS_drift}
The condition for correct drift velocity, given in Eq.~(\ref{eq:drift}) due to a constant force, can be deduced directly from Eq.~(\ref{eq:vel_ansatz}) or from Eq.~({\ref{eq:vel_velm1}) to be
\begin{subequations}
\begin{eqnarray}
\langle v^n\rangle & = & v_d \; \; \Rightarrow \nonumber \\
2\mu_1+\mu_2-\mu_6 & = & 1\, . \label{eq:vel_cond_6a}
\end{eqnarray}
Inserting this drift condition into Eq.~(\ref{eq:Gamma_0}), and enforcing the condition Eq.~(\ref{eq:vel_cond_4}), gives
\begin{eqnarray}
c_1(5-3c_2)\mu_1^2+(\mu_6-2c_1\mu_1)(1+\mu_6) & = & \frac{1-c_2}{c_3}-\frac{1}{2}\, , \nonumber \\  \label{eq:vel_cond_6b}
\end{eqnarray}\label{eq:vel_cond_6}\noindent
\end{subequations}
which, given a value for $\mu_6$, is a simple second order polynomial in $\mu_1$.

\subsection{Antisymmetric on-site cross-correlation}
\label{sec:OS_anti_S}

In addition to the on-site covariance condition Eq.~(\ref{eq:on-site_cond}), one may impose the known antisymmetry of the covariance, expressed for continuous-time in Eq.~(\ref{eq:Corr_vr}), by the condition
\begin{eqnarray}
&&\langle r^nv^{n+1}\rangle+\langle r^{n+1}v^n\rangle \; = \; 0\, .   \label{eq:on-site_cross_cond_a}
\end{eqnarray}
Inserting Eq.~(\ref{eq:vel_ansatz}) combined with Eq.~(\ref{eq:Stormer_lin_complete}) into Eq.~(\ref{eq:on-site_cross_cond_a}) yields
\begin{eqnarray}
&& \frac{\mu_1+\mu_3}{\Delta{t}}\langle r^nr^n\rangle + \frac{2\mu_2+(1-c_2)\mu_6}{\Delta{t}}\langle r^nr^{n+1}\rangle \nonumber \\
&&+\frac{\mu_1+\mu_3+2c_1X\mu_6}{\Delta{t}}\langle r^{n-1}r^{n+1}\rangle + \frac{\mu_5+(1-c_2)\mu_7}{m}\langle\beta^nr^n\rangle\nonumber \\
&&+\frac{\mu_4+2c_1X\mu_7}{m}\langle\beta^nr^{n+1}\rangle \; = \; 0\, ,  \label{eq:on-site_cross_cond_b}
\end{eqnarray}
where inserting the known covariances, Eq.~(\ref{eq:rbr_corr}), and applying the core on-site conditions of Eqs.~(\ref{eq:vel_cond_0}), (\ref{eq:vel_cond_12}), and (\ref{eq:vel_cond_3}), as well as the condition of Eq.~(\ref{eq:vel_cond_4}), further yields
\begin{eqnarray}
\langle r^nv^{n+1}\rangle&+&\langle r^{n+1}v^n\rangle  \nonumber \\
&=& -c_1\frac{1-X}{\Delta{t}}\left((1-c_2)\mu_1+\mu_2\right)\frac{k_BT}{\kappa} \nonumber \\ 
& = & -\frac{c_3}{2}\Omega_0\Delta{t}((1-c_2)\mu_1+\mu_2)\frac{k_BT}{\sqrt{\kappa m}}\, .  \label{eq:on-site_asymmetry}
\end{eqnarray}
Enforcing the condition, Eq.~(\ref{eq:on-site_cross_cond_a}), of antisymmetry of the cross-correlation implies that
\begin{eqnarray}
\mu_2 & = & -(1-c_2)\mu_1\, , \label{eq:vel_cond_7}
\end{eqnarray}
where we notice that the antisymmetry in the cross-correlation between the configurational coordinate and the on-site velocity is given solely by the relationship between $\mu_1$ and $\mu_2$. Also notice that we have not imposed the drift condition Eq.~(\ref{eq:vel_cond_6a}) in order to arrive at the condition, Eq.~(\ref{eq:vel_cond_7}).

\begin{figure}[t]
\centering
\scalebox{0.5}{\centering \includegraphics[trim={2.0cm 5.0cm 1cm 6.0cm},clip]{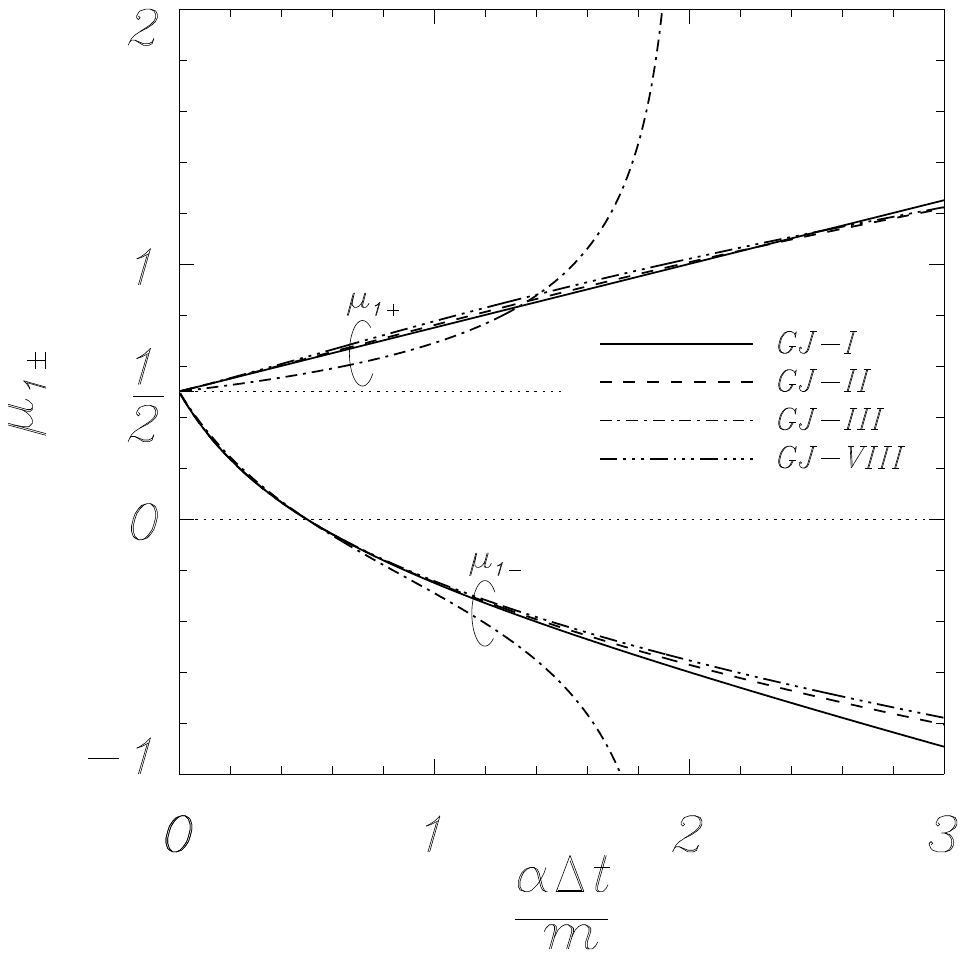}}
\caption{Lead coefficients, $\mu_{1_\pm}$, of Eq.~(\ref{eq:gamma_1_roots}), for the on-site velocities, $v_\pm^n$ from Eq.~(\ref{eq:vpm_condensed}) and $w_\pm^n$ from Eq.~(\ref{eq:wpm_condensed}), as a function of reduced time step for methods given in Eqs.~(\ref{eq:c2_GJ_1_2_3}) and (\ref{eq:c2_GJ8}).
}
\label{fig:fig_12}
\end{figure}

\subsection{Constructing on-site velocities}
\label{sec:OS_OS_vel}
Combining the conditions of Eqs.~(\ref{eq:vel_cond_6a}) and (\ref{eq:vel_cond_7}) yields $\mu_6=2c_1\mu_1-1$, which can be inserted into Eq.~(\ref{eq:vel_cond_6b}) to give a quadratic equation in $\mu_1$,
\begin{eqnarray}
c_1(5-3c_2)\mu_1^2-2c_1\mu_1+\frac{1}{2}-\frac{1-c_2}{c_3} & = & 0\, , \label{eq:gamma_1_equation}
\end{eqnarray}
and the two resulting values of $\mu_1$ are (Fig.~\ref{fig:fig_12})
\begin{eqnarray}
\mu_{1_\pm} & = & \frac{1\pm\sqrt{\frac{1}{c_1}\frac{\alpha\Delta{t}}{m}}\sqrt{3(1-c_2)+2(1-c_3)}}{5-3c_2}\, . \label{eq:gamma_1_roots}
\end{eqnarray}
We notice that
\begin{eqnarray}
\mu_{1_\pm} & \rightarrow & \frac{1}{2}+\frac{\alpha\Delta{t}}{2m}\left(\pm\sqrt{3+c_2^{\prime\prime}(0)}-\frac{3}{2}\right) \; \; {\rm for} \; \frac{\alpha\Delta{t}}{m}\rightarrow0 \, , \nonumber \\ \label{eq:gamma_1_limit0}
\end{eqnarray}
where $c_2^{\prime\prime}(0)$ is the curvature of $c_2(\frac{\alpha\Delta{t}}{m})$ for $\frac{\alpha\Delta{t}}{m}\rightarrow0$.

We also notice that for $c_3=c_1$, i.e., the GJ-I method \cite{GJF1,2GJ,GJ}, $\mu_{1_+}=\frac{1}{2c_1}$, which represents the traditional on-site velocity, $v_1^n$, associated with this method.

Based on Eq.~(\ref{eq:gamma_1_roots}) we now outline two pairs of options for on-site velocities associated with GJ methods with one-time-step attenuation parameter $c_2$.

\begin{figure}[t]
\centering
\scalebox{0.525}{\centering \includegraphics[trim={3.25cm 14.0cm 2cm 4.5cm},clip]{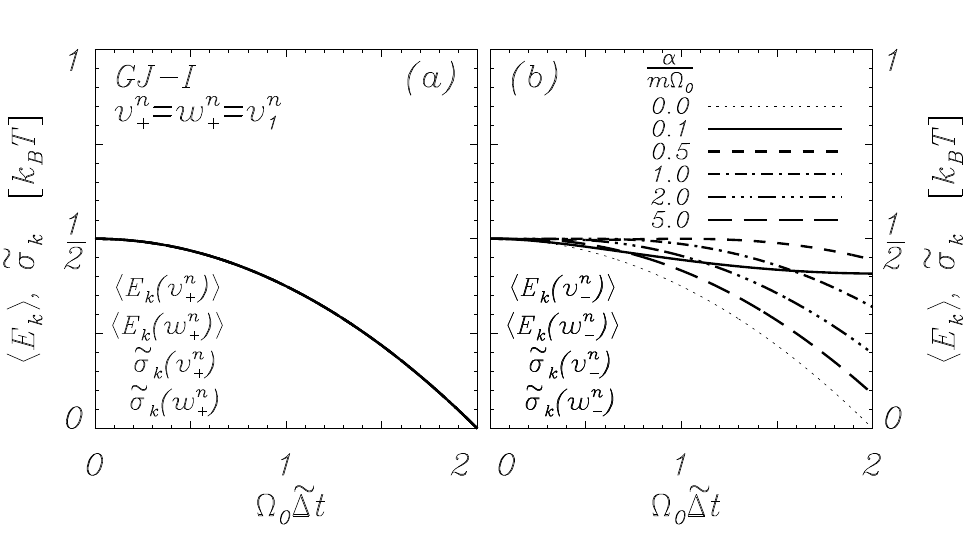}}
\caption{Kinetic statistics for a noisy harmonic oscillator with damping parameter as shown in the figure for the on-site velocities, $v_\pm^n$ from Eq.~(\ref{eq:vpm_condensed}) and $w_\pm^n$ from Eq.~(\ref{eq:wpm_condensed}), as a function of reduced time step for the GJ-I method, Eq.~(\ref{eq:c2_GJ1}). Simulation results of Eqs.~(\ref{eq:vnpm_Split_GJ}) and (\ref{eq:vnpm2_Split_GJ}) are indistinguishable from the predicted results of Eq.~(\ref{eq:on-site_energy}). Fluctuations, $\sigma_k=\sqrt{2}\,\widetilde{\sigma}_k$ are in agreement with the predicted, $\widetilde{\sigma}_k=\langle{E_k}\rangle$. Simulation details given in Sec.~\ref{sec:HS_one_numerical}. Panels (a) and (b) display data for the quantities listed in the lower left corner of each panel.
}
\label{fig:fig_13}
\end{figure}

\begin{figure}[t]
\centering
\scalebox{0.525}{\centering \includegraphics[trim={3.25cm 14.0cm 2cm 4.5cm},clip]{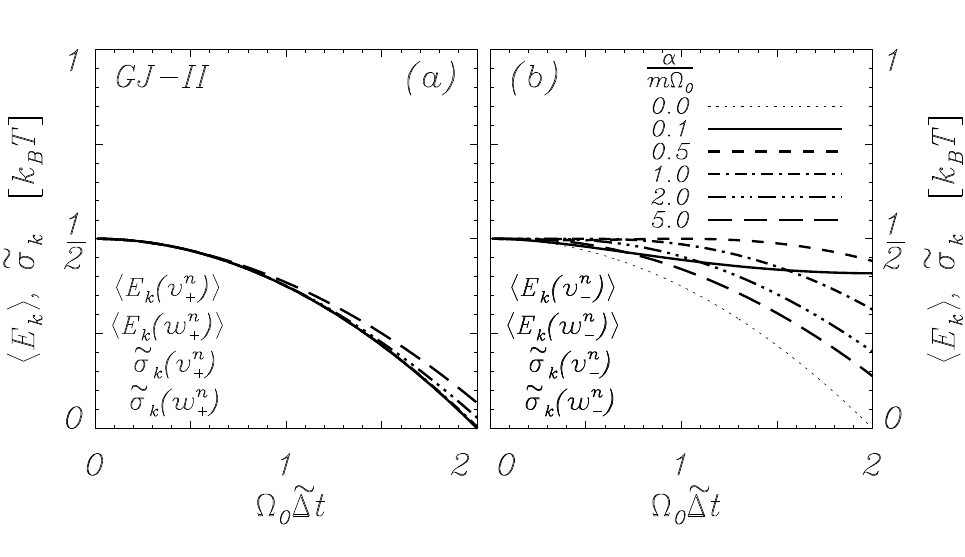}}
\caption{Kinetic statistics for a noisy harmonic oscillator with damping parameter as shown in the figure for the on-site velocities, $v_\pm^n$ from Eq.~(\ref{eq:vpm_condensed}) and $w_\pm^n$ from Eq.~(\ref{eq:wpm_condensed}), as a function of reduced time step for the GJ-II method, Eq.~(\ref{eq:c2_GJ2}). Simulation results of Eqs.~(\ref{eq:vnpm_Split_GJ}) and (\ref{eq:vnpm2_Split_GJ}) are indistinguishable from the predicted results of Eq.~(\ref{eq:on-site_energy}). Fluctuations, $\sigma_k=\sqrt{2}\,\widetilde{\sigma}_k$ are in agreement with the predicted, $\widetilde{\sigma}_k=\langle{E_k}\rangle$. Simulation details given in Sec.~\ref{sec:HS_one_numerical}. Panels (a) and (b) display data for the quantities listed in the lower left corner of each panel.
}
\label{fig:fig_14}
\end{figure}

\begin{figure}[t]
\centering
\scalebox{0.525}{\centering \includegraphics[trim={3.25cm 14.0cm 2cm 4.5cm},clip]{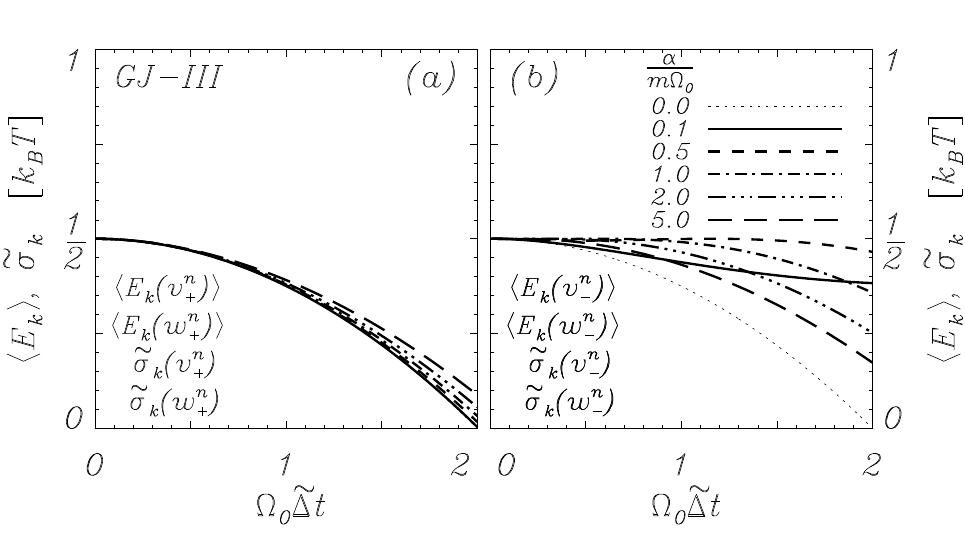}}
\caption{Kinetic statistics for a noisy harmonic oscillator with damping parameter as shown in the figure for the on-site velocities, $v_\pm^n$ from Eq.~(\ref{eq:vpm_condensed}) and $w_\pm^n$ from Eq.~(\ref{eq:wpm_condensed}), as a function of reduced time step for the GJ-III method, Eq.~(\ref{eq:c2_GJ3}). Simulation results of Eqs.~(\ref{eq:vnpm_Split_GJ}) and (\ref{eq:vnpm2_Split_GJ}) are indistinguishable from the predicted results of Eq.~(\ref{eq:on-site_energy}). Fluctuations, $\sigma_k=\sqrt{2}\,\widetilde{\sigma}_k$ are in agreement with the predicted, $\widetilde{\sigma}_k=\langle{E_k}\rangle$. Simulation details given in Sec.~\ref{sec:HS_one_numerical}. Panels (a) and (b) display data for the quantities listed in the lower left corner of each panel.
}
\label{fig:fig_15}
\end{figure}

\begin{figure}[t]
\centering
\scalebox{0.475}{\centering \includegraphics[trim={2.5cm 1.0cm 0.0cm 4.0cm},clip]{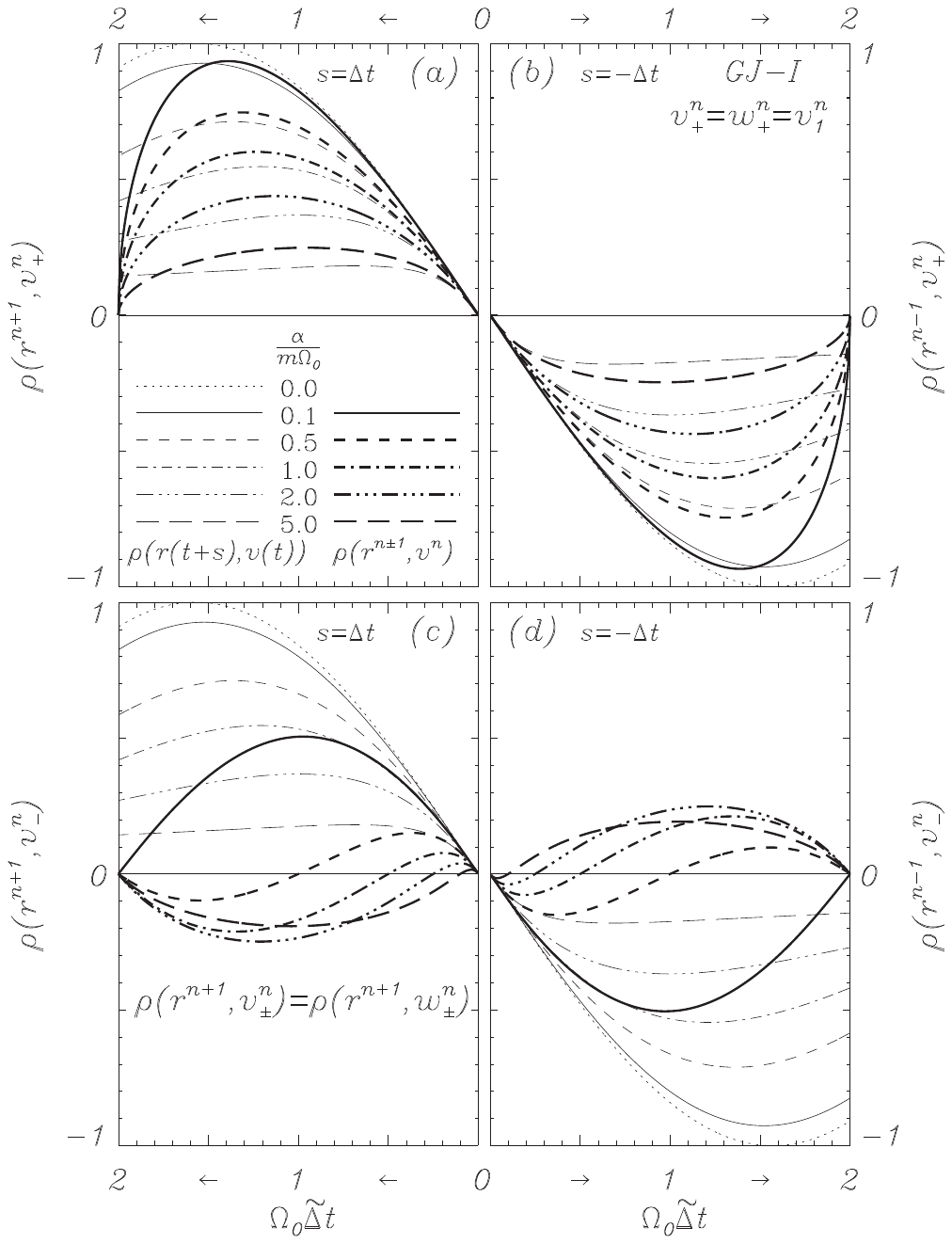}}
\caption{Cross-correlations between the configurational coordinate and the on-site velocities, $v_\pm^{n}$, for the GJ-I method with several different damping parameters. Thick curves represent the analytical result, Eq.~(\ref{eq:OS_AS_cross-corr}), through a correlation expression similar to Eq.~(\ref{eq:cross_corr}), as well as simulations of Eq.~(\ref{eq:vnpm_Split_GJ}), which coincide with the analysis. Thin curves represent the continuous-time expectation, Eq.~(\ref{eq:Cont_corr_vr}) with $s=\Delta{t}$. Simulation details given in Sec.~\ref{sec:HS_one_numerical}.
}
\label{fig:fig_16}
\end{figure}

\begin{figure}[t]
\centering
\scalebox{0.475}{\centering \includegraphics[trim={2.5cm 1.0cm 0.0cm 4.0cm},clip]{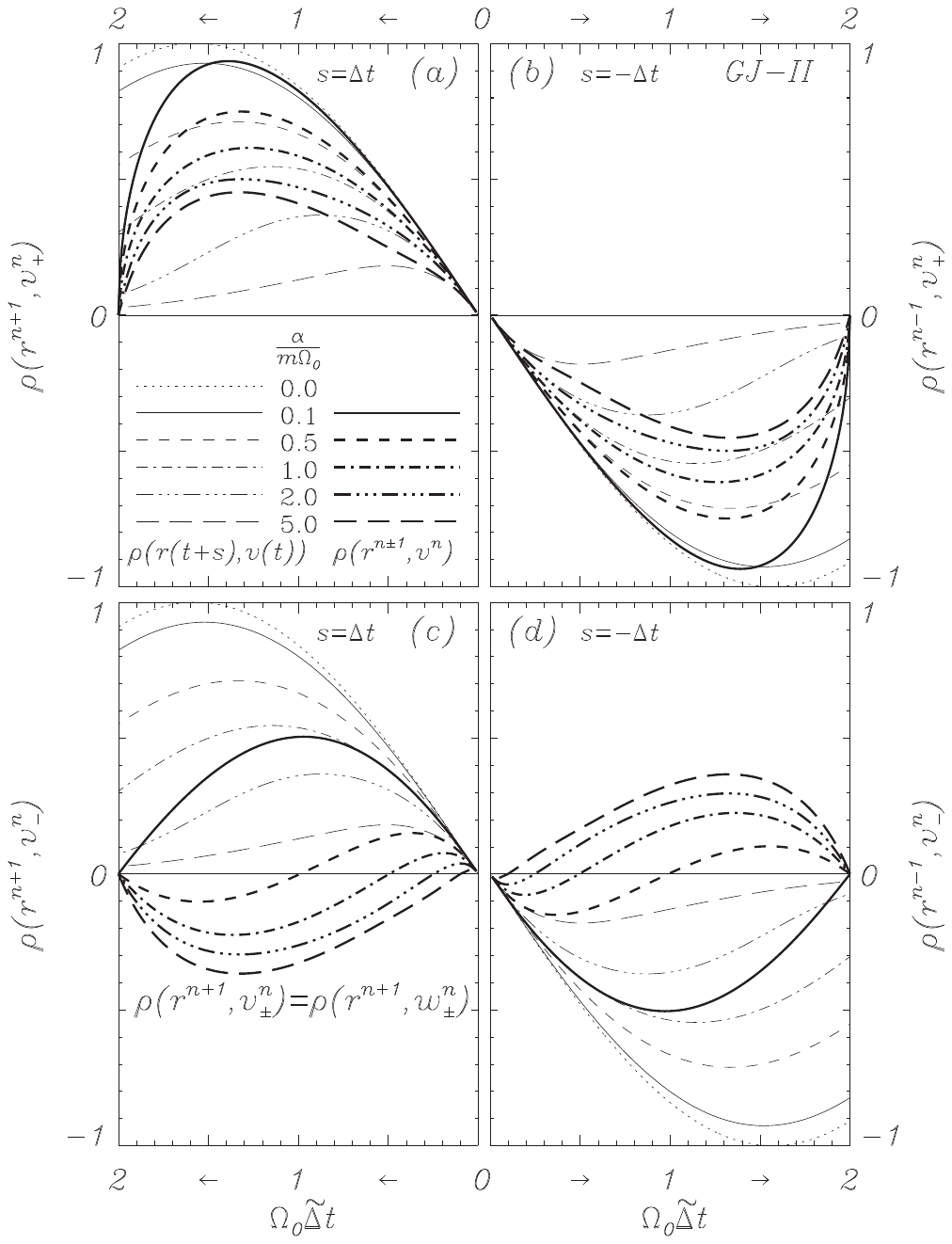}}
\caption{Cross-correlations between the configurational coordinate and the on-site velocities, $v_\pm^{n}$, for the GJ-II method with several different damping parameters. Thick curves represent the analytical result, Eq.~(\ref{eq:OS_AS_cross-corr}), through a correlation expression similar to Eq.~(\ref{eq:cross_corr}), as well as simulations of Eq.~(\ref{eq:vnpm_Split_GJ}), which coincide with the analysis. Thin curves represent the continuous-time expectation, Eq.~(\ref{eq:Cont_corr_vr}) with $s=\Delta{t}$. Simulation details given in Sec.~\ref{sec:HS_one_numerical}.
}
\label{fig:fig_17}
\end{figure}

\begin{figure}[t]
\centering
\scalebox{0.475}{\centering \includegraphics[trim={2.5cm 1.0cm 0.0cm 4.0cm},clip]{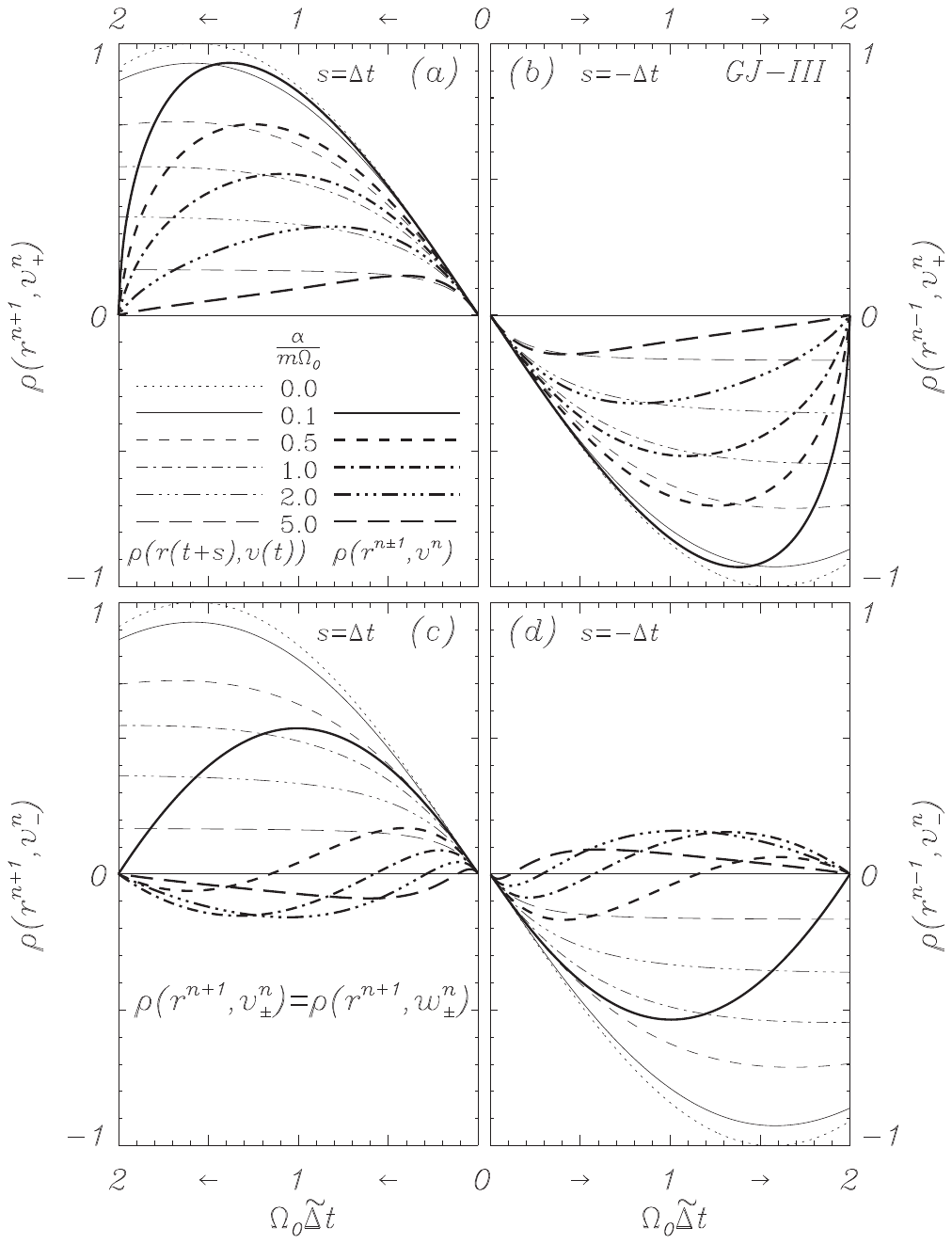}}
\caption{Cross-correlations between the configurational coordinate and the on-site velocities, $v_\pm^{n}$, for the GJ-III method with several different damping parameters. Thick curves represent the analytical result, Eq.~(\ref{eq:OS_AS_cross-corr}), through a correlation expression similar to Eq.~(\ref{eq:cross_corr}), as well as simulations of Eq.~(\ref{eq:vnpm_Split_GJ}), which coincide with the analysis. Thin curves represent the continuous-time expectation, Eq.~(\ref{eq:Cont_corr_vr}) with $s=\Delta{t}$. Simulation details given in Sec.~\ref{sec:HS_one_numerical}.
}
\label{fig:fig_18}
\end{figure}

\begin{figure}[t]
\centering
\scalebox{0.475}{\centering \includegraphics[trim={2.5cm 1.0cm 0.0cm 4.0cm},clip]{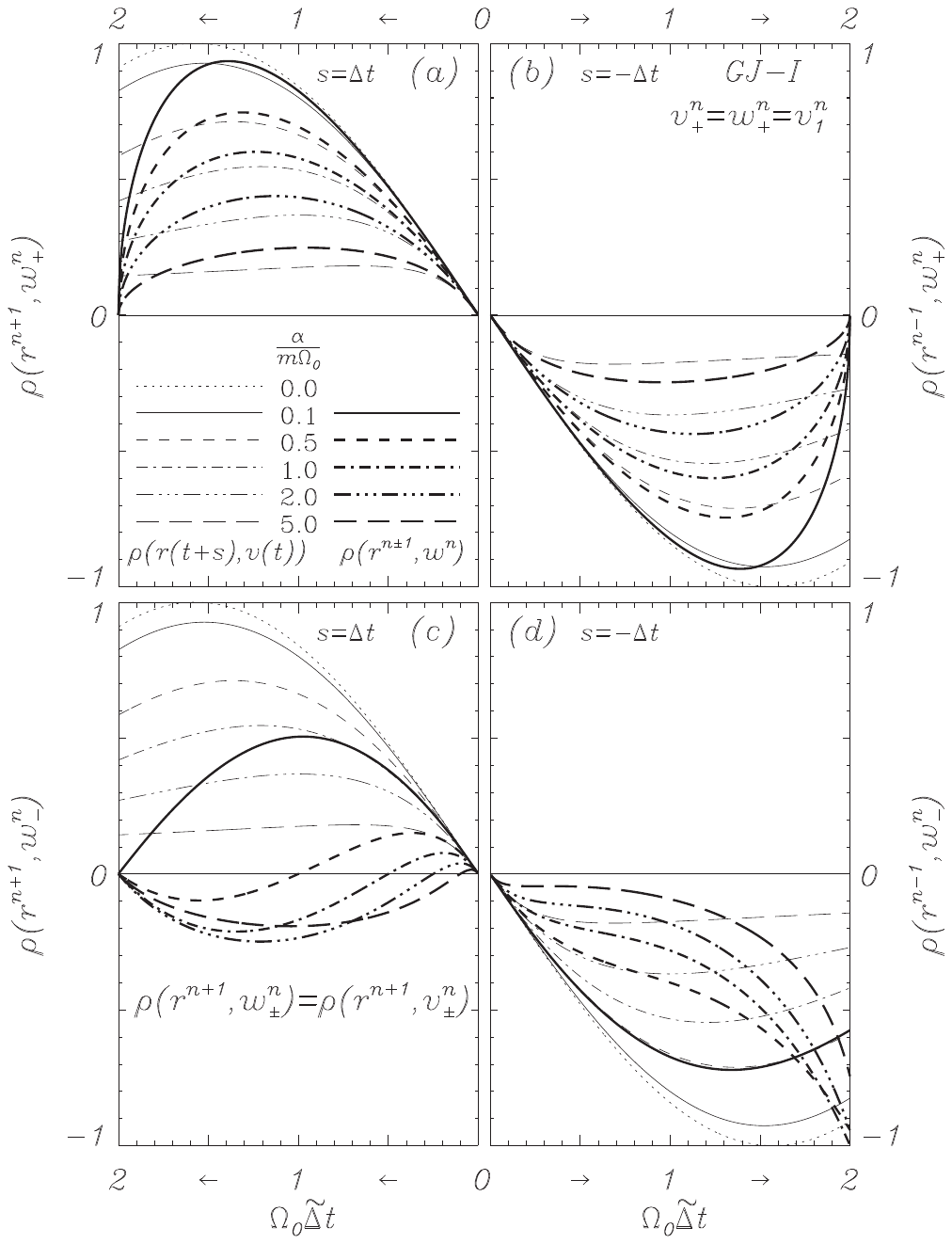}}
\caption{Cross-correlations between the configurational coordinate and the on-site velocities, $w_\pm^{n}$, for the GJ-I method with several different damping parameters. Thick curves represent the analytical result, Eqs.~(\ref{eq:OS_NS1_cross_corr}) and (\ref{eq:OS_NS2_cross_corr}), through a correlation expression similar to Eq.~(\ref{eq:cross_corr}), as well as simulations of Eq.~(\ref{eq:vnpm2_Split_GJ}), which coincide with the analysis. Thin curves represent the continuous-time expectation, Eq.~(\ref{eq:Cont_corr_vr}) with $s=\Delta{t}$. Simulation details given in Sec.~\ref{sec:HS_one_numerical}.
}
\label{fig:fig_19}
\end{figure}

\begin{figure}[t]
\centering
\scalebox{0.475}{\centering \includegraphics[trim={2.5cm 1.0cm 0.0cm 4.0cm},clip]{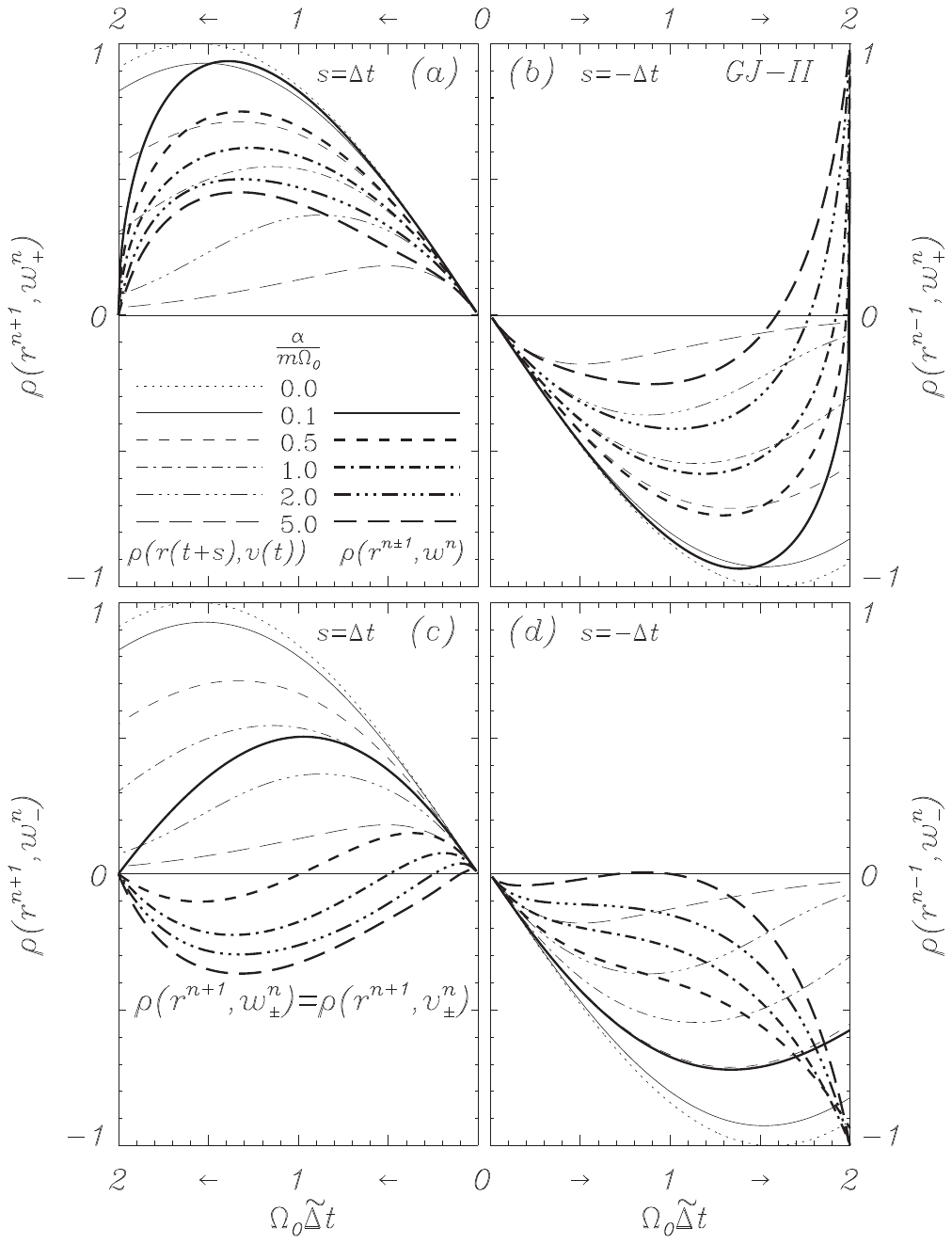}}
\caption{Cross-correlations between the configurational coordinate and the on-site velocities, $w_\pm^{n}$, for the GJ-II method with several different damping parameters. Thick curves represent the analytical result, Eqs.~(\ref{eq:OS_NS1_cross_corr}) and (\ref{eq:OS_NS2_cross_corr}), through a correlation expression similar to Eq.~(\ref{eq:cross_corr}), as well as simulations of Eq.~(\ref{eq:vnpm2_Split_GJ}), which coincide with the analysis. Thin curves represent the continuous-time expectation, Eq.~(\ref{eq:Cont_corr_vr}) with $s=\Delta{t}$. Simulation details given in Sec.~\ref{sec:HS_one_numerical}.
}
\label{fig:fig_20}
\end{figure}

\begin{figure}[t]
\centering
\scalebox{0.475}{\centering \includegraphics[trim={2.5cm 1.0cm 0.0cm 4.0cm},clip]{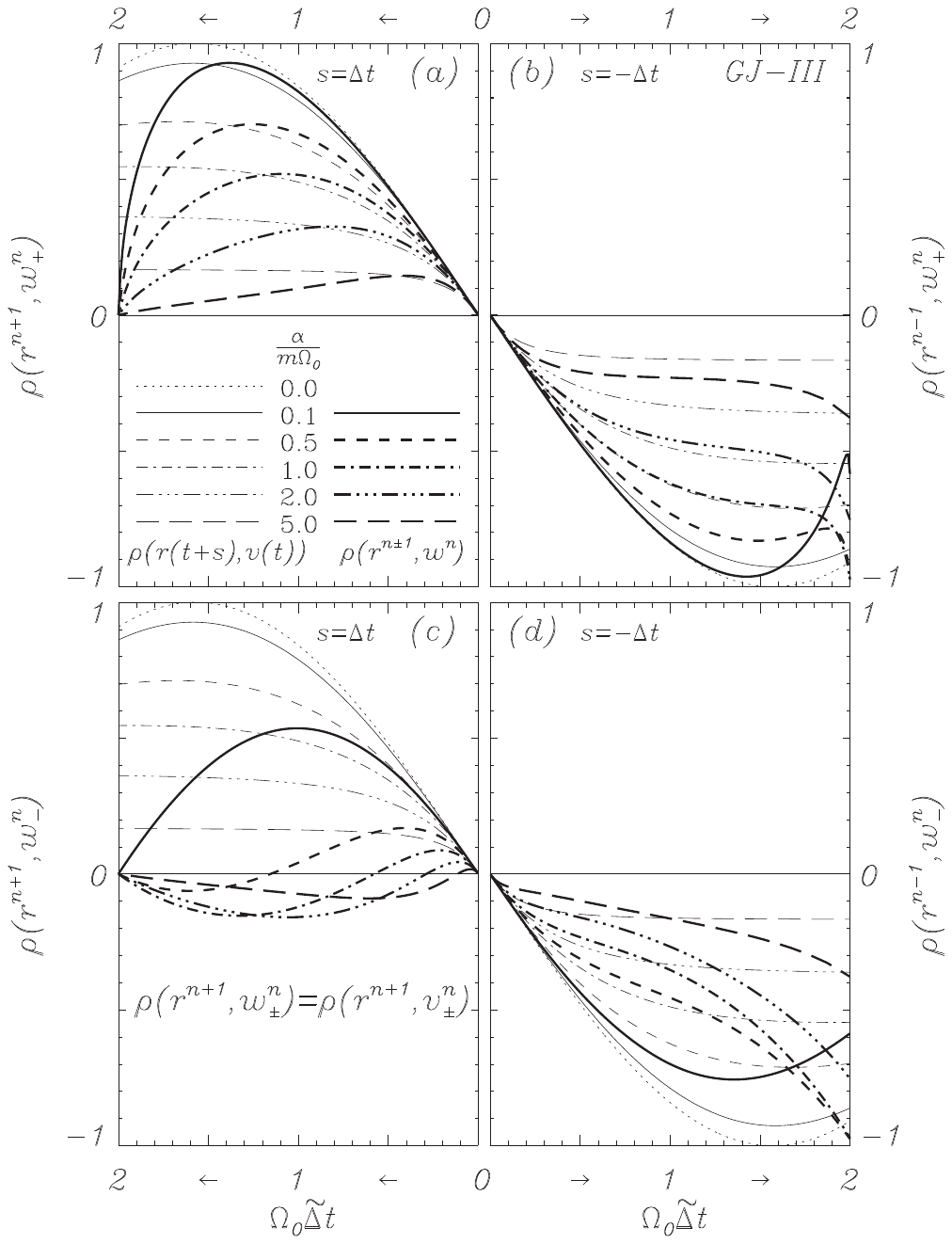}}
\caption{Cross-correlations between the configurational coordinate and the on-site velocities, $w_\pm^{n}$, for the GJ-III method with several different damping parameters. Thick curves represent the analytical result, Eqs.~(\ref{eq:OS_NS1_cross_corr}) and (\ref{eq:OS_NS2_cross_corr}), through a correlation expression similar to Eq.~(\ref{eq:cross_corr}), as well as simulations of Eq.~(\ref{eq:vnpm2_Split_GJ}), which coincide with the analysis. Thin curves represent the continuous-time expectation, Eq.~(\ref{eq:Cont_corr_vr}) with $s=\Delta{t}$. Simulation details given in Sec.~\ref{sec:HS_one_numerical}.
}
\label{fig:fig_21}
\end{figure}

\subsubsection{Enforcing correct drift and correlation antisymmetry}

Requiring that the on-site velocity has correct linear measure of drift velocity, Eq.~(\ref{eq:vel_cond_6a}), and antisymmetric covariance, Eq.~(\ref{eq:vel_cond_7}), yields the finite-difference coefficients
\begin{subequations}\begin{eqnarray}
\mu_{2_\pm} & = & -(1-c_2)\mu_{1_\pm} \label{eq:gamma_2_full} \\
\mu_{3_\pm} & = &1-(1+2c_2)\mu_{1_\pm} \label{eq:gamma_3_full} \\
\mu_{6_\pm} & = &2c_1\mu_{1_\pm}-1 \, , \label{eq:gamma_6_full}
\end{eqnarray}\noindent
\end{subequations}
where $\mu_{1\pm}$ is given by the two options in Eq.~(\ref{eq:gamma_1_roots}), and the associated values, $\mu_{4_\pm}$, $\mu_{5_\pm}$, and $\mu_{7_\pm}$, are given by Eqs.~(\ref{eq:vel_cond_12}) and (\ref{eq:vel_cond_3}). The finite-difference expression for the resulting velocity is thereby given from Eq.~(\ref{eq:vel_ansatz}). However, a more condensed, and algorithmically favorable, expression can be written
\begin{eqnarray}
v^n_\pm & = & 2\sqrt{c_1c_3}\,\mu_{1_\pm}v_1^{n}
-\mu_{6_\pm}(\sqrt{c_3}v_1^{n-\frac{3}{2}}-\frac{c_3}{2m}\frac{\beta^{n-1}}{1-c_2})\, , \nonumber \\ \label{eq:vpm_condensed}
\end{eqnarray}
where $v_1^n$ and $v_1^{n-\frac{3}{2}}$ are given by Eqs.~(\ref{eq:OS_def_GJ}) and (\ref{eq:HS_def_GJ}). Including this pair of velocities into the GJ method, Eq.~(\ref{eq:A_Split_GJ}), can read
\begin{subequations}
\begin{eqnarray}
&& v^{n+\frac{1}{4}} \; = \; v_1^n+\sqrt{\frac{c_3}{c_1}}\frac{\Delta{t}}{2m}f^n \nonumber \\
&& r^{n+\frac{1}{2}} \; = \; r^n+\sqrt{\frac{c_3}{c_1}}\frac{\Delta{t}}{2}v^{n+\frac{1}{4}} \nonumber\\
v_*^{n-\frac{1}{2}} & = &  \sqrt{c_3}v_1^{n-\frac{1}{2}}-\frac{c_3}{2m}\frac{1}{1-c_2}\,\beta^{n} \label{eq:vnpm_vs2}\\
&& v_1^{n+\frac{1}{2}} \; = \; \sqrt{c_1}\,v^{n+\frac{1}{4}} + \frac{\sqrt{c_3}}{2m}\,\beta^{n+1}\nonumber\\
&& v^{n+\frac{3}{4}} \; = \; \frac{c_2}{\sqrt{c_1}}\,v_1^{n+\frac{1}{2}} + \sqrt{\frac{c_3}{c_1}}\frac{1}{2m}\,\beta^{n+1}\nonumber\\
&& r^{n+1} \; = \; r^{n+\frac{1}{2}}+\sqrt{\frac{c_3}{c_1}}\frac{\Delta{t}}{2}v^{n+\frac{3}{4}}\nonumber\\
&& v_1^{n+1} \; = \; v^{n+\frac{3}{4}}+\sqrt{\frac{c_3}{c_1}}\frac{\Delta{t}}{2m}f^{n+1} \nonumber\\
v_\pm^{n+1} & = & 2\sqrt{c_1c_3}\mu_{1_\pm}v_1^{n+1}-\mu_{6_\pm}v_*^{n-\frac{1}{2}} \, ,  \label{eq:vnpm_vpm2}
\end{eqnarray}\label{eq:vnpm_Split_GJ}\noindent
\end{subequations}
where only the labeled equations differ from Eq.~(\ref{eq:A_Split_GJ}), and where the single value, $v_*^{n-\frac{1}{2}}$, in Eq.~(\ref{eq:vnpm_vs2}) is used in Eq.~(\ref{eq:vnpm_vpm2}) to generate both on-site velocities, $v_\pm^{n+1}$.

The antisymmetric covariance is in this case given by
\begin{eqnarray}
\langle r^{n+1}v_\pm^{n}\rangle & = & -\langle r^{n-1}v_\pm^{n}\rangle  \label{eq:OS_AS_cross-corr} \\
& = & 2c_1c_3\mu_{1_\pm}\Omega_0\Delta{t}\left(1-\frac{c_3}{4c_1}\Omega_0^2\Delta{t}^2\right)\frac{k_BT}{\sqrt{\kappa m}}\, ,  \nonumber 
\end{eqnarray}
which conforms to the continuous-time result, Eq.~(\ref{eq:Corr_vr}), in the limit
\begin{eqnarray}
\langle r^{n\pm1}v_-^{n}\rangle & \rightarrow & \langle r^{n\pm1}v_+^{n}\rangle \; \rightarrow \; \pm\Omega_0\Delta{t}\frac{k_BT}{\sqrt{\kappa m}} \label{eq:vp_corr_limit}
\end{eqnarray}
for $\Omega_0\Delta{t}\rightarrow0$.

Notice that Eq.~(\ref{eq:gamma_1_roots}) yields the inequality $\mu_{1_-}^2<\mu_{1_+}^2$, which by Eq.~(\ref{eq:on-site_energy}) implies that errors in statistical kinetic measures can be considerably smaller using $v_-^n$ than using $v_+^n$, which will always exhibit time-step-dependent depression in kinetic measures since Eq.~(\ref{eq:gamma_1_roots}) also shows that $0<\mu_{1_+}$ (for convex potentials).
This statistical advantage of $v_-^n$ is exemplified for key methods in Figs.~\ref{fig:fig_13}-\ref{fig:fig_15} for both the kinetic energy and its fluctuations. We can further observe from Fig.~\ref{fig:fig_12} that typical methods will have $\mu_{1_-}\approx0$ for $\frac{\alpha\Delta{t}}{m}\approx\frac{1}{2}$, offering the possibility of using the on-site velocity $v_-^n$ to measure the kinetic energy, temperature, and fluctuations (almost) correctly. See Appendix~\ref{sec:Appendix_A} for direct use of this feature.

Consistent with earlier comment, we note that for the GJ-I method, where $c_3=c_1$ implies $\mu_{1_+}=\frac{1}{2c_1}$, the functional parameter $\mu_6$ vanishes, and the algorithm simplifies considerably (see Appendix~\ref{sec:Appendix_A}). In particular, the velocity measure $v_+^n$ coincides with $v_1^n$, leaving Eqs.~(\ref{eq:vnpm_vs2}) and (\ref{eq:vnpm_vpm2}) superfluous if $v_+^n$ is the desired GJ-I choice for an on-site velocity with correct drift and antisymmetric cross-correlation. 

Figures~\ref{fig:fig_16}-\ref{fig:fig_18} show the results of Eq.~(\ref{eq:OS_AS_cross-corr}) through a correlation expression similar to Eq.~(\ref{eq:cross_corr}) along with validating simulations of Eq.~(\ref{eq:vnpm_Split_GJ}). Simulation results are indistinguishable from those of Eq.~(\ref{eq:OS_AS_cross-corr}). Also shown are the continuous-time expectations, Eq.~(\ref{eq:Cont_corr_vr}) with $s=\Delta{t}$, which are in asymptotic agreement with the discrete-time results for $\frac{\alpha\Delta{t}}{m}\rightarrow0$. As is the case for the comparable half-step velocity $v_-^{n+\frac{1}{2}}$ above, due to the related sign-change in $\mu_{1_-}$, seen in Eq.~(\ref{eq:gamma_1_roots}) and exemplified in Fig.~\ref{fig:fig_12}, the on-site velocity $v_-^n$ can exhibit peculiar behavior in its correlation with the trajectory. However, this may not necessarily imply irrelevance of this velocity, since a discrete-time velocity has no impact on the evolution of the simulation, and since the velocity is only used to extract information, which usually pertains to kinetic energy, kinetic fluctuations, and perhaps the correlations, $\rho_{rv}(0)$ or $\rho_{E_pE_k}(0)$; all well represented by the velocity, $v_-^n$. That said, in comparison, the velocity $v_+^n$ shows results much closer to the expected cross-correlation from continuous time.

\subsubsection{Constraining $\mu_6=0$ $\Leftrightarrow$ $\mu_7=0$}
The two conditions, Eqs.~(\ref{eq:vel_cond_6a}) and (\ref{eq:vel_cond_7}), are generally mutually incompatible if $\mu_6=0$. Thus, a choice must be made about which of the two features is preferred. \\

\noindent
{\underline{\bf Correct on-site drift}} (antisymmetric cross-correlation is not enforced):\\
Interestingly, the condition Eq.~(\ref{eq:vel_cond_6a}) with $\mu_6=0$, i.e., Eq.~(\ref{eq:vel_cond_6b}) with $\mu_6=0$, yields precisely the same second order polynomial, Eq~(\ref{eq:gamma_1_equation}), as the one that appears when inserting the condition Eq.~(\ref{eq:vel_cond_7}) into Eq.~(\ref{eq:vel_cond_6b}). Thus, this velocity measure is also given by $\mu_{1_\pm}$ from Eq.~(\ref{eq:gamma_1_roots}) along with
\begin{eqnarray}
\mu_{2_\pm} & = & 1-2\mu_{1_\pm} \label{eq:gamma_2_drift}\\
\mu_{3_\pm} & = & \mu_{1_\pm}-1 \, , \label{eq:gamma_3_drift}
\end{eqnarray}
given by Eqs.~(\ref{eq:vel_cond_0}) and (\ref{eq:vel_cond_6a}).
The corresponding noise coefficients, $\mu_{4_\pm}$ and $\mu_{5_\pm}$, are given by Eqs.~(\ref{eq:vel_cond_1}) and (\ref{eq:vel_cond_3}), while $\mu_7=0$ follows from $\mu_6=0$ (see Eq.~(\ref{eq:vel_cond_2})). The finite-difference expression for the resulting velocity is thereby given from Eq.~(\ref{eq:vel_ansatz}). However, a more condensed, and algorithmically favorable, expression can be written
\begin{eqnarray}
w^n_\pm & = & 2\sqrt{c_1c_3}\mu_{1_\pm}v_1^n\nonumber \\
&&-(2c_1\mu_{1_\pm}-1)(\sqrt{c_3}v_1^{n-\frac{1}{2}} -\frac{c_3}{2m}\frac{\beta^n}{1-c_2}) \, ,  \label{eq:wpm_condensed}
\end{eqnarray}
where $v_1^n$ and $v_1^{n-\frac{1}{2}}$ are given by Eqs.~(\ref{eq:OS_def_GJ}) and (\ref{eq:HS_def_GJ}). Notice that, given $\mu_{1_\pm}$ being the same for the two velocities, $v_\pm^n$ and $w_\pm^n$, the leading factor ($2c_1\mu_{1_\pm}-1$) of the last term in Eq.~(\ref{eq:wpm_condensed}) is given by the value for $\mu_{6_\pm}$ in Eq.~(\ref{eq:gamma_6_full}), even if $\mu_{6_\pm}=0$ for $w_\pm^n$. Including the pair of velocities, $w_\pm^n$, into the GJ method, Eq.~(\ref{eq:A_Split_GJ}), can read
\begin{subequations}
\begin{eqnarray}
&& v^{n+\frac{1}{4}} \; = \; v_1^n+\sqrt{\frac{c_3}{c_1}}\frac{\Delta{t}}{2m}f^n \nonumber \\
&& r^{n+\frac{1}{2}} \; = \; r^n+\sqrt{\frac{c_3}{c_1}}\frac{\Delta{t}}{2}v^{n+\frac{1}{4}} \nonumber\\
&& v_1^{n+\frac{1}{2}} \; = \; \sqrt{c_1}\,v^{n+\frac{1}{4}} + \frac{\sqrt{c_3}}{2m}\,\beta^{n+1}\nonumber\\
v_*^{n+\frac{1}{2}} & = &  \sqrt{c_3}v_1^{n+\frac{1}{2}}-\frac{c_3}{2m}\frac{1}{1-c_2}\beta^{n+1}\label{eq:vnpm2_vs2}\\
&& v^{n+\frac{3}{4}} \; = \; \frac{c_2}{\sqrt{c_1}}\,v_1^{n+\frac{1}{2}} + \sqrt{\frac{c_3}{c_1}}\frac{1}{2m}\,\beta^{n+1}\nonumber\\
&& r^{n+1} \; = \; r^{n+\frac{1}{2}}+\sqrt{\frac{c_3}{c_1}}\frac{\Delta{t}}{2}v^{n+\frac{3}{4}}\nonumber\\
&& v_1^{n+1} \; = \; v^{n+\frac{3}{4}}+\sqrt{\frac{c_3}{c_1}}\frac{\Delta{t}}{2m}f^{n+1} \nonumber\\
w_\pm^{n+1} & = & 2\sqrt{c_1c_3}\mu_{1_\pm}v_1^{n+1}-(2c_1\mu_{1_\pm}-1)v_*^{n+\frac{1}{2}} \!\!\! , \label{eq:vnpm2_vpm2}
\end{eqnarray}\label{eq:vnpm2_Split_GJ}\noindent
\end{subequations}
where only the labeled equations differ from Eq.~(\ref{eq:A_Split_GJ}), and where the single value, $v_*^{n+\frac{1}{2}}$, in Eq.~(\ref{eq:vnpm2_vs2}) is used in Eq.~(\ref{eq:vnpm2_vpm2}) to generate both on-site velocities, $w_\pm^{n+1}$.

We find the identity of the covariances
\begin{eqnarray}
\langle r^{n+1}w_\pm^{n}\rangle & = & \langle r^{n+1}v_\pm^{n}\rangle\, ,  \label{eq:OS_NS1_cross_corr}
\end{eqnarray}
where $\langle r^{n+1}v_\pm^{n}\rangle$ is given in Eq.~(\ref{eq:OS_AS_cross-corr}). The complementary covariance is
\begin{eqnarray}
&&\langle r^{n}w_\pm^{n+1}\rangle \; =  \label{eq:OS_NS2_cross_corr}\\
&&-\frac{1+2c_1\mu_{1_\pm}}{2}c_3\Omega_0\Delta{t}\left(1-\frac{c_3\mu_{1_\pm}}{1+2c_1\mu_{1_\pm}}\Omega_0^2\Delta{t}^2\right)\frac{k_BT}{\sqrt{\kappa m}}\, . \nonumber
\end{eqnarray}
Clearly, the covariance between $r^{n\pm1}$ and $w_\pm^n$ is not, in general, antisymmetric, even if, by design, $\langle r^nw_\pm^n\rangle=0$. The discrepancy in antisymmetry is found from Eq.~(\ref{eq:on-site_asymmetry}) to be
\begin{eqnarray}
\langle r^nw_\pm^{n+1}\rangle+\langle r^{n+1}w_\pm^n\rangle & = & \frac{c_3}{2}\,\Omega_0\Delta{t}\,(2c_1\mu_{1_\pm}-1) \frac{k_BT}{\sqrt{\kappa m}} \, . \nonumber \\ \label{eq:on-site_drift_asymmetry}
\end{eqnarray}
Given that Eq.~(\ref{eq:on-site_drift_asymmetry}) shows a second order, in $\Delta{t}$, deviation from antisymmetry of the covariance, and given Eqs.~(\ref{eq:vp_corr_limit}) and (\ref{eq:OS_NS2_cross_corr}), we see that also the covariances in Eq.~(\ref{eq:OS_NS2_cross_corr}) conform to the continuous-time limit, Eq.~(\ref{eq:Corr_vr}),
\begin{eqnarray}
\langle r^{n\pm1}w_-^{n}\rangle & \rightarrow & \langle r^{n\pm1}w_+^{n}\rangle \; \rightarrow \; \pm\Omega_0\Delta{t}\frac{k_BT}{\sqrt{\kappa m}} \label{eq:wp_corr_limit}
\end{eqnarray}
for $\Omega_0\Delta{t}\rightarrow0$.

Since $\mu_{1_\pm}$ apply to both on-site velocities, $v_\pm^n$ and $w_\pm^n$, the noted observation below Eq.~(\ref{eq:vp_corr_limit}), arguing for why the statistical errors of $v_-^n$ are less than those of $v_+^n$, also applies to $w_\pm^n$, as illustrated in Figs.~\ref{fig:fig_13}-\ref{fig:fig_15}. It follows that since the typical methods show $\mu_{1_-}\approx0$ for $\frac{\alpha\Delta{t}}{m}\approx\frac{1}{2}$ (see Fig.~\ref{fig:fig_12}), $w_-^n$ offers the possibility of measuring the kinetic energy, temperature, and fluctuations (almost) correctly in that vicinity.

The special case of the GJ-I method, for which $c_3=c_1$ implies $\mu_{1_+}=\frac{1}{2c_1}$, which, in turn, implies that $\mu_{6_+}=0$, we see that $w_+^n=v_+^n=v_1^n$. Thus, the velocity measure $w_+^n$ coincides with $v_1^n$, leaving Eqs.~(\ref{eq:vnpm2_vs2}) and (\ref{eq:vnpm2_vpm2}) superfluous if $w_+^n$ is the desired GJ-I choice for an on-site velocity with correct drift and antisymmetric cross-correlation.\\

\noindent
{\underline{\bf Correct on-site antisymmetry}} (Correct drift velocity measure not enforced): \\
This velocity is characterized by Eq.~(\ref{eq:vel_cond_7}) and, from Eq.~(\ref{eq:vel_cond_0}),
\begin{eqnarray}
\mu_3 & = & -\mu_1-\mu_2 \; = \; -c_2\mu_1\, ,  \label{eq:gamma_3_acor}
\end{eqnarray}
where $\mu_1$ is found from inserting Eq.~(\ref{eq:vel_cond_7}) into Eq.~(\ref{eq:Gamma_0}) with $\mu_6=0$; thus
\begin{eqnarray}
\mu_1^2 & = & \frac{1}{4{c_1c_3}} \, . \label{eq:gamma_1_acor}
\end{eqnarray}
The corresponding noise coefficients, $\mu_4$ and $\mu_5$, are given by Eqs.~(\ref{eq:vel_cond_1}) and (\ref{eq:vel_cond_3}). Notice that, since the correct drift measure is not enforced, Eq.~(\ref{eq:gamma_1_acor}) contains the possibility of a negative drift-velocity measure in response to a positive force. Clearly, $\mu_1$ should be chosen positive in this case.

The on-site velocity found here is the on-site velocity, $v_1^n$ from Eqs.~(\ref{eq:OS_def_GJ}) and, e.g., (\ref{eq:A_Split_GJ}) or (\ref{eq:D2_Split_GJ}), associated with the GJ methods of Ref.~\cite{GJ}. Reemphasizing previous comments, the GJ-I method, $c_3=c_1$, yields the correct drift velocity, even if not specifically enforced here.\\

Figures~\ref{fig:fig_19}-\ref{fig:fig_21} show similar results as Figs.~\ref{fig:fig_16}-\ref{fig:fig_18} described above, but for the velocities $w_\pm^n$. Shown are results of Eqs.~(\ref{eq:OS_NS1_cross_corr}) and (\ref{eq:OS_NS2_cross_corr}) through a correlation expression similar to Eq.~(\ref{eq:cross_corr}) along with validating simulations of Eq.~(\ref{eq:vnpm2_Split_GJ}). Numerical results are indistinguishable from the analytical predictions. The lack of antisymmetry in the correlation is clearly visible for methods GJ-II, GJ-III, and GJ-I for $w_-^n$, whereas the results for GJ-I confirm that $v_+^n=w_+^n=v_1^n$ for this method, yielding the antisymmetric correlation behavior of $v_1^n$. We further observe that $\rho(r^{n+1},v_\pm^n)=\rho(r^{n+1},w_\pm^n)$. Also shown are the continuous-time expectations, Eq.~(\ref{eq:Cont_corr_vr}) with $s=\Delta{t}$, which are in asymptotic agreement with the discrete-time results for $\frac{\alpha\Delta{t}}{m}\rightarrow0$. As is the case for the velocity $v_-^{n}$ discussed above, the velocity $w_-^n$ can exhibit peculiar behavior in its correlation with the trajectory, in addition to the lack of antisymmetry, given the related sign-change in $\mu_{1_-}$ seen in Fig.~\ref{fig:fig_12}. The comments above for $v_-^{n}$ apply here as well; namely that the velocity $w_-^n$ has no impact on the evolution of the simulation, and that it is only used to extract information, which can very well be both useful and reliable for what is needed from a simulation.

\section{Discussion}
\label{sec:Discussion}
We have systematically identified viable options for correctly including one of the most basic, yet often overlooked, expectations for a velocity measure in Langevin simulations; namely the drift velocity. We have aimed to do this while also imposing correct Maxwell-Boltzmann statistics. The applied velocity ansatz for the analysis and development involves up to seven or eight parameters consistent with the natural structure of stochastic Verlet-type algorithms, and the straight-forward, yet cumbersome, determination of these shows what options are possible. The expanded ansatz and analysis, compared to Ref.~\cite{GJ}, reaffirm that it is not generally possible to design a discrete-time, on-site velocity that produces correct, time-step independent statistics; although Appendix~\ref{sec:Appendix_A} exemplifies one of the special cases where it is possible for an on-site velocity to respond with, e.g., correct kinetic energy/temperature. However, we are able to define general-use, on-site definitions that have correct drift velocity as well as improved kinetic measures over previous expressions. For the half-step velocities, we have been able to identify specific measures that contain correct drift as well as correct Maxwell-Boltzmann statistics. These come in two forms, one using only one new stochastic value per degree of freedom per time step, and another using two (the latter was previously identified in a different form in Ref.~\cite{Josh_3}).

While not initially apparent, all the velocity definitions can be conveniently included into the existing GJ formalism as simple {\it add-ons} without significantly altering coding structure or complexity. Thus, the new velocity options can easily be implemented for evaluation into existing codes, including LAMMPS \cite{Plimpton,Plimpton2}, which is already offering the GJ-I method. We here give the complete set of new velocities embedded in the GJ method of Eq.~(\ref{eq:A_Split_GJ}):
\begin{subequations}
\begin{eqnarray}
&& v^{n+\frac{1}{4}} \; = \; v_1^n+\sqrt{\frac{c_3}{c_1}}\frac{\Delta{t}}{2m}f^n \nonumber \\
&& r^{n+\frac{1}{2}} \; = \; r^n+\sqrt{\frac{c_3}{c_1}}\frac{\Delta{t}}{2}v^{n+\frac{1}{4}} \nonumber\\
&& v_1^{n+\frac{1}{2}} \; = \; \sqrt{c_1}\,v^{n+\frac{1}{4}} + \frac{\sqrt{c_3}}{2m}\,\beta^{n+1}\nonumber\\
v_*^{n+\frac{1}{2}} & = &  \sqrt{c_3}v_1^{n+\frac{1}{2}}-\frac{c_3}{2m}\frac{1}{1-c_2}\beta^{n+1}\label{eq:final_vs}\\
v_\pm^{n+\frac{1}{2}} & = & \sqrt{c_3}\gamma_{1_\pm}v_1^{n+\frac{1}{2}}-\gamma_{3_\pm}v_*^{n-\frac{1}{2}} \label{eq:final_upm}\\
v_{d2}^{n+\frac{1}{2}} & = & \sqrt{c_3}v_1^{n+\frac{1}{2}}+\frac{\sqrt{c_3}}{\sqrt{2}m}\sqrt{\frac{1-c_3}{1-c_2}}\,\beta^{n+\frac{1}{2}} \label{eq:final_d2}\\
&& v^{n+\frac{3}{4}} \; = \; \frac{c_2}{\sqrt{c_1}}\,v_1^{n+\frac{1}{2}} + \sqrt{\frac{c_3}{c_1}}\frac{1}{2m}\,\beta^{n+1}\nonumber\\
&& r^{n+1} \; = \; r^{n+\frac{1}{2}}+\sqrt{\frac{c_3}{c_1}}\frac{\Delta{t}}{2}v^{n+\frac{3}{4}}\nonumber\\
&& v_1^{n+1} \; = \; v^{n+\frac{3}{4}}+\sqrt{\frac{c_3}{c_1}}\frac{\Delta{t}}{2m}f^{n+1} \nonumber\\
v_\pm^{n+1} & = & 2\sqrt{c_1c_3}\mu_{1_\pm}v_1^{n+1}-\mu_{6_\pm}v_*^{n-\frac{1}{2}} \label{eq:vpm}\\
w_\pm^{n+1} & = & 2\sqrt{c_1c_3}\mu_{1_\pm}v_1^{n+1}-\mu_{6_\pm}v_*^{n+\frac{1}{2}}\, ,  \label{eq:wpm}
\end{eqnarray}\label{eq:Final_Split_GJ}\noindent
\end{subequations}
where $\mu_{6_\pm}$ is given by Eq.~(\ref{eq:gamma_6_full}) for both Eqs.~(\ref{eq:vpm}) and (\ref{eq:wpm}), per comment after Eq.~(\ref{eq:wpm_condensed}). Incorporating the new velocities can equally well be accomplished within the compact form given in Eq.~(\ref{eq:A_Compact_GJ}) by placing Eqs.~(\ref{eq:final_vs}), (\ref{eq:final_upm}), and (\ref{eq:final_d2}) after Eq.~(\ref{eq:A_Cu}); and Eqs.~(\ref{eq:vpm}) and (\ref{eq:wpm}) after Eq.~(\ref{eq:A_Cv}). The half-step velocity $v_{d2}^{n+\frac{1}{2}}$ is simple to express and it is, from Figs.~\ref{fig:fig_8}ab, \ref{fig:fig_10}, and \ref{fig:fig_11}, apparently well correlated with the configurational coordinate, especially for the GJ-I and GJ-III methods. The velocities that arise from the parameters $\gamma_{1_-}$ and $\mu_{1_-}$ are peculiar in both the leading coefficient and the cross correlations with the configurational coordinate. However, the on-site velocities are perfectly uncorrelated with the trajectory at the same time, and the on-site velocities $v_-^n$ and $w_-^n$ can have significantly less error in kinetic measures, such as kinetic temperature and fluctuations, than other on-site velocities. It is therefore possible that these velocities can be useful for statistical simulations despite their unusual appearance.  Particular interest may be given to the GJ-I method, for which the relationship $c_3=c_1$ significantly simplifies the method; see Appendix~\ref{sec:Appendix_A}. Further, in that case $v_+^n=w_+^n=v_1^n$. One particularly simple case is, again, GJ-I if one decides to use the half-step velocity $v_{d2}^{n+\frac{1}{2}}$ and not the on-site velocities, $w_-^n$ and $v_-^n$. This will eliminate the use for Eqs.~(\ref{eq:final_vs}), (\ref{eq:final_upm}), (\ref{eq:vpm}), and (\ref{eq:wpm}) since, $v_+^n=w_+^n=v_1^n$. The only cost from including $v_{d2}^{n+\frac{1}{2}}$ is the additional stochastic number, $\beta^{n+\frac{1}{2}}$.

The analysis in this work is linear in order to systematically formulate the developments. While it is likely that the analytically perfect behavior of these velocities may be perturbed by nonlinearities in the potential, we have not here performed  evaluations of how the new velocity definitions perform for a variety of nonlinear and complex systems. Instead, we have provided the options and their rationale together with a convenient path to implement these options into existing codes. It is our hope that these methods for accurate kinetic measures can thereby be tested and evaluated broadly in many different contexts and applications.

\section{Acknowledgments}
N/A

\section{Data Availability Statement}
The data presented and discussed in the current study are available from the author on reasonable request.\\

\section{Funding and/or Competing Interests}
No funding was received for conducting this study.
The author has no relevant financial or non-financial interests to disclose.

\appendix
\section{The Augmented GJ-I Thermostat}
\label{sec:Appendix_A}

From the general expressions of Eq.~(\ref{eq:Final_Split_GJ}), summarizing the multi-velocity Verlet-based algorithms of this work, we exemplify the simple, practical, and efficient GJ-I choice, where $c_3=c_1$ results in an unscaled time step (see Eq.~(\ref{eq:scaling})).
Inserting $c_3=c_1$ into Eq.~(\ref{eq:Final_Split_GJ}), and using the relationships of Eq.~(\ref{eq:c1c3}) and (\ref{eq:c2_GJ1}) for $c_1=c_3$ and $c_2$,  yields
\begin{subequations}
\begin{eqnarray}
v^{n+\frac{1}{4}} & = & v_1^n+\frac{\Delta{t}}{2m}f^n \nonumber \\
r^{n+\frac{1}{2}} & = & r^n+\frac{\Delta{t}}{2}v^{n+\frac{1}{4}} \nonumber\\
v_*^{n-\frac{1}{2}} & = &  8\sqrt{c_1}\,(1-c_1)v_1^{n-\frac{1}{2}} -\frac{2c_1}{m}\beta^{n} \nonumber \\
v_1^{n+\frac{1}{2}} & = & \sqrt{c_1}\,\left(v^{n+\frac{1}{4}} + \frac{1}{2m}\,\beta^{n+1}\right)\nonumber\\
v_{d2}^{n+\frac{1}{2}} & = & \sqrt{c_1}\,\left(v_1^{n+\frac{1}{2}}+\frac{1}{2m}\,\beta^{n+\frac{1}{2}}\right) \label{eq:final_d2_GJ-I}\\
v^{n+\frac{3}{4}} & = &  \frac{c_2}{\sqrt{c_1}}\,v_1^{n+\frac{1}{2}} + \frac{1}{2m}\,\beta^{n+1}\nonumber\\
r^{n+1} & = & r^{n+\frac{1}{2}}+\frac{\Delta{t}}{2}v^{n+\frac{3}{4}} \label{eq:r_GJ-I}\\
v_1^{n+1} & = & v^{n+\frac{3}{4}}+\frac{\Delta{t}}{2m}f^{n+1} \label{eq:v1_GJ-I} \\
v_-^{n+1} & = & \frac{1}{4-3c_1}\left((5c_1-4)v_1^{n+1}+v_*^{n-\frac{1}{2}}\right)\label{eq:vpm_GJ-I} \, ,
\end{eqnarray}\label{eq:Final_Split_GJ-I}\noindent
\end{subequations}
where
we from Eqs.~(\ref{eq:c1c3}) and (\ref{eq:c2_GJ1}) have
\begin{eqnarray}
c_3 \; = \; c_1 & = & \frac{1}{1+\displaystyle\frac{\alpha\Delta{t}}{2m}}\, .
\end{eqnarray}
The labeled equations in Eq.~(\ref{eq:Final_Split_GJ-I}) indicate the desired variables; i.e., the half-step velocity, $v_{d2}^{n+\frac{1}{2}}$, of Eq.~(\ref{eq:final_d2_GJ-I}) with correct drift in a linear potential and correct kinetic energy/temperature statistical measure for linear and harmonic potentials; the statistically correct configurational coordinate, $r^n$, in Eq.~(\ref{eq:r_GJ-I}); the on-site velocity $v_1^n=v_+^n=w_+^n$ of Eq.~(\ref{eq:v1_GJ-I}) with correct drift velocity for flat potentials (but time-step-dependent errors for harmonic potentials); and the complementing on-site velocity $v_-^n$ of Eq.~(\ref{eq:vpm_GJ-I}), also with correct linear surface drift. Generally, the latter velocity $v_-^n$ also has time-step dependent errors in kinetic energy measures, but these errors are smaller than those of $v_1^n$ because $|\mu_{1_-}|<|\mu_{1_+}|$; see Eqs.~(\ref{eq:gamma_1_roots}) and (\ref{eq:on-site_energy}). For simplicity we have here omitted the half-step velocities $v_\pm^{n+\frac{1}{2}}$, but these can easily be included. We have also excluded the on-site velocity $w_-^n$, since this does not possess the expected antisymmetric cross-correlation with the configurational coordinate (see Fig.~\ref{fig:fig_19}cd and Eqs.~(\ref{eq:OS_AS_cross-corr}), (\ref{eq:OS_NS1_cross_corr}), and (\ref{eq:OS_NS2_cross_corr})). Thus, the velocity $v_-^n$, which does have antisymmetric cross-correlation with $r^n$ (see Fig.~\ref{fig:fig_16}cd and Eq.~(\ref{eq:OS_AS_cross-corr})), is intuitively a more attractive variable than $w_-^n$.\\

The on-site velocity $v_-^n$ may become particularly useful for simulations in which each degree of freedom has its friction $\alpha$ adjusted such that $\mu_{1_-}=0$, which, from Eqs.~(\ref{eq:v2_corr_1}), (\ref{eq:Gamma_2}), and (\ref{eq:on-site_energy}), ensures that the on-site statistical measure of kinetic energy (temperature) for linear systems is correct in the entire stability range. For the GJ-I method outlined here, this is accomplished for
\begin{eqnarray}
\mu_{1_-} & = & 0 \; \; \Rightarrow \; \; c_1 \; = \; \frac{4}{5}  \; ; \; \; c_2 \; = \; \frac{3}{5}   \; ; \; \; \frac{\alpha\Delta{t}}{m} \; = \; \frac{1}{2} \, . 
\label{eq:GJ-Ip_param}
\end{eqnarray}
A friction-adjusted GJ-I algorithm for $\mu_{1_-}=0$ is then found from Eqs.~(\ref{eq:Final_Split_GJ-I}) and (\ref{eq:GJ-Ip_param}):
\begin{subequations}
\begin{eqnarray}
v^{n+\frac{1}{4}} & = & v_1^n+\frac{\Delta{t}}{2m}f^n \nonumber \\
r^{n+\frac{1}{2}} & = & r^n+\frac{\Delta{t}}{2}v^{n+\frac{1}{4}} \nonumber\\
v_-^{n+1} & = & \frac{2}{\sqrt{5}}v_1^{n-\frac{1}{2}} -\sqrt{\frac{k_BT}{m}}{\cal N}^{n}\label{eq:vpm_GJ-Ip}\\
v_1^{n+\frac{1}{2}} & = &  \frac{2}{\sqrt{5}}\,\left(v^{n+\frac{1}{4}} + \frac{1}{2}\sqrt{\frac{k_BT}{m}}\,{\cal N}^{n+1}\right)\nonumber\\
v_{d2}^{n+\frac{1}{2}} & = &  \frac{2}{\sqrt{5}}\,\left(v_1^{n+\frac{1}{2}}+ \frac{1}{2}\sqrt{\frac{k_BT}{m}}\,{\cal N}^{n+\frac{1}{2}}\right) \label{eq:final_d2_GJ-Ip}\\
v^{n+\frac{3}{4}} & = &  \frac{3}{2\sqrt{5}}\,v_1^{n+\frac{1}{2}} + \frac{1}{2}\sqrt{\frac{k_BT}{m}}\,{\cal N}^{n+1}\nonumber\\
r^{n+1} & = & r^{n+\frac{1}{2}}+\frac{\Delta{t}}{2}v^{n+\frac{3}{4}}\label{eq:r_GJ-Ip}\\
v_1^{n+1} & = & v^{n+\frac{3}{4}}+\frac{\Delta{t}}{2m}f^{n+1} \label{eq:v1_GJ-Ip}  \, . 
\end{eqnarray}\label{eq:Final_Split_GJ-Ip}\noindent
\end{subequations}
Here, ${\cal N}^n$ and ${\cal N}^{n+\frac{1}{2}}$ are both standard Normal random variables, such that $\sqrt{2\,\alpha\Delta{t}\,k_BT}\,{\cal N}^n=\beta^n$ and $\sqrt{2\,\alpha\Delta{t}\,k_BT}\,{\cal N}^{n+\frac{1}{2}}=\beta^{n+\frac{1}{2}}$, where $\beta^n$ is given by Eq.~(\ref{eq:FD_discrete}) and $\beta^{n+\frac{1}{2}}$ is given in and around Eq.~(\ref{eq:beta_half}). We note that $\mu_{1_-}=0$ $\Rightarrow$ $\mu_{6_-}=-1$, which from Eq.~(\ref{eq:Final_Split_GJ}) implies that $w_-^n=v_-^{n+1}$ in this special case.
Langevin simulations with the particular algorithm of Eq.~(\ref{eq:Final_Split_GJ-Ip}) will therefore, for linear systems, result in the important combination of time-step independent statistical results for averages of potential and kinetic energies, as well as the covariance of the two
\begin{subequations}
\begin{eqnarray}
\left\langle E_p(r^n)\right\rangle & = & \frac{k_BT}{\kappa} \\
\left\langle E_k(v_-^n)\right\rangle & = & \frac{k_BT}{m} \\
\left\langle E_p(r^n)E_k(v_-^n)\right\rangle & - & \left\langle E_p(r^n)\right\rangle\left\langle E_k(v_-^n)\right\rangle \; = \; 0  \, .
\end{eqnarray}\label{eq:Final}\noindent
\end{subequations}
The correct numerical evaluation of all three expressions in Eq.~(\ref{eq:Final}) may be important for calculations of thermodynamic quantities that rely on accurate calculations of both configurational coordinates as well as their cross-correlations; such as, e.g., specific heat, thermal pressure coefficient, instantaneous pressure, and pressure fluctuations in NVT simulations in molecular dynamics (see, e.g., Ref.~\cite{AllenTildesley}).


\end{document}